\RequirePackage{fix-cm}
\documentclass[final,10pt,onecolumn,floatfix,prd,authoryear,nofootinbib,aps,eqsecnum]{revtex4-2}
\usepackage{bm}
\usepackage{newtxmath,newtxtext}
\usepackage{graphicx}
\usepackage{textcomp}
\usepackage{booktabs}
\usepackage{dcolumn}
\usepackage{ragged2e}
\usepackage{hyperref}
\usepackage{xcolor}

\definecolor{ReviewGreen}{RGB}{0,110,48}
\definecolor{ReviewBlue}{RGB}{0,40,125}
\definecolor{ReviewPurple}{RGB}{116,35,150}

\usepackage{epsfig}
\usepackage{appendix}
\usepackage{orcidlink}
\usepackage{multirow}
\usepackage{tcolorbox}
\usepackage[section]{placeins}
\makeatletter
\long\def\@makecaption#1#2{%
  \par
  \vskip\abovecaptionskip
  \begingroup
    \small\rmfamily
    \noindent
    \@make@capt@title{#1}{#2}\par
  \endgroup
  \vskip\belowcaptionskip
}
\makeatother
\hypersetup{colorlinks=true, linkcolor=black, urlcolor=black, citecolor=black}
\newcommand{\bq}{\begin{equation}}
\newcommand{\eq}{\end{equation}}
\newcommand{\bqn}{\begin{eqnarray}}
\newcommand{\eqn}{\end{eqnarray}}
\newcommand{\nb}{\nonumber}
\newcommand{\lb}{\label}

\begin{document}

\title{Strong Coupling of Cosmological Perturbations in mLQC-I: The Hybrid Approach}
\date{Working revision: November 17, 2026}

\author{Jamal Saeed}
\email{Jamal$\_$Saeed1@baylor.edu}

\author{Rui Pan}
\email{rui$\_$pan1@baylor.edu}

\author{Anzhong Wang}
\email{anzhong$\_$wang@baylor.edu}
\thanks{Corresponding author}

\affiliation{GCAP-CASPER, Department of Physics and Astronomy, Baylor University, Waco, TX 76798-7316, USA}

\begin{abstract}
\begingroup
We investigate nonlinear perturbative consistency of scalar cosmological
perturbations in modified loop quantum cosmology I (mLQC-I) within the hybrid
approach. We express the classical quadratic and cubic Hamiltonians in a
common canonical representation before imposing the background constraint,
and extend the hybrid regularization to cubic order through a minimal
sector-preserving prescription. During a pre-bounce interval close to the
quantum-geometrical de Sitter phase, the nonlinear coefficients develop a
large enhancement. The $\alpha_I^3/\kappa$ contribution survives the assembled
leading ultraviolet cubic kernel for each fixed nondegenerate triangle in
the selected frequency assignment. The enhancement persists in the
Starobinsky, chaotic, polynomial-chaotic, and power-law-plateau examples,
while the formal suppression scale depends strongly on the potential
normalization. Within an assumed factorized quartic continuation, the
selected second-order-lapse contact term does not cancel the enhanced cubic
exchange in an analytic scattering family. The scalar quartic Legendre term is derived explicitly; its leading Hamiltonian contact contribution cancels the selected lapse contact at $V=K_\phi$, without establishing cancellation of the full amplitude. The omitted quartic sectors have
not been proved subleading, and the tiny formal ultraviolet scales do not
locate a physical cutoff. We also examine perturbations initialized directly
at the bounce. For the instantaneous quadratic initial state, the local cubic
diagnostic first reaches unity in the equilateral configuration at
$K_{\mathrm{pert},B}^{(3)}\simeq0.6483\,m_{\rm P}$; refined shape sampling
supports this ordering. Evolving the same state over
$0\le t/t_{\rm P}\le10$ gives a slightly lower equilateral crossing,
$K\simeq0.64626\,m_{\rm P}$. Additional independent frequency-assignment tests do not lower these thresholds on the sampled domain. A smooth fourth-order adiabatic ultraviolet tail can
be attached without changing the finite modes used in this test. These
results identify a pre-bounce nonlinear enhancement and a promising local
at-bounce-initial state with a finite-time evolution check. Broader evolution
and shape tests, renormalized backreaction, observational mode mapping on a
consistently normalized background, and a complete fourth-order construction
remain necessary to
assess the full interacting theory.
\par\endgroup
\end{abstract}

\maketitle
\section{Introduction}
\lb{sec:intro}

Loop quantum cosmology (LQC) replaces the classical big-bang singularity by a quantum bounce and provides a setting in which cosmological perturbations can be propagated through a Planck-curvature epoch \cite{Ashtekar:2011ni,Agullo:2023rqq}. Two widely used frameworks for treating cosmological perturbations are the dressed-metric and hybrid approaches. In both approaches, the homogeneous background is loop quantized, while the inhomogeneous perturbations are quantized in a Fock representation. In the dressed-metric approach, the perturbations propagate as quantum fields on an effective metric dressed by the quantum background \cite{Agullo:2012sh,Agullo:2012fc,Agullo:2013ai}, whereas in the hybrid approach the truncated background-plus-perturbation system is quantized as a constrained system, combining a polymeric quantization of the homogeneous sector with a Fock quantization of the perturbations \cite{Fernandez-Mendez:2012poe,Fernandez-Mendez:2013jqa,Fernandez-Mendez:2014raa}. 
The two formulations are related at the classical level, but their effective theories need not coincide once the background variables entering the perturbative Hamiltonian are regularized; this point has been analyzed explicitly in Refs.~\cite{Li:2022evi,Iteanu:2022zha}.

Modified loop quantum cosmologies (mLQCs) arise from regularization ambiguities in the gravitational Hamiltonian constraint. In standard LQC, the Euclidean and Lorentzian parts of the gravitational Hamiltonian are first related at the classical level in spatially flat FLRW spacetime, and only the resulting combined expression is quantized. A different, bottom-up route is to quantize these two contributions separately, more closely following their treatment in full LQG. This possibility was first investigated by Yang, Ding, and Ma in 2009, leading to two alternative quantizations that were later referred to as mLQC-I and mLQC-II \cite{Yang:2009fp}. The quantum dynamics of mLQC-I in the improved \(\bar{\mu}\) scheme was subsequently developed more systematically in Ref. \cite{Assanioussi:2018hee}. An independent top-down route from full LQG initially yielded a closely related effective Hamiltonian in a fixed-graph, \(\mu_0\)-type scheme \cite{Dapor:2017rwv}. {Dynamical-lattice constructions subsequently yielded related improved cosmological dynamics \cite{Han:2021cwb}. This relationship motivates comparison with mLQC-I, but does not identify their effective Hamiltonians or finite-area corrections.} {The effective dynamics of mLQC-I and mLQC-II were studied systematically} in Refs. \cite{Li:2018opr,Li:2018fco}, where the notation mLQC-I and mLQC-II was adopted. A distinctive feature of the ordering studied in the present paper is the intrinsically asymmetric bounce in mLQC-I: the post-bounce phase approaches classical general relativity, whereas the remote pre-bounce phase approaches a quantum-geometrical de Sitter spacetime \footnote{A complementary ordering of the two asymptotic phases in the framework of mLQC-I can arise in the Bianchi~I extension of the FLRW universe through the quantum damping of cosmological shear \cite{Gan:2025uvt,Gan:2026ivx}. In this case, the remote pre-bounce phase is asymptotically described by classical general relativity, whereas the post-bounce phase approaches a quantum-inflationary regime characterized by an effective Planck-scale cosmological constant. Since the strong-coupling enhancement found below is tied to evolution through the quantum-geometrical de Sitter pre-bounce phase of the ordering studied here, it is natural to ask whether the reversed ordering removes or substantially weakens this enhancement. We plan to investigate this possibility separately.}. Slow-roll inflation was subsequently shown to arise generically for kinetic-energy-dominated initial data at the bounce \cite{Li:2019ipm}, while the background evolution exhibits remarkable universal properties for a broad class of inflationary potentials \cite{Saeed:2024xhk}. A comprehensive review of the background dynamics and phenomenological implications of mLQC-I and mLQC-II is given in Ref. \cite{Li:2021mop}.

The distinctive pre-bounce dynamics of mLQC-I naturally raises the question of how quantum-geometrical effects are transmitted to cosmological perturbations and, ultimately, to observable quantities. Cosmological perturbations in mLQC-I have therefore been investigated in both the dressed-metric and hybrid approaches. Within the dressed-metric framework, the primordial power spectrum was first studied in Ref. \cite{Agullo:2018wbf}, followed by a systematic comparison of the scalar and tensor primordial spectra of LQC, mLQC-I, and mLQC-II in Ref. \cite{Li:2019qzr}. Within the hybrid framework, the perturbative construction was extended to mLQC-I in Ref. \cite{Gomar:2020orw}, and the corresponding time-dependent scalar and tensor masses and their relation to those of the dressed-metric formulation were analyzed in Ref. \cite{Garcia-Quismondo:2020wna}. The primordial scalar spectra of the three loop cosmologies were subsequently compared within the hybrid approach in Ref. \cite{Li:2020mfi}. These studies showed that the two perturbative frameworks can differ appreciably in the Planck-curvature regime, particularly for mLQC-I, making both the quantization prescription and the choice of the initial state potentially important. The pre-bounce perturbation dynamics and the selection of stable initial states were investigated further in Ref. \cite{Pan:2025ppa}. More recently, in our companion linear analysis (Paper I), we constructed high-accuracy mode functions in the hybrid approach, established the ultraviolet regularity of the relevant initial states, and studied gravitational particle creation throughout the pre-bounce, bounce, and post-bounce evolution \cite{SaeedPanWang2026PaperI}.

 Linear regularity, however, is not sufficient to establish the consistency of a
 perturbative description.  Near a nonsingular bounce, time-dependent background
 coefficients can enhance higher-order interactions even when the quadratic mode
 equation is regular.  The appropriate question is therefore whether cubic and higher-order contributions remain parametrically subleading to the quadratic dynamics throughout the regime in which the effective description is used.  Closely
 related strong-coupling consistency tests have played an important role in
 other nonsingular and nonstandard cosmologies
 \cite{JoyceKhoury2011,KoehnEtAl2016,DehghaniEtAl2025}.  In standard LQC, the
 second-order extension of the dressed-metric framework also showed that
 non-Gaussianity can be enhanced around the bounce while perturbation theory can
 remain under control for the states considered there \cite{Agullo:2017eyh} \footnote{For studies of primordial non-Gaussianity in LQC and related bouncing scenarios, see Refs. \cite{Agullo:2015aba,Agullo:2017eyh,Zhu:2017onp,Wu:2018sbr,Agullo:2020cvg,Agullo:2020fbw,Agullo:2021oqk}.}.
 These results make it essential to separate the existence of a regular
 background from the independent question of nonlinear perturbative control.

Beyond the question of perturbative consistency itself, recent work has emphasized that the observational interpretation of bounce-generated non-Gaussianity depends crucially on its momentum-space structure. Roshna and Sreenath showed for standard LQC that rapid primordial oscillations can be strongly averaged by the CMB transfer functions \cite{Roshna:2023}, whereas van Tent, Delgado, and Durrer demonstrated that smoother bouncing bispectrum templates can survive the projection and be tightly constrained by Planck data \cite{vanTent:2023}. In our companion nonlinear analysis (Paper~II), we investigate primordial non-Gaussianity in mLQC-I within the hybrid approach, including the finite-momentum bispectrum and correlation-function perturbativity for both bounce-defined and remote pre-bounce de Sitter initial states \cite{PanSaeedWang2026PaperII}.

The present paper addresses a complementary and logically prior question: whether the nonlinear hybrid Hamiltonian itself remains perturbatively controlled throughout the mLQC-I evolution. Our emphasis is therefore on the hierarchy between the quadratic and higher-order Hamiltonians and on the possible onset of strong coupling, rather than on the bispectrum as an observable. The ordering of operations is central to this analysis.  We first express the
classical quadratic and cubic Hamiltonians in one and the same hybrid canonical
representation, without using the classical Friedmann constraint to simplify
their coefficients.  We then regularize the background structures at the
operator/effective level and only afterward impose the effective background
constraint.  Schematically,
\bq
H_{\rm cl}=H_{\rm cl}^{(0)}+H_{\rm cl}^{(2)}+H_{\rm cl}^{(3)}+\cdots,
\lb{eq:Hcl-expansion}
\eq
is mapped according to
\bq
H_{\rm cl}^{(0)},\;H_{\rm cl}^{(2)},\;H_{\rm cl}^{(3)}
\quad\longrightarrow\quad
H_{\rm eff}^{(0)},\;H_{\rm eff}^{(2)},\;H_{\rm eff}^{(3)}.
\lb{eq:strategy}
\eq
This construction reproduces the quadratic hybrid mass of \cite{Fernandez-Mendez:2012poe,Fernandez-Mendez:2013jqa,Fernandez-Mendez:2014raa} on the
effective constraint surface and defines a minimal sector-preserving extension of
the same prescription to the inverse odd powers that first appear at cubic
order.

\begingroup
The analysis identifies two complementary regimes. In the pre-bounce
quantum de Sitter interval, the chosen nonlinear continuation develops large
background-dependent coefficients. The enhancement survives the assembled
leading ultraviolet cubic kernel for fixed nondegenerate shapes in the
selected frequency assignment. A test of enhanced cubic exchange against the
selected $n_{2I}^2$ contact term shows non-cancellation within an assumed
factorized quartic continuation \footnote{It should be noted that it does not establish completeness of the
leading four-point amplitude: omitted scalar-shift, transverse-shift, tensor,
and canonical-completion terms still require joint power counting.}. We explicitly derive the scalar quartic Legendre contribution fixed by the cubic Hamiltonian. Its leading contact term cancels the selected lapse contribution at $V/K_\phi=1$ in the Hamiltonian representation. This contact-sector cancellation is distinct from a complete amplitude calculation, which requires matching exchange and derivative-contact conventions and the omitted sectors. The very
small formal suppression scales also lie outside the weakly coupled overlap
required to interpret the ultraviolet reduction as a physical cutoff.

If the perturbative initial state is instead specified directly at the bounce, as commonly done for the homogenious and isotropic background \cite{Ashtekar:2011ni,Agullo:2023rqq}, this
pre-bounce preparation interval is bypassed. For the chosen instantaneous
quadratic state, the exact finite-mode cubic diagnostic admits a local
below-unity range, with an independently reproduced first equilateral
crossing at approximately $0.6483m_{\rm P}$. Refined shape grids support the
equilateral ordering at that scale. This selected diagnostic is neither an
operator norm. Additional conjugation-assignment tests support this threshold on the sampled domain. Its finite modes can
be preserved while adding a normalized fourth-order adiabatic ultraviolet
tail, but this construction alone does not establish interacting control
of the tail. A finite evolution check of the same state over
$0\le t/t_{\rm P}\le10$ finds a modest initial increase: the refined
equilateral crossing is $K\simeq0.64626m_{\rm P}$, about $0.31\%$ below
the bounce-only value. The finite triangle scan at five momentum scales
also attains its sampled maxima at the equilateral shape. However, broader evolution
and continuous shape tests, full renormalized backreaction, consistent
observational mode mapping, and the complete fourth-order hybrid theory
remain open.
\par\endgroup

This paper is therefore deliberately restricted to the hybrid formulation and to the strong-coupling problem. A parallel analysis in the dressed-metric approach is sufficiently distinct at nonlinear order to merit a separate companion study.

The remainder of the paper is organized as follows. In Sec.~II, we write the classical quadratic and cubic Hamiltonians in a common set of hybrid canonical variables, keeping the background constraint off shell. In Sec.~III, we implement the mLQC-I effective quantization, recover the quadratic hybrid theory, and construct the corresponding cubic Hamiltonian together with its consistency checks. Sec.~IV develops the perturbativity diagnostics and analyzes the strong-coupling problem for states prepared in the remote contracting phase, including the ultraviolet cubic kernel, potential dependence, and the minimal quartic consistency test. In Sec.~V, we consider the alternative prescription in which the perturbative state is specified directly at the bounce and examine its cubic perturbative window, ultraviolet completion, backreaction diagnostics, and mapping to observable scales. Sec.~VI summarizes our conclusions and discusses the remaining consistency requirements and future directions. Some useful algebraic relations and details of the canonical completion are collected in an Appendix.

Throughout this manuscript we use $\hbar=c=1$ and $\kappa=8\pi G$.
We denote the (unreduced) Planck mass and Planck time by
$m_{\rm P}\equiv G^{-1/2}$ and $t_{\rm P}\equiv G^{1/2}$, respectively.
We write $v=a^3$ and
$\Omega\equiv-\kappa\gamma a\pi_a/6$.
For spatial derivatives in the transformed hybrid Hamiltonian we adopt the
fiducial Euclidean convention
\bq
\partial^2\equiv\delta^{ij}\partial_i\partial_j,
\quad
(\partial f)^2\equiv\delta^{ij}\left(\partial_i f\right)\,\left(\partial_j f\right),
 \lb{eq:spatial-derivative-convention}
\eq
and all repeated spatial indices in those expressions are raised and lowered
with $\delta_{ij}$.

\section{Classical Hamiltonians in the hybrid canonical variables}
\lb{sec:classical}

For completeness and to make the present paper as self-contained as possible,
we reproduce in Secs.~II and III the parts of the canonical construction and
hybrid effective dynamics that are also developed in the corresponding
sections of Paper~II \cite{PanSaeedWang2026PaperII}.  Since the two papers are
based on the same nonlinear hybrid framework, some overlap is unavoidable.
We have nevertheless restricted the repeated material to the minimum needed
to establish the notation, canonical variables, and effective Hamiltonians
required for the strong-coupling analysis below.

\subsection{Quadratic Hamiltonian}
\lb{subsec:H2cl}

The quadratic Hamiltonian in the Mukhanov--Sasaki (MS) canonical pair
$(\nu,\pi_\nu)$ can be written as \cite{Fernandez-Mendez:2012poe,Fernandez-Mendez:2013jqa,Fernandez-Mendez:2014raa}
\bq
H_{\rm cl}^{(2)}
=\frac{N}{2a}\int d^3x\left[
\pi_\nu^2+(\partial\nu)^2+m_{\rm cl}^2\nu^2
\right].
\lb{eq:H2cl}
\eq
We use the off-shell classical mass in the canonical variables
$(a,\pi_a;\phi,p_\phi)$,
\bqn
m_{\rm cl}^2
=-\frac{27p_\phi^4}{2\pi_a^2a^6}
+\frac{5\kappa p_\phi^2}{2a^4}
+\frac{9p_\phi^2V}{\pi_a^2}
-\frac{12a p_\phi V_{,\phi}}{\pi_a}
-\frac{\kappa^2\pi_a^2}{72a^2}
+a^2V_{,\phi\phi}-\frac{\kappa}{2}a^2V.
\lb{eq:mcl-pia}
\eqn
No use of the classical background Hamiltonian constraint is made in
Eq.~(\ref{eq:mcl-pia}).  This is the form that will be carried into the
effective construction.
Using the classical relation
\bq
\pi_a=-\frac{6\Omega}{\kappa\gamma a},
\lb{eq:pia-Omega}
\eq
Eq.~(\ref{eq:mcl-pia}) can equivalently be written as
\bqn
m_{\rm cl}^{2}
=-\frac{3\kappa^2\gamma^2p_\phi^4}
{8v^{4/3}\Omega^2}
+\frac{5\kappa p_\phi^2}{2v^{4/3}}
+\frac{\kappa^2\gamma^2v^{2/3}p_\phi^2V}{4\Omega^2}
+\frac{2\kappa\gamma v^{2/3}p_\phi}{\Omega}V_{,\phi}
-\frac{\Omega^2}{2\gamma^2v^{4/3}}
+v^{2/3}V_{,\phi\phi}
-\frac{\kappa}{2}v^{2/3}V.
\lb{eq:mcl-Omega}
\eqn

\subsection{Cubic Hamiltonian of Agullo, Bolliet, and Sreenath}
\lb{subsec:H3Agullo}

\begingroup
A convenient off-shell third-order Hamiltonian for scalar perturbations in
the spatially flat gauge is given in Ref.~\cite{Agullo:2017eyh}.  The
background Friedmann constraint is not used to simplify either the quadratic
or cubic coefficients at this stage, which is essential for applying the
mLQC-I regularization consistently.  We also retain explicitly the momentum
redefinition made before the displayed cubic Hamiltonian in that reference,
writing
\par\endgroup
\bq
\delta\bar p_\phi
=\delta p_\phi^{\rm old}
-\frac{3p_\phi^2}{a\pi_a}\delta\phi.
\lb{eq:pbar-def}
\eq
Thus every $\delta p_\phi$ appearing in the displayed cubic Hamiltonian of
Ref.~\cite{Agullo:2017eyh} is understood here as
$\delta\bar p_\phi$.
Before transforming to the hybrid variables, the cubic Hamiltonian reads
\bqn
H_{\rm cl}^{(3)}=N\int d^3x\Bigg\{&&
\left(
\frac{9\kappa p_\phi^3}{4a^4\pi_a}
-\frac{27p_\phi^5}{2a^6\pi_a^3}
-\frac{3a^2p_\phi V_{,\phi\phi}}{2\pi_a}
+\frac{a^3V_{,\phi\phi\phi}}{6}
\right)(\delta\phi)^3
-\frac{3p_\phi}{2a^4\pi_a}(\delta\bar p_\phi)^2\delta\phi
-\frac{9p_\phi^3}{a^5\pi_a^2}\delta\bar p_\phi(\delta\phi)^2
\nb\\
&&-\frac{3a^2p_\phi}{2\pi_a}\delta\phi\,(\vec\partial\delta\phi)^2 
+\frac{3p_\phi^2}{Na\pi_a}(\delta\phi)^2\partial^2\chi
+\frac{3a^2p_\phi}{2N^2\kappa\pi_a}\delta\phi(\partial^2\chi)^2
+\frac{3p_\phi^2}{Na\pi_a}\delta\phi\,\partial_i\chi\,\partial^i\delta\phi
\nb\\
&&+\frac{1}{N}\delta\bar p_\phi\,\partial_i\delta\phi\,\partial^i\chi
-\frac{3a^2p_\phi}{2N^2\kappa\pi_a}
\delta\phi\,\partial_i\partial_j\chi\,\partial^i\partial^j\chi
\Bigg\}.
\lb{eq:H3-Agullo}
\eqn
In Eq.~(\ref{eq:H3-Agullo}) the derivatives and index contractions still
follow the original convention of Ref.~\cite{Agullo:2017eyh},
namely the physical FLRW spatial metric.  They will be converted to the
fiducial Euclidean convention below.
The scalar shift potential $\chi$ follows from the first-order constraints.
After restoring the momentum variable in Eq.~(\ref{eq:pbar-def}), its Fourier
mode is
\bqn
\widetilde\chi_{\bf k}
=\frac{3N\kappa}{k^2a}
\Bigg[\left(
\frac{p_\phi}{2}
-\frac{a^5V_{,\phi}}{\kappa\pi_a}
-\frac{3p_\phi^3}{\kappa a^2\pi_a^2}
\right)\delta\widetilde\phi_{\bf k}
-\frac{p_\phi}{\kappa a\pi_a}
\delta\widetilde{\bar p}_{\phi{\bf k}}
\Bigg].
\lb{eq:chi-bar}
\eqn

\subsection{Canonical transformation to the hybrid pair}
\lb{subsec:canonical-transform}

The MS canonical pair used in the hybrid approach is related
to the variables above by
\bq
\nu=a\,\delta\phi,
\lb{eq:nu-transform}
\eq
and
\bq
\pi_\nu
=\frac{\delta\bar p_\phi}{a}
-\frac{\kappa\pi_a}{6a}\nu.
\lb{eq:pinu-transform}
\eq
Equivalently,
\bq
\delta\phi=\frac{\nu}{a},
 \quad
\delta\bar p_\phi
=a\left(\pi_\nu+\frac{\kappa\pi_a}{6a}\nu\right).
\lb{eq:inverse-transform}
\eq
Using Eq.~(\ref{eq:pia-Omega}), we define the convenient shorthand
\bq
{\cal P}_\nu
\equiv
\pi_\nu-\frac{\Omega}{\gamma v^{2/3}}\nu,
\lb{eq:Pnu-def}
\eq
so that
\bq
\delta\bar p_\phi=v^{1/3}{\cal P}_\nu.
\lb{eq:pbar-Pnu}
\eq
The canonical transformation on the full background-plus-perturbation
phase space requires a little care.  In particular, the momentum
redefinition in Eq.~(\ref{eq:pbar-def}) and the transformation in
Eqs.~(\ref{eq:nu-transform}) and (\ref{eq:pinu-transform}) should be
regarded as a single transformation from the original perturbative
pair $(\delta\phi,\delta p_\phi^{\rm old})$ to $(\nu,\pi_\nu)$.  Explicitly, we have
\begin{equation}
 \nu=a\,\delta\phi,
  \quad
 \pi_\nu
 =
 \frac{\delta p_\phi^{\rm old}}{a}
 -\frac{3p_\phi^2}{a^2\pi_a}\,\delta\phi
 -\frac{\kappa\pi_a}{6}\,\delta\phi,
 \label{eq:full-pert-canonical-map}
\end{equation}
or equivalently,
\begin{equation}
 \delta\phi=\frac{\nu}{a},
  \quad
 \delta p_\phi^{\rm old}
 =
 a\pi_\nu
 +
 \left(
 \frac{\kappa\pi_a}{6}
 +\frac{3p_\phi^2}{a^2\pi_a}
 \right)\nu .
 \label{eq:full-pert-canonical-map-inverse}
\end{equation}

Because the coefficients in this transformation depend on the homogeneous
variables, completing it canonically on the full phase space induces
quadratic perturbative redefinitions of the homogeneous variables.
As shown in Appendix~\ref{app:canonical-completion}, these shifts modify
the background Hamiltonian $H_{\rm cl}^{(0)}$ at second perturbative
order and therefore contribute to the quadratic hybrid Hamiltonian.
Their insertion into $H_{\rm cl}^{(2)}$, however, starts at fourth order,
whereas their insertion into $H_{\rm cl}^{(3)}$ starts at fifth order.
Furthermore, after gauge fixing and imposition of the first-order
constraints the reduced perturbative Hamiltonian starts at quadratic
order, $H_{\rm pert}=H^{(2)}+H^{(3)}+\cdots$, so there is no independent
first-order contribution that could combine with the quadratic
background shifts to generate an additional cubic term.  Consequently,
within the minimal extension in which $\nu=a\,\delta\phi$ remains linear,
the full canonical completion generates no additional contribution to
the cubic Hamiltonian in Eq.~(\ref{eq:H3cl-hybrid}), given below.
We also define
\bq
\vartheta\equiv\frac{\chi}{N}.
\lb{eq:vartheta-def}
\eq
After converting all spatial contractions to the convention of
Eq.~(\ref{eq:spatial-derivative-convention}), the classical cubic Hamiltonian
becomes
\bqn
H_{\rm cl}^{(3)}
=N\int d^3x\Bigg\{&&
{\cal A}\nu^3
+{\cal B}\nu{\cal P}_\nu^2
-{\cal C}\nu^2{\cal P}_\nu
+{\cal B}\nu(\partial\nu)^2
-{\cal D}\left[
\nu^2\partial^2\vartheta
+\nu\,\partial_i\vartheta\,\partial^i\nu
\right]
+\frac{1}{v^{2/3}}{\cal P}_\nu\,
\partial_i\nu\,\partial^i\vartheta
\nb\\
&&+{\cal E}\nu\left[
(\partial_i\partial_j\vartheta)(\partial^i\partial^j\vartheta)
-(\partial^2\vartheta)^2
\right]
\Bigg\},
\lb{eq:H3cl-hybrid}
\eqn
where
\bqn
{\cal A}
\equiv-\frac{3\kappa^2\gamma p_\phi^3}{8v^2\Omega}
+\frac{\kappa^3\gamma^3p_\phi^5}{16v^2\Omega^3}
+\frac{\kappa\gamma p_\phi}{4\Omega}V_{,\phi\phi}
+\frac{1}{6}V_{,\phi\phi\phi}, 
\lb{eq:A-cl} \quad
{\cal B}\equiv \frac{\kappa\gamma p_\phi}{4v^{2/3}\Omega},
 \quad 
{\cal C}\equiv \frac{\kappa^2\gamma^2p_\phi^3}{4v^{4/3}\Omega^2},
\quad
{\cal D}\equiv\frac{\kappa\gamma p_\phi^2}{2v^{4/3}\Omega},
 \quad
{\cal E}\equiv\frac{\gamma p_\phi}{4v^{2/3}\Omega}.\quad
\eqn
The corresponding Fourier-space shift variable is
\bqn
\vartheta_{\bf k}
=\frac{3\kappa}{k^2}\Bigg\{
v^{-2/3}\left[
\frac{p_\phi}{2}
+\frac{\gamma v^2}{6\Omega}V_{,\phi}
-\frac{\kappa\gamma^2p_\phi^3}{12\Omega^2}
\right]\nu_{\bf k}
+\frac{\gamma p_\phi}{6\Omega}
{\cal P}_{\nu{\bf k}}
\Bigg\}.
\lb{eq:vartheta-cl}
\eqn
Equations~(\ref{eq:H3cl-hybrid})--(\ref{eq:vartheta-cl}) are the classical
cubic Hamiltonian written entirely in the same canonical representation used
for the hybrid quadratic theory, before any loop regularization is applied.

\section{mLQC-I effective quantization in the hybrid approach}
\lb{sec:effective}

\subsection{Background effective Hamiltonian}
\lb{subsec:H0eff}

Paper~I uses the classical homogeneous Hamiltonian
\bq
H_{\rm cl}^{(0)}
=-\frac{3\Omega^2}{\kappa v\gamma^2}
+\frac{p_\phi^2}{2v}+vV,
\lb{eq:H0cl}
\eq
and the mLQC-I effective replacement
\bqn
\Omega^2\rightarrow\Omega_{\rm I}^2
&=&-\frac{v^2\gamma^2}{\lambda^2}
\left\{
\sin^2(\lambda b)
-\frac{\gamma^2+1}{4\gamma^2}\sin^2(2\lambda b)
\right\}. ~~~~~~~
\lb{eq:OmegaI}
\eqn
Thus, we find
\bq
H_{\rm eff}^{(0)}
=-\frac{3\Omega_{\rm I}^2}{\kappa v\gamma^2}
+\frac{p_\phi^2}{2v}+vV.
\lb{eq:H0eff}
\eq
It is useful to define
\bq
\Lambda_{\rm I}\equiv\frac{v}{2\lambda}\sin(2\lambda b).
\lb{eq:LambdaI}
\eq
Then, the perturbative prescription adopted in Paper~I is
\bq
\frac{1}{\Omega^2}\longrightarrow\frac{1}{\Omega_{\rm I}^2},
 \quad
\frac{1}{\Omega}\longrightarrow
\frac{\Lambda_{\rm I}}{\Omega_{\rm I}^2}.
\lb{eq:PaperI-prescription}
\eq
The first rule is tied to the mLQC-I background regularization, while the
second is chosen so that the perturbative operator respects the background
superselection structure \cite{Li:2020mfi}.

\subsection{Quadratic effective Hamiltonian and consistency check}
\lb{subsec:H2eff}

Applying Eqs.~(\ref{eq:OmegaI}) and (\ref{eq:PaperI-prescription}) directly
to the off-shell expression (\ref{eq:mcl-Omega}) gives
\bqn
m_{\rm eff,raw}^{2}
=-\frac{3\kappa^2\gamma^2p_\phi^4}
{8v^{4/3}\Omega_{\rm I}^2}
+\frac{5\kappa p_\phi^2}{2v^{4/3}}
+\frac{\kappa^2\gamma^2v^{2/3}p_\phi^2V}
{4\Omega_{\rm I}^2}
+\frac{2\kappa\gamma v^{2/3}p_\phi\Lambda_{\rm I}}
{\Omega_{\rm I}^2}V_{,\phi}-\frac{\Omega_{\rm I}^2}{2\gamma^2v^{4/3}}+v^{2/3}V_{,\phi\phi}
-\frac{\kappa}{2}v^{2/3}V.
\lb{eq:meff-raw}
\eqn
The effective mass used in Paper~I is
\bqn
m_{\rm eff}^{2}
=\frac{\kappa p_\phi^2}{6v^{4/3}}
\left(19-\frac{3\kappa\gamma^2p_\phi^2}{\Omega_{\rm I}^2}\right)+v^{2/3}\left[
V_{,\phi\phi}
+\frac{2\kappa\gamma p_\phi\Lambda_{\rm I}}
{\Omega_{\rm I}^2}V_{,\phi}
-\frac{2\kappa}{3}V
\right].
\lb{eq:meff-PaperI}
\eqn
The two expressions differ only by the effective background constraint:
\bq
m_{\rm eff,raw}^{2}-m_{\rm eff}^{2}
=
\frac{\kappa\left(2\Omega_{\rm I}^2
+3\kappa\gamma^2p_\phi^2\right)}
{12v^{1/3}\Omega_{\rm I}^2}
H_{\rm eff}^{(0)}.
\lb{eq:quadratic-check}
\eq
Consequently, once the effective background constraint
$H_{\rm eff}^{(0)}=0$ is imposed \emph{after} quantization, the off-shell
canonical construction reproduces precisely the mass of Paper~I.  This
provides an important consistency check on the ordering of the procedure \footnote{It should be noted that it does not establish uniqueness of the nonlinear regularization or closure of the full nonlinear constraint algebra.}.
Thus, the quadratic effective Hamiltonian reads   
\bq
H_{\rm eff}^{(2)}
=\frac{N}{2v^{1/3}}\int d^3x\left[
\pi_\nu^2+(\partial\nu)^2+m_{\rm eff}^2\nu^2
\right].
\lb{eq:H2eff}
\eq

\subsection{Cubic extension of the hybrid prescription}
\lb{subsec:H3-quantization}

At cubic order one encounters odd inverse powers of the background
momentum that are not explicitly required in the quadratic theory.
Rather than introducing an independent polymerization prescription for
each such power, we adopt the same factor-ordering and
superselection-sector prescription used in the hybrid quantization of
the mLQC-I background and its perturbations
\cite{Garcia-Quismondo:2019dwa,Li:2020mfi}.  The same cubic extension is
also used in Paper~II, where it provides the interaction Hamiltonian for the
bispectrum and correlation-function perturbativity analysis
\cite{PanSaeedWang2026PaperII}.  We present the construction here as well
because it is independently required for the Hamiltonian-level strong-coupling
test carried out in the present paper.

For convenience, we define the background operators
\bq
\lb{eq3.10}
\widehat X_{\rm I}\equiv\widehat{\Omega_{\rm I}^{2}},
 \quad
\widehat L_{\rm I}\equiv\widehat{\Lambda_{\rm I}},
\eq
where $\widehat X_{\rm I}$ is the operator representing the squared
background momentum and $\widehat L_{\rm I}$ is the corresponding odd
operator chosen so as to preserve the same step-four superselection
sectors; the underlying background-operator construction and its
superselection structure are discussed in
Refs.~\cite{Garcia-Quismondo:2019dwa,Li:2020mfi}.
Guided by these constructions, we extend the same ordering principle to
arbitrary odd integer powers.  Even integer powers are represented by
powers of $\widehat X_{\rm I}$, whereas odd integer powers are represented
symmetrically as
\bq
\lb{eq3.11}
\widehat{\Omega^{\,2\ell+1}}
=
\left|\widehat X_{\rm I}\right|^{\ell/2}
\widehat L_{\rm I}
\left|\widehat X_{\rm I}\right|^{\ell/2},
 \quad
\ell\in{\mathbb Z}.
\eq

 In particular,
for $\ell=-1$ we have
\bq
\lb{eq3.12}
\widehat{\Omega^{-1}}
=
\left|\widehat X_{\rm I}\right|^{-1/2}
\widehat L_{\rm I}
\left|\widehat X_{\rm I}\right|^{-1/2}.
\eq
For sharply peaked semiclassical background states, this reduces to
\bq
\lb{eq3.13}
\frac{1}{\Omega}
\longrightarrow
\frac{\Lambda_{\rm I}}{\Omega_{\rm I}^{2}},
\eq
which reproduces the prescription already used in the hybrid mLQC-I
construction and in Paper~I
\cite{Li:2020mfi,SaeedPanWang2026PaperI}.

The cubic Hamiltonian additionally contains the inverse odd power
$\Omega^{-3}$.  Taking $\ell=-2$ in the same operator prescription gives
\bq
\lb{eq3.14}
\widehat{\Omega^{-3}}
=
\left|\widehat X_{\rm I}\right|^{-1}
\widehat L_{\rm I}
\left|\widehat X_{\rm I}\right|^{-1}.
\eq
Consequently, for sharply peaked semiclassical background states, we find
\begin{equation}
 {
 \frac{1}{\Omega^3}
 \longrightarrow
 \frac{\Lambda_{\rm I}}{\Omega_{\rm I}^{4}},
 }
 \lb{eq:Omega3-prescription}
\end{equation}

which fixes the inverse-cubic term appearing in the coefficient
${\cal A}$.

The same prescription also fixes the positive odd power of $\Omega$
that occurs in the hybrid momentum combination.  Indeed, setting
$\ell=0$ we find
\bq
\lb{eq3.16}
\widehat{\Omega}=\widehat L_{\rm I},
\eq
and therefore, at the effective level, we obtain
\begin{equation}
 {
 {\cal P}_{\nu{\rm I}}
 \equiv
 \pi_\nu
 -
 \frac{\Lambda_{\rm I}}{\gamma v^{2/3}}\nu .
 }
 \lb{eq:PnuI-def}
\end{equation}
The power $v^{-2/3}=a^{-2}$ follows directly from the canonical transformation in Eq.~(\ref{eq:pinu-transform}).
Thus, replacements
$\Omega^{-1}\rightarrow\Lambda_{\rm I}/\Omega_{\rm I}^{2}$,
$\Omega^{-3}\rightarrow\Lambda_{\rm I}/\Omega_{\rm I}^{4}$, and
$\Omega\rightarrow\Lambda_{\rm I}$ are not treated as independent
polymerization choices.  They follow from the same odd-power
factor-ordering prescription.

A useful organizational rule is to retain the composite shift variable
$\vartheta_{\rm I}$ in factorized form.  In particular, products
generated by $\vartheta_{\rm I}^{\,2}$ should not be expanded and then
reinterpreted as independent polymerizations of higher inverse powers.
The factors entering $\vartheta_{\rm I}$ are quantized first according
to the preceding prescription, and their products are subsequently
formed.  This preserves the operator factorization inherited from the
classical expression.

We stress that this construction does not establish a unique
quantization of the cubic theory.  Rather, it is the extension selected
by maintaining the same factor ordering and superselection-sector
prescription used in the hybrid mLQC-I quantization underlying the
quadratic theory.

 \subsection{Effective cubic Hamiltonian}
 \lb{subsec:H3eff}

 With the preceding prescription, we define
 \bqn
 {\cal A}_{\rm I}
 &=&-\frac{3\kappa^2\gamma p_\phi^3}{8v^2}
 \frac{\Lambda_{\rm I}}{\Omega_{\rm I}^2}
 +\frac{\kappa^3\gamma^3p_\phi^5}{16v^2}
 \frac{\Lambda_{\rm I}}{\Omega_{\rm I}^4}+\frac{\kappa\gamma p_\phi}{4}
 \frac{\Lambda_{\rm I}}{\Omega_{\rm I}^2}V_{,\phi\phi}
 +\frac{1}{6}V_{,\phi\phi\phi},
 \nb \\
 {\cal B}_{\rm I}
 &=&\frac{\kappa\gamma p_\phi}{4v^{2/3}}
 \frac{\Lambda_{\rm I}}{\Omega_{\rm I}^2},\quad
 {\cal C}_{\rm I}
 =\frac{\kappa^2\gamma^2p_\phi^3}
 {4v^{4/3}\Omega_{\rm I}^2},\quad
 {\cal D}_{\rm I}
 =\frac{\kappa\gamma p_\phi^2}{2v^{4/3}}
 \frac{\Lambda_{\rm I}}{\Omega_{\rm I}^2},\quad
 {\cal E}_{\rm I}
 =\frac{\gamma p_\phi}{4v^{2/3}}
 \frac{\Lambda_{\rm I}}{\Omega_{\rm I}^2}.\quad
 \eqn
 Then, the effective shift variable is
 \bqn
 \vartheta_{{\rm I},{\bf k}}
 =\frac{3\kappa}{k^2}\Bigg\{&&
 \left[
 \frac{p_\phi}{2v^{2/3}}
 +\frac{\gamma v^{4/3}}{6}
 \frac{\Lambda_{\rm I}}{\Omega_{\rm I}^2}V_{,\phi}
 -\frac{\kappa\gamma^2p_\phi^3}
 {12v^{2/3}\Omega_{\rm I}^2}
 \right]\nu_{\bf k}
 +\frac{\gamma p_\phi}{6}
 \frac{\Lambda_{\rm I}}{\Omega_{\rm I}^2}
 {\cal P}_{\nu{\rm I},{\bf k}}
 \Bigg\}.
 \lb{eq:vartheta-I}
 \eqn
 Hence, we find that the resulting effective cubic Hamiltonian takes the compact form
 \bqn
 H_{\rm eff}^{(3)}
 =N\int d^3x\Bigg\{&&
 {\cal A}_{\rm I}\nu^3
 +{\cal B}_{\rm I}\nu{\cal P}_{\nu{\rm I}}^2
 -{\cal C}_{\rm I}\nu^2{\cal P}_{\nu{\rm I}}
 +{\cal B}_{\rm I}\nu(\partial\nu)^2
 -{\cal D}_{\rm I}\left[
 \nu^2\partial^2\vartheta_{\rm I}
 +\nu\,\partial_i\vartheta_{\rm I}\,\partial^i\nu
 \right]
 +\frac{1}{v^{2/3}}{\cal P}_{\nu{\rm I}}
 \partial_i\nu\,\partial^i\vartheta_{\rm I}
 \nb\\
 &&+{\cal E}_{\rm I}\nu\left[
 (\partial_i\partial_j\vartheta_{\rm I})
 (\partial^i\partial^j\vartheta_{\rm I})
 -(\partial^2\vartheta_{\rm I})^2
 \right]
 \Bigg\}.
 \lb{eq:H3eff}
 \eqn
 The full effective Hamiltonian through cubic order is therefore
 \bq
 H_{\rm eff}
 =H_{\rm eff}^{(0)}+H_{\rm eff}^{(2)}+H_{\rm eff}^{(3)}+\cdots.
 \lb{eq:Heff-total}
 \eq
 The classical limit is immediate:
 \bq
 \Omega_{\rm I}^2\rightarrow\Omega^2,
  \quad
 \Lambda_{\rm I}\rightarrow\Omega,
  \quad
 H_{\rm eff}^{(3)}\rightarrow H_{\rm cl}^{(3)}.
 \lb{eq:classical-limit}
 \eq
The effective cubic Hamiltonian in Eq.~(\ref{eq:H3eff}) is the same
interaction Hamiltonian employed in Paper~II.  There it is used to compute
the primordial bispectrum and nonlinear corrections to the two-point
function, whereas here it is used to test the hierarchy of the nonlinear
Hamiltonians and the possible onset of strong coupling.  Since the
consistency of this Hamiltonian is essential to both analyses, we refer the consideraitons of
the Hermiticity, 
and classical low-curvature checks
to 
\cite{PanSaeedWang2026PaperII}.

\section{Consistency conditions and strong-coupling analysis}
\lb{sec:strong-coupling}

Before Eq.~(\ref{eq:H3eff}) can be used for any nonlinear observable, the
perturbative expansion must first be shown to remain under control.  We
therefore test the cubic interactions against
the quadratic dynamics along the mLQC-I background.  Since the quadratic
Hamiltonian contains momentum pairs whereas the cubic Hamiltonian couples
momentum triangles, a ratio of the total integrated Hamiltonians is not the
most useful diagnostic.  Schematically, the appropriate perturbative
validity condition is instead
\bq
\frac{|{\cal L}^{(3)}|_{R_{\rm cg}}}
{{\cal Q}_2(\eta,R_{\rm cg})}\ll1,
\lb{eq:schematic-strong-coupling}
\eq
where $R_{\rm cg}$ denotes a comoving coarse-graining scale and the precise
definitions are given below.

\subsection{Momentum-space representation and perturbativity diagnostics}
\label{subsec:momentum_strong_coupling}

The quadratic and cubic effective Hamiltonians derived above are written in
position space, whereas the mode functions obtained in Paper I are most
naturally expressed in Fourier space. In this subsection we establish the
connection between the two descriptions and introduce the quantities that
will be used below to test perturbative control and estimate the
strong-coupling scale.
We adopt the Fourier convention
\begin{equation}
 F(\eta,{\bf x})
 =
 \int_{\bf k} F_{\bf k}(\eta)e^{i{\bf k}\cdot{\bf x}},
  \quad
 \int_{\bf k}
 \equiv
 \int\frac{d^3k}{(2\pi)^3},
 \label{eq:Fourier_convention}
\end{equation}
so that
\begin{equation}
 \partial_i\longrightarrow ik_i,
  \quad
 \partial^2\longrightarrow -k^2 .
 \label{eq:derivative_fourier}
\end{equation}
For a real classical perturbation one has
$F_{-{\bf k}}=F_{\bf k}^{*}$.
At the quantum level we write, for example,
\begin{equation}
 \hat\nu(\eta,{\bf x})
 =
 \int_{\bf k}
 \left[
 \nu_k(\eta)\hat a_{\bf k}e^{i{\bf k}\cdot{\bf x}}
 +
 \nu_k^{*}(\eta)\hat a_{\bf k}^{\dagger}
 e^{-i{\bf k}\cdot{\bf x}}
 \right],
 \label{eq:nu_mode_expansion}
\end{equation}
\begingroup
where the mode function $\nu_k(\eta)$ is the one studied in Paper~I.
We use $[\hat a_{\bf k},\hat a^\dagger_{\bf q}]
=(2\pi)^3\delta^{(3)}({\bf k}-{\bf q})$ and the Wronskian normalization
$\nu_k\nu_k^{*\prime}-\nu_k^*\nu_k'=i$.
For comparison with the mode equation, it is convenient to choose
conformal time, corresponding to
\begin{equation}
 N=a=v^{1/3}.
 \label{eq:conformal_lapse}
\end{equation}
The quadratic Hamiltonian then takes the canonical form
\begin{equation}
 H_{\rm eff}^{(2)}
 =
 \frac12\int d^3x
 \left[
 \pi_\nu^2
 +(\partial\nu)^2
 +m_{\rm eff}^2\nu^2
 \right].
 \label{eq:H2_position}
\end{equation}
Its momentum-space representation is therefore
\begin{equation}
 H_{\rm eff}^{(2)}
 =
 \frac12\int_{\bf k}
 \left[
 \pi_{\nu,{\bf k}}\pi_{\nu,-{\bf k}}
 +
 \omega_k^2
 \nu_{\bf k}\nu_{-{\bf k}}
 \right],
 \label{eq:H2_momentum}
\end{equation}
where  $\omega_k^2\equiv k^2+m_{\rm eff}^2$. 
Then, Hamilton's equation gives, at linear order,
\begin{equation}
 {
 \pi_{\nu,{\bf k}}=\nu_{\bf k}' ,
 }
 \label{eq:pi_nu_mode}
\end{equation}
where a prime denotes a derivative with respect to conformal time. Thus,
both $\nu_{\bf k}$ and $\pi_{\nu,{\bf k}}$ are directly available from the
mode functions obtained in Paper I.
It is also useful to introduce
\begin{equation}
 \Gamma_{\rm I}
 \equiv
 \frac{\Lambda_{\rm I}}{\gamma v^{2/3}},
 \label{eq:GammaI_def}
\end{equation}
so that
\begin{equation}
 {\cal P}_{\nu{\rm I},{\bf k}}
 =
 \pi_{\nu,{\bf k}}
 -
 \Gamma_{\rm I}\nu_{\bf k}
 =
 \nu_{\bf k}'
 -
 \Gamma_{\rm I}\nu_{\bf k}.
 \label{eq:PnuI_fourier}
\end{equation}
The auxiliary variable $\vartheta_{\rm I}$ appearing in the cubic
Hamiltonian is correspondingly given by
\begin{align}
 \vartheta_{{\rm I},{\bf k}}
 =
 \frac{3\kappa}{k^2}
 \Bigg\{
 \left[
 \frac{p_\phi}{2v^{2/3}}
 +
 \frac{\gamma v^{4/3}}{6}
 \frac{\Lambda_{\rm I}}{\Omega_{\rm I}^2}
 V_{,\phi}
 -
 \frac{\kappa\gamma^2p_\phi^3}
 {12v^{2/3}\Omega_{\rm I}^2}
 \right]\nu_{\bf k}
 +
 \frac{\gamma p_\phi}{6}
 \frac{\Lambda_{\rm I}}{\Omega_{\rm I}^2}
 {\cal P}_{\nu{\rm I},{\bf k}}
 \Bigg\}.
 \label{eq:varthetaI_fourier}
\end{align}
The expression in Eq.~(\ref{eq:varthetaI_fourier}) applies to $k\neq0$.
The homogeneous $k=0$ mode belongs to the background sector and is not part
of the perturbative shift variable.

We next transform the cubic Hamiltonian. Let us first define
\bqn 
 \int_{123}
 \equiv
 \int_{{\bf k}_1}
 \int_{{\bf k}_2}
 \int_{{\bf k}_3},
 \quad
 \delta_{123}
 \equiv
 (2\pi)^3
 \delta^{(3)}
 \left(
 {\bf k}_1+{\bf k}_2+{\bf k}_3
 \right).
 \label{eq:triangle_notation}
\eqn 
For compactness, we also use
\begin{equation}
 \nu_i\equiv\nu_{{\bf k}_i},
  \quad
 {\cal P}_i\equiv
 {\cal P}_{\nu{\rm I},{\bf k}_i},
  \quad
 \vartheta_i\equiv
 \vartheta_{{\rm I},{\bf k}_i}.
 \label{eq:short_momentum_notation}
\end{equation}
The position-space expression obtained above can then be written as
\begin{equation}
 {
 H_{\rm eff}^{(3)}
 =
 N\int_{123}
 \delta_{123}\,
 {\cal V}_{123},
 }
 \label{eq:H3_momentum_general}
\end{equation}
where
\begin{align}
 {\cal V}_{123}
 =&{}
 {\cal A}_{\rm I}\nu_1\nu_2\nu_3
 +
 {\cal B}_{\rm I}\nu_1{\cal P}_2{\cal P}_3
 -
 {\cal C}_{\rm I}\nu_1\nu_2{\cal P}_3
 -
 {\cal B}_{\rm I}
 ({\bf k}_2\cdot{\bf k}_3)
 \nu_1\nu_2\nu_3
 +
 {\cal D}_{\rm I}k_3^2
 \nu_1\nu_2\vartheta_3
  \notag\\
 &+{\cal D}_{\rm I}
 ({\bf k}_2\cdot{\bf k}_3)
 \nu_1\vartheta_2\nu_3
 -
 \frac{1}{v^{2/3}}
 ({\bf k}_2\cdot{\bf k}_3)
 {\cal P}_1\nu_2\vartheta_3
 +
 {\cal E}_{\rm I}
 \left[
 ({\bf k}_2\cdot{\bf k}_3)^2
 -
 k_2^2k_3^2
 \right]
 \nu_1\vartheta_2\vartheta_3 .
 \label{eq:cubic_vertex_kernel}
\end{align}
The delta function in Eq.~(\ref{eq:H3_momentum_general}) imposes the
usual triangle condition
\begin{equation}
 {\bf k}_1+{\bf k}_2+{\bf k}_3=0.
 \label{eq:triangle_condition}
\end{equation}
Equation~(\ref{eq:cubic_vertex_kernel}) is written in an unsymmetrized
form that follows directly from the position-space Hamiltonian. Since
the integration measure and $\delta_{123}$ are invariant under
permutations of the three momenta, one may equivalently replace
${\cal V}_{123}$ by the completely symmetrized kernel
\begin{equation}
 {\cal V}_{123}^{\rm sym}
 =
 \frac{1}{3!}
 \sum_{\sigma\in S_3}
 {\cal V}_{\sigma(1)\sigma(2)\sigma(3)} ,
 \label{eq:sym_vertex}
\end{equation}
which will be particularly convenient for the full-kernel analysis below.

Before proceeding, it is useful to clarify the relation between the
Hamiltonian formulation used here and the Lagrangian criteria commonly
employed in analyses of strong coupling in bouncing cosmologies
\cite{KoehnEtAl2016,JoyceKhoury2011,DehghaniEtAl2025}. To this end, we first
write the Hamiltonian $H$ schematically as
\begin{equation}
 H=H^{(2)}+H^{(3)}+\cdots .
 \label{eq:H_expansion_sc}
\end{equation}
Then, solving the momentum--velocity relation perturbatively as
$\pi_\nu=\pi_\nu^{(1)}+\pi_\nu^{(2)}+\cdots$, with
$\pi_\nu^{(1)}=\nu'$, we find that the terms involving $\pi_\nu^{(2)}$ cancel in the
Legendre transform at cubic order. Consequently, we have
\begin{equation}
 {
 {\cal L}^{(3)}
 =
 -{\cal H}^{(3)}
 \big|_{\pi_\nu=\nu'}
 }
  \quad
 \text{to cubic order},
 \label{eq:L3_H3_relation}
\end{equation}
whereas
\begin{equation}
 {\cal L}^{(2)}
 =
 \frac12
 \left[
 (\nu')^2
 -(\partial\nu)^2
 -m_{\rm eff}^2\nu^2
 \right].
 \label{eq:L2_nu}
\end{equation}
Thus the mode functions of Paper I can be inserted directly into the
cubic Hamiltonian, with $\pi_{\nu,{\bf k}}=\nu_{\bf k}'$, to obtain the
corresponding cubic Lagrangian at the order required here.

A useful first diagnostic of perturbative control is the ratio between
the magnitude of cubic and quadratic terms. In position space such a
comparison should be made in terms of the typical real-space amplitude
of the perturbations rather than a single Fourier coefficient
\cite{DehghaniEtAl2025}. We therefore introduce a coarse-graining radius
$R_{\rm cg}$ and the associated characteristic comoving scale
$k_{\rm cg}\equiv R_{\rm cg}^{-1}$.  \begingroup
For a time-independent isotropic window function
$\widetilde W(kR_{\rm cg})$, we choose
$\int d\ln k\,|\widetilde W(kR_{\rm cg})|^2=1$ on its support and define
\par\endgroup
\bqn 
 \nu_{\rm cg}^2(\eta;R_{\rm cg})
 =
 \int_0^\infty\frac{dk}{k}
 |\widetilde W(kR_{\rm cg})|^2
 \Delta_\nu^2(k,\eta),
 \Delta_\nu^2(k,\eta)
 \equiv
 \frac{k^3}{2\pi^2}|\nu_k(\eta)|^2,
 \label{eq:nu_coarse}
\eqn 
or similarly,
\begin{align}
 {\Pi_{\rm cg}^{\,2}}
 &=
 \int_0^\infty\frac{dk}{k}
 |\widetilde W(kR_{\rm cg})|^2
 \frac{k^3}{2\pi^2}|\nu_k'|^2 ,
 \label{eq:nup_coarse}
 \\
 (\partial\nu)_{\rm cg}^2
 &=
 \int_0^\infty\frac{dk}{k}
 |\widetilde W(kR_{\rm cg})|^2
 k^2
 \frac{k^3}{2\pi^2}|\nu_k|^2 ,
 \label{eq:gradnu_coarse}
 \\
 {\cal P}_{\rm cg}^2
 &=
 \int_0^\infty\frac{dk}{k}
 |\widetilde W(kR_{\rm cg})|^2
 \frac{k^3}{2\pi^2}
 \left|
 \nu_k'-\Gamma_{\rm I}\nu_k
 \right|^2 .
 \label{eq:P_coarse}
\end{align}

For $\vartheta_{\rm I}$ we analogously define
\begin{equation}
 (\partial^n\vartheta_{\rm I})_{\rm cg}^2
 =
 \int_0^\infty\frac{dk}{k}
 |\widetilde W(kR_{\rm cg})|^2
 k^{2n}
 \frac{k^3}{2\pi^2}
 |\vartheta_{{\rm I},k}|^2,
 \label{eq:vartheta_coarse}
\end{equation}
with $n=0$ defining $\vartheta_{\rm cg}$ itself.  For a narrow window centered on $k_{\rm cg}=R_{\rm cg}^{-1}$, these
expressions reduce, up to factors of order unity, to
\bqn 
 \nu_{\rm cg}
 \sim
 \left(\frac{k_{\rm cg}^3}{2\pi^2}\right)^{1/2}
 |\nu_{k_{\rm cg}}|,
  \quad
 {\Pi_{\rm cg}}
 \sim
 \left(\frac{k_{\rm cg}^3}{2\pi^2}\right)^{1/2}
 |\nu_{k_{\rm cg}}'|,
  \quad (\partial\nu)_{\rm cg}
 \sim
 k_{\rm cg}\nu_{\rm cg} .
 \label{eq:narrow_window_estimates}
\eqn 
\begingroup
These relations provide a narrow-window estimate connecting the mode
functions of Paper~I to typical real-space amplitudes. Here
${\Pi_{\rm cg}^{\,2}}$ denotes the variance of the coarse-grained derivative
field, not the square of the derivative of the root-mean-square (RMS) amplitude
$\nu_{\rm cg}$. Products of such amplitudes below are order-of-magnitude
estimates, not exact expectation values of cubic operators; an odd moment
vanishes in a centered Gaussian state.
\par\endgroup

Since the signed quadratic Lagrangian can pass through zero as a result
of cancellations between its kinetic, gradient, and mass terms, it is
more useful for order-of-magnitude estimates to introduce the positive
quadratic norm
\begin{equation}
 {\cal Q}_2(\eta,R_{\rm cg})
 \equiv
 \frac12
 \left[
 {\Pi_{\rm cg}^{\,2}}
 +(\partial\nu)_{\rm cg}^2
 +|m_{\rm eff}^2|\nu_{\rm cg}^2
 \right].
 \label{eq:Q2_norm}
\end{equation}

\begingroup
For analytic estimates, a clearly dominant quadratic contribution may approximate this norm. All quoted numerical ratios use the full positive norm and, for triangles, its arithmetic mean over the three legs.
\par\endgroup

Writing the cubic Hamiltonian density inside the braces of
$H_{\rm eff}^{(3)}$ as
\begin{equation}
 {\mathfrak h}^{(3)}
 =
 \sum_A{\mathfrak h}^{(3)}_A ,
 \label{eq:h3_terms}
\end{equation}
we define the contribution of each cubic operator to the perturbativity
parameter by
\begin{equation}
 {
 {\cal R}_A(\eta,R_{\rm cg})
 \equiv
 \frac{|{\cal L}^{(3)}_A|_{R_{\rm cg}}}
 {{\cal Q}_2(\eta,R_{\rm cg})}
 =
 \frac{N\left|{\mathfrak h}^{(3)}_A\right|_{R_{\rm cg}}}
 {{\cal Q}_2(\eta,R_{\rm cg})} .
 }
 \label{eq:RA_definition}
\end{equation}
For $N=a$, the individual estimates are schematically
\begin{align}
 N|{\mathfrak h}^{(3)}_{\cal A}|_{R_{\rm cg}}
 &\sim
 a|{\cal A}_{\rm I}|\nu_{\rm cg}^3,
 \nonumber\\
 N|{\mathfrak h}^{(3)}_{\cal B_1}|_{R_{\rm cg}}
 &\sim
 a|{\cal B}_{\rm I}|\nu_{\rm cg}{\cal P}_{\rm cg}^2,
 \nonumber\\
 N|{\mathfrak h}^{(3)}_{\cal C}|_{R_{\rm cg}}
 &\sim
 a|{\cal C}_{\rm I}|\nu_{\rm cg}^2{\cal P}_{\rm cg},
 \nonumber\\
 N|{\mathfrak h}^{(3)}_{\cal B_2}|_{R_{\rm cg}}
 &\sim
 a|{\cal B}_{\rm I}|
 \nu_{\rm cg}(\partial\nu)_{\rm cg}^2,
 \nonumber\\
 N|{\mathfrak h}^{(3)}_{{\cal D}_1}|_{R_{\rm cg}}
 &\sim
 a|{\cal D}_{\rm I}|
 \nu_{\rm cg}^2(\partial^2\vartheta_{\rm I})_{\rm cg},
 \nonumber\\
 N|{\mathfrak h}^{(3)}_{{\cal D}_2}|_{R_{\rm cg}}
 &\sim
 a|{\cal D}_{\rm I}|
 \nu_{\rm cg}
 (\partial\vartheta_{\rm I})_{\rm cg}
 (\partial\nu)_{\rm cg},
 \nonumber\\
 N|{\mathfrak h}^{(3)}_{\cal P}|_{R_{\rm cg}}
 &\sim
 a v^{-2/3}
 {\cal P}_{\rm cg}
 (\partial\vartheta_{\rm I})_{\rm cg}
 (\partial\nu)_{\rm cg} .
 \label{eq:cubic_amplitude_estimates}
\end{align}
For the last term in $H_{\rm eff}^{(3)}$, it is preferable to retain
its tensor structure,
\begin{equation}
 {\cal S}_{\vartheta}
 \equiv
 (\partial_i\partial_j\vartheta_{\rm I})
 (\partial^i\partial^j\vartheta_{\rm I})
 -
 (\partial^2\vartheta_{\rm I})^2 ,
 \label{eq:S_vartheta}
\end{equation}
and estimate
\begin{equation}
 N|{\mathfrak h}^{(3)}_{\cal E}|_{R_{\rm cg}}
 \sim
 a|{\cal E}_{\rm I}|
 \nu_{\rm cg}
 |{\cal S}_{\vartheta}|_{R_{\rm cg}} .
 \label{eq:Eterm_estimate}
\end{equation}
For a narrow band of modes,
$|{\cal S}_{\vartheta}|_{R_{\rm cg}}$ is at most of order
$k_{\rm cg}^4\vartheta_{\rm cg}^2$, with an additional angular dependence determined
by the momentum triangle. It should therefore not be replaced
prematurely by the difference of two identical scalar estimates.

\begingroup
A conservative sufficient smallness test for the separate cubic
contributions in this canonical representation is
\par\endgroup
\begin{equation}
 {\cal R}_A(\eta,R_{\rm cg})\ll 1
 \label{eq:pert_control_each}
\end{equation}
\begingroup
for all relevant cubic operators \footnote{This tests the chosen cubic diagnostic;
it is not a representation-independent necessary condition for weak
coupling. Constraint identities, integrations by parts, and field
redefinitions can cancel or redistribute separate terms, and cubic
smallness alone does not establish control at higher orders.}. Examining
individual $\mathcal R_A$ helps identify the source of an enhancement.
The full result,
\begin{equation}
 {\cal R}_3(\eta,R_{\rm cg})
 =
 \frac{
 |{\cal L}^{(3)}|_{R_{\rm cg}}
 }{
 {\cal Q}_2(\eta,R_{\rm cg})
 },
 \label{eq:R3_total}
\end{equation}
can subsequently be evaluated while retaining possible cancellations
among different interaction terms. This distinction is important in
bouncing cosmologies, where apparently singular individual
coefficients need not lead to singular physical interactions
\cite{KoehnEtAl2016,DehghaniEtAl2025}.

For the numerical analysis of general momentum triangles, the
single-scale definition above must be extended to configurations with
$k_1$, $k_2$, and $k_3$ not necessarily equal.  We do this in a
permutation-symmetric manner.  For each leg of the triangle, we define
\begin{equation}
 w_i\equiv
 \left(\frac{k_i^3}{2\pi^2}\right)^{1/2},
  \quad i=1,2,3,
 \label{eq:wi-triangle}
\end{equation}
and the corresponding narrow-band quadratic contribution
\begin{equation}
 {\cal Q}_{2,i}
 \equiv
 \frac12 w_i^2
 \left[
 |\nu'_{k_i}|^2+
 \left(k_i^2+|m_{\rm eff}^2|\right)|\nu_{k_i}|^2
 \right].
 \label{eq:Q2i-triangle}
\end{equation}
We then associate with the triangle the symmetric quadratic norm
\begin{equation}
 {
 {\cal Q}_{2,\triangle}
 \equiv
 \frac13\sum_{i=1}^{3}{\cal Q}_{2,i}.
 }
 \label{eq:Q2-triangle}
\end{equation}
This definition treats the three participating modes on an equal footing.
In the equilateral limit it equals the quadratic norm per logarithmic
wavenumber evaluated at that mode. Equation~(\ref{eq:Q2_norm}), in contrast,
is its window integral. The two agree in the normalized delta-window limit,
or when the integrand is constant across the window.
\par\endgroup

The corresponding cubic quantity is obtained from the completely
symmetrized kernel in Eq.~(\ref{eq:sym_vertex}) by making the narrow-band
replacements
\begin{equation}
 \nu_i\rightarrow w_i\nu_{k_i},
  \quad
 {\cal P}_i\rightarrow
 w_i{\cal P}_{\nu{\rm I},k_i},
  \quad
 \vartheta_i\rightarrow
 w_i\vartheta_{{\rm I},k_i}.
 \label{eq:triangle-narrow-band}
\end{equation}
We then define
\begin{equation}
 {
 {\cal R}_3^\triangle(\eta;k_1,k_2,k_3)
 \equiv
 \frac{
 a\,|{\cal V}^{\rm sym}_{123}|_{\rm cg}
 }{
 {\cal Q}_{2,\triangle}
 } .
 }
 \label{eq:R3-triangle}
\end{equation}

\begingroup
Permutation symmetrization preserves the chosen frequency assignment. For equilateral ultraviolet legs, denoting an unstarred positive-frequency mode by $+$ and its conjugate by $-$, the two independent coefficients up to conjugation and permutation are
\begin{equation}
g_{+++}=\frac{3\alpha_I}{4}-\frac{3\alpha_I^3}{8\kappa},\qquad
g_{++-}=-\frac{7\alpha_I}{12}+\frac{\alpha_I^3}{8\kappa}.
\end{equation}
At $\alpha_I^2=2\kappa$, the former vanishes whereas the latter equals $-\alpha_I/3$. Thus, cancellation of the selected contraction does not imply smallness of every interaction matrix element.
\par\endgroup

\begingroup
The individual operator contributions
${\cal R}_A^\triangle$ are defined by retaining the corresponding term in
${\cal V}^{\rm sym}_{123}$. These substitutions select the product of the
three unstarred mode functions, including their derivatives. They include
all displayed cubic terms for that frequency assignment, but not the
independent assignments with one or more conjugated modes. 
The bounce-state analysis below additionally evaluates the independent conjugation assignments. Thus
${\cal R}_3^\triangle$ is a state-dependent three-mode kernel diagnostic,
not an operator norm or a bound on every interaction matrix element.
\par\endgroup

\begingroup
This prescription is invariant under permutations of $(k_1,k_2,k_3)$.
Its equilateral limit uses the same per-logarithmic-scale narrow-band
normalization as above, rather than an exact finite-window integral.
The arithmetic average in
Eq.~(\ref{eq:Q2-triangle}) also ensures that a configuration
containing a very soft leg is normalized by the {arithmetic mean of the quadratic
contributions} of the three participating modes rather than by the
smallest individual contribution.
The same quantities can be used to define a first, state-dependent
estimate of the scale at which the perturbative expansion ceases to be
small.  In the ultraviolet, where the propagation is luminal,
\begin{equation}
 \omega_k\simeq k,
  \quad
 c_s^2=1,
 \label{eq:UV_luminal}
\end{equation}
there is no enhancement associated with a small sound speed. For each
interaction one may determine the comoving scale $k_{{\rm pert},A}$ at
which
\begin{equation}
 {\cal R}_A\!\left(\eta,R_{\rm cg}=k_{{\rm pert},A}^{-1}\right)\sim1,
 \label{eq:kpert_definition}
\end{equation}
and define the corresponding physical scale by
\begin{equation}
 \Lambda_{{\rm pert},A}(\eta)
 \equiv
 \frac{k_{{\rm pert},A}(\eta)}{a(\eta)} .
 \label{eq:LambdapertA_definition}
\end{equation}
The most restrictive interaction determines the preliminary
perturbative-breakdown scale
\begin{equation}
 {
 \Lambda_{\rm pert}(\eta)
 =
 \min_A\Lambda_{{\rm pert},A}(\eta).
 }
 \label{eq:Lambda_pert_definition}
\end{equation}
%
%
Because $\Lambda_{\rm pert}$ is obtained after inserting the mode amplitudes,
it is state dependent and should not be identified with a physical EFT
strong-coupling cutoff.  Canonically normalized ultraviolet vertices can
instead be used to define candidate operator-level suppression scales.  As
shown in Sec.~IV.D, these scales must still be checked against the domain of
validity of the ultraviolet reduction, coherent cancellations in the full
kernel, and possible redundant interactions before a final physical cutoff
$\Lambda_{\rm sc}$ can be assigned.  Agreement between the state-dependent
and operator-level diagnostics at the level of parametric scaling would
provide an important consistency check.

An independent measure of the characteristic background energy scale is
also required. Following the strategy commonly used in analyses of
nonsingular bouncing cosmologies, we define the dynamical scale
\begin{equation}
 E_{\rm dyn}(\eta)
 \equiv
 \max
 \left\{
 |H|,
 |\dot H|^{1/2},
 \left|
 \frac{\ddot\phi}{\dot\phi}
 \right|,
 \ldots
 \right\}.
 \label{eq:E_background}
\end{equation}
\begingroup
Freezing the background also requires the retained interaction coefficients to vary little during a mode oscillation. For each nonzero coefficient $c_A(t)$ entering the local reduction, the condition is
\begin{equation}
\frac{|\dot c_A|}{|c_A|p}\ll1,\qquad p=k/a.
\end{equation}
\begingroup
Here dots denote proper-time derivatives. The ratio
$\ddot\phi/\dot\phi$ is used only where $\dot\phi\ne0$ and is relevant
when the retained reduction divides by $\dot\phi$. Its divergence at an
inflaton turning point need not signal a divergence of the exact mode
Hamiltonian; the regular variables must be examined there. At the bounce,
$H=0$, while $|\dot H|^{1/2}$ generally remains finite.
\par\endgroup  In addition, the bounce
occurs at a finite inflaton energy density, so we monitor separately the scale
\begin{equation}
 E_{\rho_\phi}(\eta)\equiv \rho_\phi(\eta)^{1/4}.
 \label{eq:E_rhophi}
\end{equation}
\begingroup
If a physical ultraviolet cutoff $\Lambda_{\rm sc}$ is established for
the retained effective theory, fluctuation frequencies and the relevant
background variation scales should remain well below it. In the
ultraviolet regime this gives the proposed hierarchy
\par\endgroup
\begingroup
\begin{equation}
 {
 E_{\rm dyn}(\eta),\quad
 \frac{k}{a(\eta)}
 \ll
 \Lambda_{\rm sc}(\eta)
 }
 \label{eq:strong_coupling_hierarchy}
\end{equation}
%
for the modes and regular background scales to which this description
applies. We monitor $E_{\rho_\phi}=\rho_\phi^{1/4}$ separately as a
background dimensional scale. A further requirement
$E_{\rho_\phi}\ll\Lambda_{\rm sc}$ needs a model-specific effective-field-theory
argument and is not assumed to be a universal necessary condition.
Before a physical cutoff is determined, $\Lambda_{\rm pert}$ indicates
where the specified state-dependent cubic diagnostic becomes large.
\par\endgroup

We emphasize that the background Hamiltonian satisfies
$H_{\rm eff}^{(0)}=0$ on the effective constraint surface. Therefore,
unlike in an unconstrained matter theory, the ratio
$H^{(2)}/H^{(0)}$ is not an appropriate diagnostic of perturbative
control in the present canonical formulation. Equations
(\ref{eq:E_background}) and (\ref{eq:E_rhophi}) provide corresponding
background scales without encountering this problem.

Finally, the criterion
${\cal R}_3\ll1$ should be regarded as the first perturbative
diagnostic rather than a complete quantum proof of weak coupling.
The cubic-to-quadratic ratio provides an estimate of the scale at which
nonlinear and loop corrections may cease to be suppressed
\cite{JoyceKhoury2011,KoehnEtAl2016,DehghaniEtAl2025}. Higher-order correlators
and genuine loop corrections to the two-point function provide additional
tests once a perturbatively controlled regime has been established.
Complementary correlation-function diagnostics provide another test.
Paper~II carries out such an analysis by computing the finite-momentum
\begingroup
bispectrum and a retained nonlinear-conversion correction to the
curvature power spectrum \cite{PanSaeedWang2026PaperII}. This correction
does not include every contribution at that perturbative order, in
particular the full contributions with two cubic insertions or one quartic
insertion and their renormalization.
\par\endgroup  That analysis is deliberately
distinct from the Hamiltonian-level criteria used here: a small correction
to a selected correlator does not by itself establish the absence of a
strong-coupling scale in the interaction Hamiltonian.
In particular, the
usual estimate
\begin{equation}
 \frac{{\cal L}^{(3)}}{{\cal L}^{(2)}}
 \sim
 f_{\rm NL}{\cal R}
 \label{eq:fNL_perturbativity}
\end{equation}
\begingroup
is a heuristic relation in a specified state, scale range, and time
regime, not a general identity or a complete loop criterion. Luminal
propagation alone does not guarantee small nonlinear interactions.
\par\endgroup

The remainder of this section implements these consistency checks in a
progressive way.  We first simplify the interaction hierarchy on the effective
background constraint surface and evaluate the state-dependent cubic
perturbativity ratios using the numerical mode functions of Paper~I.  We then
extract the strict-ultraviolet operator scaling, test the complete symmetrized
cubic kernel before the ultraviolet reduction is imposed, and follow the
enhanced interaction through a minimal quartic contact-exchange check.  \begingroup
These steps test whether an apparent enhancement survives assembling
more interaction terms and momentum configurations. Their restricted
scope must be retained when assessing a physical nonlinear obstruction.
A complete
fourth-order hybrid construction is still required for a robust physical cutoff
within a fully specified nonlinear hybrid prescription.
\par\endgroup 

\subsection{Analytic hierarchy of the cubic interactions}
\lb{subsec:cubic-hierarchy}

Before evaluating the individual perturbativity parameters numerically,
it is useful to simplify the interaction coefficients on the effective
background constraint surface.  This step is performed only after the
quantization prescription has been fixed, in accordance with the
procedure adopted above.
Let us first define the inflaton kinetic-energy density by
\bqn 
 K_\phi\equiv\frac12\dot\phi^{\,2},
  \quad
 \sigma_{\rm I}\equiv\frac{\Lambda_{\rm I}}{v}
 =\frac{\sin(2\lambda b)}{2\lambda},\quad
 \rho_\phi= K_\phi+V
 =\frac12\dot\phi^{\,2}+V,
\eqn 
here and throughout the strong-coupling analysis, $\rho_\phi$ denotes the
inflaton energy density.  It is kept distinct from the effective
pre-bounce cosmological-constant density $\rho_\Lambda$ that supports the
asymptotic de Sitter phase of mLQC-I.  In particular, the inverse-background
factors entering the perturbative Hamiltonian are controlled by
$\rho_\phi$, not by $\rho_\phi+\rho_\Lambda$.
The effective background constraint gives
\begin{equation}
 \Omega_{\rm I}^{2}
 =
 \frac{\kappa\gamma^2}{3}v^2\rho_\phi .
\end{equation}
\begingroup
These inverse relations apply where $\rho_\phi\ne0$ on the domain
of the inverse-background prescription. We use the proper-time identity
$p_\phi=v\dot\phi$. Consequently, we find
\par\endgroup
\bqn 
 \frac{1}{\Omega_{\rm I}^{2}}
 =
 \frac{3}{\kappa\gamma^2v^2\rho_\phi},
  \quad
 \frac{\Lambda_{\rm I}}{\Omega_{\rm I}^{2}}
 =
 \frac{3\sigma_{\rm I}}
 {\kappa\gamma^2v\rho_\phi},
\quad
 \frac{\Lambda_{\rm I}}{\Omega_{\rm I}^{4}}
 =
 \frac{9\sigma_{\rm I}}
 {\kappa^2\gamma^4v^3\rho_\phi^2}.
\eqn 
The cubic coefficients therefore reduce exactly to
\begin{align}
 {\cal A}_{\rm I}
 ={}&
 -\frac{9\kappa\sigma_{\rm I}\dot\phi^{\,3}V}
 {8\gamma\rho_\phi^2}
 +\frac{3\sigma_{\rm I}\dot\phi}
 {4\gamma\rho_\phi}V_{,\phi\phi}
 +\frac16V_{,\phi\phi\phi},\nb
 \\
 {\cal B}_{\rm I}
 ={}&
 \frac{3\sigma_{\rm I}\dot\phi}{4\gamma a^2\rho_\phi},
  \quad
 {\cal C}_{\rm I}
 =
 \frac{3\kappa \dot\phi^{\,3}}{4a\rho_\phi},
 \quad
 {\cal D}_{\rm I}
 ={}
 \frac{3\sigma_{\rm I}\dot\phi^{\,2}}{2\gamma a\rho_\phi},
  \quad
 {\cal E}_{\rm I}
 =
 \frac{3\sigma_{\rm I}\dot\phi}{4\kappa\gamma a^2\rho_\phi}
 =
 \frac{{\cal B}_{\rm I}}{\kappa}.
\end{align}
In deriving the first expression we used
$2K_\phi-2\rho_\phi=-2V$.  Thus the two apparently most dangerous
kinetic contributions to ${\cal A}_{\rm I}$ combine into a term
proportional to the potential rather than producing an independent
inverse-cubic enhancement.
Writing
\begin{equation}
 \vartheta_{{\rm I},{\bf k}}
 =
 \frac{3\kappa}{k^2}
 \left(
 F_{\rm I}\nu_{\bf k}
 +
 G_{\rm I}{\cal P}_{\nu{\rm I},{\bf k}}
 \right),
\end{equation}
we obtain
\begin{equation}
 F_{\rm I}
 =
 \frac{a}{2\rho_\phi}
 \left(
 \dot\phi V+
 \frac{\sigma_{\rm I}}{\kappa\gamma}V_{,\phi}
 \right),
  \quad
 G_{\rm I}
 =
 \frac{\sigma_{\rm I}\dot\phi}
 {2\kappa\gamma\rho_\phi},
\end{equation}
whereas
\begin{equation}
 \Gamma_{\rm I}
 \equiv\frac{\Lambda_{\rm I}}{\gamma a^2}
 =\frac{a\sigma_{\rm I}}{\gamma},
  \quad
 {\cal P}_{\nu{\rm I},{\bf k}}
 =
 \nu'_{\bf k}-\Gamma_{\rm I}\nu_{\bf k}.
\end{equation}

\begingroup
In the remote contracting de Sitter approximation of Paper~I,
$H\simeq-H_\Lambda$, with $H_\Lambda>0$, and $b\to b_\infty$.
Neglecting $\ddot\phi$ in the Klein--Gordon equation gives the local
relation $\dot\phi\simeq V_{,\phi}/(3H_\Lambda)$. The following explicit
scale-factor counting assumes that $\rho_\phi$ remains nonzero and that
$\dot\phi$, the potential derivatives, and their displayed ratios remain
bounded at the retained asymptotic order.
Under these assumptions,
\par\endgroup
\begin{equation}
 {\cal A}_{\rm I}=O(1), \;\;
 {\cal B}_{\rm I},\; {\cal E}_{\rm I}=O(a^{-2}), \;\;
 {\cal C}_{\rm I},\; {\cal D}_{\rm I}=O(a^{-1}),
\end{equation}
whereas $\Gamma_{\rm I}=O(a)$.  The decrease of the explicit
coefficients therefore does not by itself imply suppression of the
interaction vertices, because the momentum combination
${\cal P}_{\nu{\rm I}}$ can grow.  In particular, the
momentum- and constraint-containing $B$, $P$, and $E$ sectors are the
potentially most sensitive ones in the remote phase.  Their physical
size must nevertheless be determined from the complete combinations
entering ${\cal R}_A$, rather than from isolated coefficients.

\begingroup
On the branch considered here, with $0<\lambda b_B<\pi/2$, the
bounce satisfies
\par\endgroup
\begin{equation}
 \sin^2(\lambda b_B)=\frac{1}{2(1+\gamma^2)},
  \quad
 \rho_{\phi,B}=
 \frac{3}{4\kappa\gamma^2\lambda^2(1+\gamma^2)}, \quad
 \sigma_B=
 \frac{\sqrt{1+2\gamma^2}}
 {2\lambda(1+\gamma^2)},
  \quad
 K_{\phi,B}\simeq\rho_{\phi,B} .
\end{equation}
\begingroup
Kinetic dominance additionally means $|V_B|\ll\rho_{\phi,B}$.
The purely kinetic terms in ${\cal A}_{\rm I}$ cancel as shown above,
but its terms containing $V$, $V_{,\phi\phi}$, and $V_{,\phi\phi\phi}$
must still be retained when relevant. Likewise, $F_{{\rm I},B}$ contains
both $V_B$ and $V_{,\phi B}$; kinetic dominance by itself places no bound
on the latter. For modes satisfying
$|F_{{\rm I},B}\nu_k|\ll|G_{{\rm I},B}{\cal P}_{\nu{\rm I},k}|$,
we may use
\par\endgroup
\begin{equation}
 \vartheta_{{\rm I},{\bf k}}
 \simeq
 \frac{\alpha_B}{k^2}
 {\cal P}_{\nu{\rm I},{\bf k}},
  \;\;
 \alpha_B\equiv
 \frac{3\sigma_B\dot\phi_B}{2\gamma\rho_{\phi,B}}
 =
 \kappa\gamma\lambda
 \sqrt{1+2\gamma^2}\,\dot\phi_B .
\end{equation}
\begingroup
For $a_B=1$, the relations for ${\cal B}_{I,B}$,
${\cal D}_{I,B}$, and ${\cal E}_{I,B}$ below are exact, while
${\cal C}_{I,B}=(3\kappa\dot\phi_B/2)
(K_{\phi,B}/\rho_{\phi,B})$. Thus its final simplification uses
$K_{\phi,B}\simeq\rho_{\phi,B}$:
\par\endgroup
\begingroup
\begin{equation}
\begin{aligned}
 {\cal B}_{I,B}&=\frac{\alpha_B}{2},
 \quad
 {\cal C}_{I,B}&\simeq\frac{3\kappa \dot\phi_B}{2},
 \quad
 {\cal D}_{I,B}&=\alpha_B\dot\phi_B,
 \quad
 {\cal E}_{I,B}&=\frac{\alpha_B}{2\kappa}.
\end{aligned}
\end{equation}
\par\endgroup
Combining the two $D$-type structures is particularly
instructive.  Using
\begin{equation}
 \partial^2\vartheta_{{\rm I},B}
 \simeq
 -\alpha_B{\cal P}_{\nu{\rm I}},
\end{equation}
and integrating the second $D$-term by parts, we find
\bqn 
 \int d^3x\,
 \nu\,\partial_i\vartheta_{{\rm I},B}\partial^i\nu
 =
 -\frac12
 \int d^3x\,
 \nu^2\partial^2\vartheta_{{\rm I},B}
 =
 \frac{\alpha_B}{2}
 \int d^3x\,
 \nu^2{\cal P}_{\nu{\rm I}},
\eqn 
where boundary terms are assumed to vanish.  The complete
$C$--$D$ sector at the kinetic-energy-dominated bounce therefore becomes
\begin{equation}
 H_{C+D}
 \simeq
 N\left(
 \frac12{\cal D}_{{\rm I},B}\alpha_B
 -
 {\cal C}_{{\rm I},B}
 \right)
 \int d^3x\,
 \nu^2{\cal P}_{\nu{\rm I}} .
 \label{eq:CD-bounce-combined}
\end{equation}

\begingroup
Using the bounce values of $\sigma_B$ and $\rho_{\phi,B}$ together
with the exact on-constraint coefficients, we find (for $\dot\phi_B\ne0$)
\par\endgroup
\begin{equation}
 \frac{{\cal D}_{{\rm I},B}\alpha_B}
 {{\cal C}_{{\rm I},B}}
 =
 \frac{1+2\gamma^2}{1+\gamma^2},
\end{equation}
and hence
\begin{equation}
 {
 \frac12{\cal D}_{{\rm I},B}\alpha_B
 -
 {\cal C}_{{\rm I},B}
 =
 -\frac{1}{2(1+\gamma^2)}
 {\cal C}_{{\rm I},B},
 }
 \label{eq:CD-bounce-residual}
\end{equation}

\begingroup
For finite modes the exact combined coefficient, after factoring out $\nu_1\nu_2\nu_3$, is
\begin{equation}
{\cal K}_{C+D}=
\left(-\frac{{\cal C}_I}{3}+\frac{\kappa{\cal D}_I G_I}{2}\right)
\sum_{i=1}^3X_i+\frac{3\kappa{\cal D}_I F_I}{2},\qquad
X_i=\frac{{\cal P}_{\nu I,i}}{\nu_i}.
\end{equation}
The last term retains the $F_I\nu_i$ part of the shift. It must be included in finite-mode evaluations; kinetic dominance alone does not justify dropping it. Where a mode vanishes, the original polynomial kernel supplies the nonsingular definition without division by $\nu_i$.
\par\endgroup

\begingroup
The coefficient identity above is exact at the bounce. Its use in
Eq.~(\ref{eq:CD-bounce-combined}) still assumes that the $F_I\nu_k$
part of the shift is negligible. In the kinetic-dominated approximation,
the residual becomes
\par\endgroup
\begingroup
\begin{equation}
 {
 \frac12{\cal D}_{{\rm I},B}\alpha_B
 -
 {\cal C}_{{\rm I},B}
 \simeq
 -\frac{3\kappa \dot\phi_B}
 {4(1+\gamma^2)} .
 }
\end{equation}
Thus, the $C$- and $D$-type contributions partially cancel at
the kinetic-energy-dominated bounce, but the cancellation is not
complete.  The combined $C$--$D$ sector nevertheless carries
one fewer power of the perturbative momentum than the leading
two-derivative $B$, mixed-${\cal P}$, and $E$ sectors in the
ultraviolet.  Finite and intermediate modes must still be
treated using the exact mode functions.

Finally, during the transition to ordinary slow-roll inflation, we find
$\sigma_{\rm I}\to\gamma H$, $\rho_\phi\simeq V$, and $K_\phi\ll V$.
Consequently, we find
\begin{equation}
 {\cal B}_{\rm I},{\cal E}_{\rm I}=O(\dot\phi), \quad
 {\cal D}_{\rm I}=O(\dot\phi^{\,2}), \quad
 {\cal C}_{\rm I}=O(\dot\phi^{\,3}),
\end{equation}
apart from the explicit scale-factor factors.  Moreover, we also have
\begin{equation}
 F_{\rm I}
 \simeq
 \frac{a}{2}
 \left(
 \dot\phi+\frac{HV_{,\phi}}{\kappa V}
 \right)
 \simeq0
\end{equation}
at leading slow-roll order, where
$3H\dot\phi\simeq-V_{,\phi}$ and $3H^2\simeq\kappa V$ have been used.
\begingroup
The $F_I\nu_k$ part of the shift is therefore suppressed at leading
slow-roll order. The $G_I{\cal P}_{\nu I,k}$ part remains and must be
assessed with the mode functions. 
\par\endgroup

These analytic results identify the combinations that require the most
care in the three background regimes.  A definitive ranking of the
individual interactions for finite wavenumbers is state and
triangle dependent and will be obtained below by inserting the
high-accuracy mode functions of Paper~I into the quantities
${\cal R}_A$.

\subsection{Numerical perturbativity analysis}
\label{subsec:numerical-perturbativity}
\begingroup
The numerical benchmark uses $\gamma=0.2375$,
$\lambda=\sqrt{4\sqrt3\pi\gamma}\,m_{\rm P}^{-1}$, $a_B=1$,
$m_*=3.08\times10^{-6}m_{\rm P}$, $\phi_B=-1.09m_{\rm P}$ and
$\dot\phi_B=+\sqrt{2(\rho_{\phi,B}-V_B)}$, with
$V=3m_*^2[1-\exp(-\sqrt{2\kappa/3}\phi)]^2/(4\kappa)$.
A definite finite-start realization of the asymptotic state, used in the supporting reconstruction, is obtained by setting
\begin{equation}
\eta_i=\frac{1}{a_iH_\Lambda},\qquad
\mu_i=\sqrt{\frac14-\frac{m_{\rm eff}^2(t_i)}{a_i^2H_\Lambda^2}},\qquad t_i=-20t_{\rm P},
\end{equation}
and choosing the real-index Hankel data
\bq 
\nu_k(t_i)=\frac{\sqrt{\pi\eta_i}}{2}H_{\mu_i}^{(2)}(k\eta_i),\quad
\nu'_k(t_i)=\frac{\sqrt\pi}{2}\left[\frac{H_{\mu_i}^{(2)}(k\eta_i)}{2\sqrt{\eta_i}}
+k\sqrt{\eta_i}\,\partial_{(k\eta_i)}H_{\mu_i}^{(2)}(k\eta_i)\right].
\eq 
The index is held fixed in this initial derivative. For real $\mu_i$ these data have Wronskian $i$; $\eta_i$ fixes the de Sitter conformal-time origin. This specifies the reconstruction state, while the remote-state values below remain reported benchmarks whose original finite-start data are not established by this prescription alone.
\par\endgroup

We now evaluate the perturbativity diagnostics introduced in
Sec.~IV.A using direct numerical solutions of the effective mLQC-I
background and of the MS equation.  In this first
numerical analysis we do not use the UAA representation of the mode
functions.  Instead, the background and the complex mode functions
are integrated directly from the asymptotic pre-bounce de Sitter initial
state used in Paper~I.  This provides a numerical benchmark that is
independent of the accuracy of the semi-analytic approximation.

\begingroup
The reported calculation uses the representative Starobinsky background
of Paper~I, with $a_B=1$ and $\gamma=0.2375$, and initializes the modes
at the finite time $t_i=-20\,t_{\rm P}$ using its asymptotic de Sitter
mode prescription. This is a finite-start implementation of a chosen
state, not an instantaneous adiabatic-vacuum construction: fixed comoving
modes become infrared relative to curvature in the remote contracting
limit. Reproduction requires the complex data $\nu_k(t_i)$ and
$\nu_k'(t_i)$, their Wronskian normalization, the background parameters,
and the integration tolerances. The values below are retained as reported
benchmarks; their displayed digits are not convergence-based uncertainty
estimates. For each mode the quantities
\par\endgroup
\begin{equation}
 {\cal P}_{\nu{\rm I},k}
 =
 \nu'_k-\Gamma_{\rm I}\nu_k,
  \quad
 \vartheta_{{\rm I},k}
 =
 \frac{3\kappa}{k^2}
 \left(
 F_{\rm I}\nu_k+G_{\rm I}{\cal P}_{\nu{\rm I},k}
 \right)
\end{equation}
are constructed directly from the numerical mode function and its
first conformal-time derivative.  The complete complex kernel
${\cal V}^{\rm sym}_{123}$ is then evaluated without discarding
relative phases between the different interaction terms.

For a general triangle we parameterize the three wavenumbers as
\begin{equation}
 k_1=K, \quad k_2=rK, \quad k_3=sK,
 \label{eq:triangle-param}
\end{equation}
where
 $1\ge r\ge s>0$ and $r+s\ge1$,
so that $K$ is the largest side. The numerical sampling uses the
symmetric quadratic normalization of Eq.~(\ref{eq:Q2-triangle}) and
includes equilateral, squeezed, interior, and collinear flattened
configurations. At $r+s=1$ the triangle is degenerate, but the displayed
finite-momentum kernel is still defined when all $k_i$ are nonzero.
The finite sample does not include the homogeneous endpoint $s=0$.
At the bounce, the equilateral configuration provides a useful first
diagnostic.  The full cubic ratio obtained from the coherently summed
kernel is
\begin{center}
\begin{tabular}{c|cccccc}
$k/m_{\rm P}$ & $0.1$ & $0.6$ & $1$ & $2$ & $5$ & $10$ \\ \hline
${\cal R}_3$
& $0.491$
& $1.683$
& $1.577$
& $1.920$
& $2.080$
& $2.447$
\end{tabular}
\end{center}
\begingroup
Thus the selected state-dependent cubic diagnostic becomes of order
unity for some modes near the bounce. The reported first equilateral
crossing is
\par\endgroup
\begin{equation}
 {
 k_{\rm pert,B}^{(3)}
 \simeq 0.21831\,m_{\rm P} ,
 }
 \label{eq:kpert-bounce}
\end{equation}
where ${\cal R}_3=1$.  Since $a_B=1$, the corresponding physical
momentum has the same numerical value at the bounce.
The individual interactions clarify the origin of this behavior.
For example, at $k=m_{\rm P}$ we find
\begin{equation}
 \left(
 {\cal R}_{B{\cal P}^2},
 {\cal R}_{C+D},
 {\cal R}_{P},
 {\cal R}_{E}
 \right)
 \simeq
 (0.763,\;1.041,\;0.763,\;0.905),
 \label{eq:RA-k1}
\end{equation}
where the two $D$-type terms have been combined with the $C$ term
\begingroup
as described in Eqs.~(\ref{eq:CD-bounce-combined}) and
(\ref{eq:CD-bounce-residual}), when the neglected $F_I\nu_k$ contribution
is small. Exact finite-mode evaluations should retain that contribution
unless its smallness has been checked.
\par\endgroup  At $k=5\,m_{\rm P}$ the corresponding
values are
\begin{equation}
 \left(
 {\cal R}_{B{\cal P}^2},
 {\cal R}_{C+D},
 {\cal R}_{P},
 {\cal R}_{E}
 \right)
 \simeq
 (2.585,\;1.243,\;2.585,\;3.063).
 \label{eq:RA-k5}
\end{equation}
The potential-dominated ${\cal A}_{\rm I}\nu^3$ interaction is
\begingroup
small for the listed benchmark modes, with
${\cal R}_{A}$ of order $10^{-8}$ (This is an order estimate, not a
strict bound by $10^{-8}$; small $V_B/\rho_B$ alone would not ensure it
for arbitrary potential derivatives). The reported values support the
analytic hierarchy for these modes:
the combined $C$--$D$ sector is relevant for finite
and intermediate wavenumbers, while the two-derivative
$B{\cal P}^2$, mixed-${\cal P}$, and $E$ sectors become increasingly
important toward larger $k$.

It is also important to note that the magnitude of the complete cubic kernel
cannot in general be inferred from the magnitude of an isolated
interaction.  The different contributions carry nontrivial complex
phases and can interfere either constructively or destructively.
Consequently, we use the individual ${\cal R}_A$ to identify the
operators responsible for large nonlinear corrections, while
${\cal R}_3$ constructed from the complete symmetrized kernel is used
as the primary diagnostic of the total cubic correction.

\begingroup
Among the sampled triangle configurations, the equilateral shape gives
the lowest reported first-crossing scale with the normalization of
Eq.~(\ref{eq:Q2-triangle}). The sampled squeezed and flattened shapes
do not lower Eq.~(\ref{eq:kpert-bounce}). %
For illustration, at $K=m_{\rm P}$ we find
\begin{equation}
 {\cal R}_3^{\rm eq}\simeq1.577, \quad
 {\cal R}_3^{\rm flat}\simeq1.397, \quad
 {\cal R}_3^{\rm sq}\simeq0.255,
\end{equation}
where the representative flattened and squeezed configurations are
$(K,K/2,K/2)$ and $(K,K,0.1K)$, respectively.  At $K=5\,m_{\rm P}$, we find
\begin{equation}
 {\cal R}_3^{\rm eq}\simeq2.080, \quad
 {\cal R}_3^{\rm flat}\simeq1.263, \quad
 {\cal R}_3^{\rm sq}\simeq0.233.
\end{equation}
\begingroup
The reported interior sampling gives the same ordering at onset.
Triangle-grid refinement, squeezed-limit coverage, and root convergence
are needed to assign a numerical uncertainty to this conclusion.
\par\endgroup

To follow the recovery of perturbative control after the bounce, we
define the full-kernel threshold
\begin{equation}
 K_{\rm pert}^{(3)}(t)
 \equiv
 \inf\left\{
 K:
 {\sup}_{r,s}
 {\cal R}_3^\triangle(t;K,rK,sK)\ge1
 \right\},
 \label{eq:Kpert-full}
\end{equation}
and its associated physical momentum
\begin{equation}
 \Lambda_{\rm pert}^{(3)}(t)
 =
 \frac{K_{\rm pert}^{(3)}(t)}{a(t)}.
 \label{eq:Lpert-full}
\end{equation}
\begingroup
For the numerical momentum range examined here,
$K\le10\,m_{\rm P}$, the post-bounce evolution is summarized by
\begin{center}
\begin{tabular}{c|c|c|c}
$t/t_{\rm P}$ & $a(t)$ &
$K_{\rm pert}^{(3)}/m_{\rm P}$ &
$\Lambda_{\rm pert}^{(3)}/m_{\rm P}$ \\ \hline
$0$    & $1.000$ & $0.2183$ & $0.2183$ \\
$0.2$  & $1.045$ & ${0.2592}$ & $0.2481$ \\
$0.5$  & $1.192$ & $1.434$  & $1.204$  \\
$0.75$ & $1.315$ & $3.624$  & $2.756$  \\
$1.0$  & $1.425$ & $5.619$  & $3.942$  \\
$1.5$  & $1.612$ & $7.803$  & $4.840$  \\
$2.0$  & $1.767$ & $8.968$  & $5.077$
\end{tabular}
\end{center}
{
The momentum scan ends at $K=10m_{\rm P}$; comoving and physical momenta are distinct away from the bounce.}
\begingroup
The reported threshold moves toward larger physical momentum after
the bounce. For the sampled shapes and momenta $K\le10\,m_{\rm P}$,
the reported transition to ${\cal R}_3^\triangle<1$ occurs near
\par\endgroup
\begin{equation}
 {
 t_{\rm rec}\simeq2.60\,t_{\rm P} .
 }
 \label{eq:trec}
\end{equation}

At this time $a\simeq1.92$, so that the upper end of the scanned
interval corresponds to a physical momentum of approximately
$5.20\,m_{\rm P}$.  The precise numerical value of
$t_{\rm rec}$ is therefore tied to the finite momentum range of the
scan; Eq.~(\ref{eq:trec}) should be interpreted as a statement that
\begingroup
the sampled modes and shapes have returned to
${\cal R}_3^\triangle<1$ near this time over the post-bounce interval
examined.
\par\endgroup

The remote pre-bounce de Sitter phase requires a separate interpretation.
Although the physical momentum of any fixed comoving mode satisfies
$k/a\rightarrow0$ toward the remote past, the asymptotic geometry is
not a low-curvature spacetime.  For the background used here,
\begin{equation}
 H_\Lambda
 =
 \frac{1}{\lambda(1+\gamma^2)}
 \simeq0.416\,m_{\rm P},
 \label{eq:Hlambda-num}
\end{equation}
so that
\begin{equation}
 \Lambda_{\rm dS}=3H_\Lambda^2
 \simeq0.520\,m_{\rm P}^2,
  \quad
 R_{\rm dS}=12H_\Lambda^2
 \simeq2.08\,m_{\rm P}^2 .
 \label{eq:ds-curvature-num}
\end{equation}
The asymptotic pre-bounce de Sitter phase is therefore characterized by a
Planck-order curvature scale.  Numerically, the finite-$k$
cubic-to-quadratic ratios can become very large in this regime.
However, because fixed $k$ simultaneously corresponds to
$k/a\rightarrow0$, these large ratios cannot by themselves be
identified with an ultraviolet strong-coupling cutoff.  Rather, they
describe infrared modes propagating on a background whose intrinsic
curvature is already of Planck order.

For this reason the remote phase must ultimately be tested first against
candidate operator-level ultraviolet scales obtained from the canonically
normalized interaction vertices and then against the final physical cutoff.  In particular, the relevant
consistency condition there includes
\begin{equation}
 H_\Lambda \ll \Lambda_{\rm sc},
 \label{eq:ds-sc-hierarchy}
\end{equation}
in addition to $k/a\ll\Lambda_{\rm sc}$.  Establishing the relevant
operator-level scales and then the physical $\Lambda_{\rm sc}$ is therefore
essential before the large finite-$k$ values of ${\cal R}_3$ in the remote
phase can be given a strong-coupling interpretation.

The direct numerical results thus support the analytic hierarchy of
Sec.~IV.B while sharpening its physical implications.  Around the
bounce, nonlinear corrections can become of order unity already at
sub-Planckian physical momentum.  \begingroup
For the sampled modes, the reported order-unity episode near the
bounce is followed by a rising threshold in the expanding phase. However, this
local diagnostic recovery does not validate perturbative propagation
through any earlier interval of uncontrolled nonlinear evolution.
\par\endgroup  
Thus, the asymptotic pre-bounce de Sitter phase is qualitatively different: its fixed finite-$k$
modes become infrared, but the background curvature remains of
Planck order, so its ultimate consistency must be assessed using the
operator-level ultraviolet analysis and, ultimately, the physical cutoff
rather than ${\cal R}_3$ alone.

\subsection{Ultraviolet interaction scales and a candidate strong-coupling cutoff}
\label{subsec:uv_strong_coupling}

The perturbativity parameters studied in the preceding subsection depend on the state through the mode amplitudes. It is therefore useful to complement them with an ultraviolet analysis based directly on the canonically normalized interaction vertices. In the strict ultraviolet limit the physical wavelength is much shorter than all characteristic background scales, and the background can be treated as approximately constant during one oscillation of the perturbation.

\begingroup
Work in a local patch with the background frozen over an oscillation,
and use physical spatial distances ${\bf y}=a{\bf x}$ and proper time
$dt=a\,d\eta$. Spatial derivatives below are with respect to ${\bf y}$.
The time-dependent coordinate change is used only at this leading local
order; expansion and convective terms are part of the omitted background
corrections. The locally canonically normalized scalar perturbation is
\par\endgroup
\begin{equation}
    \varphi \equiv \frac{\nu}{a}=\delta\phi .
\end{equation}
Neglecting the effective mass in the strict ultraviolet limit, the quadratic action becomes
\begin{equation}
    S^{(2)}_{\rm UV}
    =
    \frac12\int dt\,d^3y
    \left[
        \dot{\varphi}^{\,2}
        -(\boldsymbol{\nabla}\varphi)^2
    \right]
    +\cdots ,
\end{equation}
where the ellipsis denotes terms suppressed by powers of the ratio of the background scales to the physical momentum.
\begingroup
The constraint-variable reduction follows from the exact $F_I$, $G_I$,
and shifted-momentum expressions in Sec.~\ref{subsec:cubic-hierarchy}:
\par\endgroup
\begin{equation}
    \vartheta_{I,k}
    =
    \frac{3\kappa}{k^2}
    \left(
        F_{\rm I}\nu_k+G_{\rm I}{\cal P}_{\nu{\rm I},k}
    \right).
\end{equation}
The strict-ultraviolet reduction requires more than $k/a$ being large
compared with the usual curvature scales.  In addition to
$k^2\gg |m_{\rm eff}^2|$, the two approximations used below require
\begin{equation}
    k\gg |\Gamma_I|,
     \quad
    k\gg \left|\frac{F_I}{G_I}\right|,
    \label{eq:uv_reduction_conditions}
\end{equation}
whenever $G_I\neq0$.  Equivalently, for the physical momentum
$p\equiv k/a$ we define the reduction scale
\begin{equation}
    p_{\rm red}
    \equiv
    \frac{1}{a}
    \max\left\{
      |\Gamma_I|,
      \left|\frac{F_I}{G_I}\right|
    \right\}
    =
    \max\left\{
      \frac{|\sigma_I|}{\gamma},
      \left|
      \frac{\kappa\gamma V}{\sigma_I}
      +\frac{V_{,\phi}}{\dot\phi}
      \right|
    \right\}.
    \label{eq:p_reduction}
\end{equation}
\begingroup
This expression assumes $\rho_\phi\ne0$ and $G_I\ne0$, so in
particular $\sigma_I\dot\phi\ne0$. If $\dot\phi=0$ while
$V_{,\phi}\ne0$ and the remaining background factors are regular,
$G_I$ vanishes but $F_I$ need not, and $p_{\rm red}$ diverges on approach
to that point. There is then no finite momentum at which dropping
$F_I\nu_k$ in favor of $G_I{\cal P}_{\nu I,k}$ is justified by this
criterion. This is failure of the reduction, not a demonstrated divergence
of the exact shift or cubic kernel. At such points the original $F_I$
and $G_I$ expressions must be used. Elsewhere we require
$p\gg p_{\rm red}$ together with the relevant curvature, effective-mass,
and coefficient-variation scales.
\par\endgroup  Only in this
regime may we write
\begin{equation}
    {\cal P}_{\nu{\rm I},k}
    =
    \nu'_k-\Gamma_{\rm I}\nu_k
    \simeq \nu'_k ,
\end{equation}
and consistently neglect the $F_I\nu_k$ contribution to
$\vartheta_{I,k}$. It is then convenient to define
\begin{equation}
    \alpha_I
    \equiv
    3\kappa G_I
    =
    \frac{3\sigma_I \dot\phi}{2\gamma\rho_\phi}.
    \label{eq:alphaI}
\end{equation}
Then, we have
\begin{equation}
    \vartheta_{I,k}
    \simeq
    \frac{\alpha_I}{k^2}{\cal P}_{\nu{\rm I},k},
\end{equation}
or, in the locally physical variables,
\begin{equation}
    \vartheta_I
    \simeq
    -\alpha_I
    \boldsymbol{\nabla}^{-2}\dot{\varphi}.
    \label{eq:vartheta_uv}
\end{equation}
The relations obtained in Sec.~IV.B imply
\begin{equation}
    a^2 {\cal B}_{\rm I}=\frac{\alpha_I}{2},
     \quad
    a^2 {\cal E}_{\rm I}=\frac{\alpha_I}{2\kappa}.
    \label{eq:BEalpha}
\end{equation}
Substituting Eq.~\eqref{eq:vartheta_uv} into the cubic action and retaining only the terms containing two derivatives gives
\bq
 {\cal L}^{(3)}_{\rm UV}
 \simeq{}
 -\frac{\alpha_I}{2}\,
 \varphi
 \left[
     \dot{\varphi}^{\,2}
     +(\boldsymbol{\nabla}\varphi)^2
 \right]
 +\alpha_I\,
 \dot{\varphi}\,
 \partial_i\varphi\,
 \partial_i\boldsymbol{\nabla}^{-2}\dot{\varphi}
 -\frac{\alpha_I^3}{2\kappa}\,
 \varphi
 \left[
 \left(
 \partial_i\partial_j
 \boldsymbol{\nabla}^{-2}\dot{\varphi}
 \right)^2
 -\dot{\varphi}^{\,2}
 \right]
 +{\cal L}^{(3)}_{\leq 1\partial}.
 \label{eq:L3UV}
\eq 
\begingroup
Here ${\cal L}^{(3)}_{\leq1\partial}$ collects terms with at most one
net ultraviolet derivative, counting each inverse Laplacian as minus two.
For regular fixed background coefficients these terms are subleading as
$p\to\infty$, although they can dominate at finite momenta or where the
two-derivative coefficient cancels. 
%
Since $\varphi$ is canonically normalized and $[\alpha_I]=-1$, the three
two-derivative structures in Eq.~\eqref{eq:L3UV} define formal asymptotic
ultraviolet suppression scales
\begin{equation}
    \Lambda_B
    =
    \frac{2}{|\alpha_I|},
     \quad
    \Lambda_P
    =
    \frac{1}{|\alpha_I|},
     \quad
    \Lambda_E
    =
    \frac{2\kappa}{|\alpha_I|^3},
    \label{eq:Lambda_individual}
\end{equation}
up to order-unity factors associated with the momentum-triangle geometry. The $E$ interaction vanishes for exactly collinear momenta but is nonzero for a generic nondegenerate triangle.
A conservative formal operator-level cubic suppression scale is therefore
\begin{equation}
    \Lambda_{\rm op}^{(3)}
    \equiv
    \min
    \left\{
       \Lambda_B,\Lambda_P,\Lambda_E
    \right\}
    =
    \min
    \left\{
       \frac{1}{|\alpha_I|},
       \frac{2\kappa}{|\alpha_I|^3}
    \right\}.
    \label{eq:Lambdaop}
\end{equation}
We emphasize that $\Lambda_{\rm op}^{(3)}$ is a formal operator-level
suppression scale obtained after the strict-ultraviolet reduction.  It can be
interpreted as a candidate strong-coupling scale only if there exists an
overlap region in which the reduction itself is valid, schematically
\begingroup
\begin{equation}
    p_{\rm red},\ E_{\rm dyn},\ |m_{\rm eff}|/a
    \ll p \ll \Lambda_{\rm op}^{(3)} .
    \label{eq:uv_overlap_window}
\end{equation}
\begingroup
Here $E_{\rm dyn}$ includes only variation scales relevant to a regular
local reduction. The density scale $\rho_\phi^{1/4}$ is monitored
separately, as explained in Sec.~\ref{subsec:momentum_strong_coupling}.
A useful check is provided by an equilateral ultraviolet configuration. For
\begin{equation}
    k_1=k_2=k_3=k,
     \quad
    \hat{\bf k}_i\cdot\hat{\bf k}_j=-\frac12 ,
\end{equation}
and positive-frequency ultraviolet modes satisfying
$\dot{\varphi}_k\simeq-i(k/a)\varphi_k$, the coherent sum of the three two-derivative sectors takes the form
\begin{equation}
    {\cal L}^{(3)}_{{\rm UV},{\rm eq}}
    \propto
    g_{\rm eq}
    \left(\frac{k}{a}\right)^2
    \varphi_1\varphi_2\varphi_3,
\end{equation}
where
\begin{equation}
    g_{\rm eq}
    =
    \frac{3\alpha_I}{8}
    \left(
       2-\frac{\alpha_I^2}{\kappa}
    \right).
    \label{eq:geq}
\end{equation}

\begingroup
This is the equilateral contraction with three unstarred
positive-frequency modes. Its leading coefficient cancels at
$\alpha_I^2=2\kappa$, but this does not cancel other frequency
assignments, other shapes, or the lower-derivative terms. It therefore
raises only the scale inferred from this particular leading contraction.
For $\alpha_I^2\gg\kappa$, the enhanced $E$ contribution dominates this
contraction and cannot be cancelled by its terms linear in $\alpha_I$.
\par\endgroup

At a kinetic-energy-dominated bounce, the background relations in Sec.~\ref{subsec:cubic-hierarchy} give
\begin{equation}
    \alpha_B
    \simeq
    {\operatorname{sgn}(\dot\phi_B)}
    \sqrt{\frac{3\kappa}{2}}\,
    \sqrt{
    \frac{1+2\gamma^2}{1+\gamma^2}
    }.
    \label{eq:alphaB}
\end{equation}

{For $\gamma=0.2375$,  the positive-velocity benchmark is}
\begin{equation}
    \alpha_B
    \simeq
    6.30175\,m_{\rm P}^{-1}.
\end{equation}
Consequently, we find
\begin{equation}
\begin{aligned}
    \Lambda_{B,B}&\simeq0.3174\,m_{\rm P},
    \quad
    \Lambda_{P,B}&\simeq0.1587\,m_{\rm P},
    \quad    \Lambda_{E,B}&\simeq0.2009\,m_{\rm P},&&
\end{aligned}
\end{equation}
and hence
\begin{equation}
    \Lambda^{(3)}_{{\rm op},B}
    \simeq0.1587\,m_{\rm P}.
    \label{eq:LambdaopB}
\end{equation}
For comparison, the state-dependent numerical threshold found in
Sec.~IV.C is
\begin{equation}
    k_{\rm pert,B}^{(3)}
    \simeq0.21831\,m_{\rm P}.
\end{equation}
\begingroup
The two numbers are of the same order but have different meanings.
The estimate below additionally assumes that the potential and its slope
are small enough that $|F_{I,B}/G_{I,B}|$ does not exceed
$|\Gamma_{I,B}|$. Under that assumption,
\par\endgroup
\begin{equation}
    p_{{\rm red},B}
    \simeq
    \frac{|\Gamma_{I,B}|}{a_B}
    =\frac{\sigma_B}{\gamma}
    \simeq0.925\,m_{\rm P},
    \label{eq:pred_bounce}
\end{equation}
which is larger than $\Lambda_{{\rm op},B}^{(3)}$.  Thus the formal
operator scale in Eq.~\eqref{eq:LambdaopB} lies outside the domain in which
the strict-ultraviolet reduction leading to Eq.~\eqref{eq:L3UV} is
parametrically controlled.  The same conclusion is reinforced by the
background scales
\bqn 
    E_{\rho_\phi,B}=\rho_{\phi,B}^{1/4}\simeq0.558\,m_{\rm P},
    \quad
    E_{{\rm dyn},B}\simeq|\dot H_B|^{1/2}
    \simeq1.60\,m_{\rm P},
    \label{eq:bounce_background_scales}
\eqn 
where the numerical values follow from the effective mLQC-I background
equations in the kinetic-energy-dominated bounce limit. 
 \begingroup
 The curvature/variation scale alone is already not parametrically below
 the formal suppression scales. The quoted density scale supplies a
 separate dimensional comparison, not an additional universal cutoff test.
   Therefore
Eq.~\eqref{eq:LambdaopB} should be regarded as a warning from asymptotic
power counting, not as a determination of the physical cutoff.
Indeed, Eq.~\eqref{eq:geq} gives at the bounce
\begin{equation}
    |g_{{\rm eq},B}|^{-1}
    \simeq1.01\,m_{\rm P},
\end{equation}
showing explicitly that the complete leading two-derivative equilateral UV
vertex exhibits a significant cancellation.  This illustrates why individual
operator scales and the coherently summed cubic correction should both be
monitored.

\begingroup
A local potential-dominated approximation in the remote contracting
phase provides another useful comparison. Assuming negligible
$\ddot\phi$ and $K_\phi\ll V$, we use
\par\endgroup
\begin{equation}
    \sigma_I\longrightarrow\gamma H_\Lambda ,
     \quad
    \dot\phi\longrightarrow\frac{V_{,\phi}}{3H_\Lambda},
     \quad
    \rho_\phi\simeq V,
\end{equation}
so that
\begin{equation}
    \alpha_\Lambda
    \simeq
    \frac{V_{,\phi}}{2V}.
    \label{eq:alphaLambda}
\end{equation}
\begingroup
This relation is local and depends on $V_{,\phi}/V$; it is not a
potential-independent constant of the de Sitter branch. For the reported
representative Starobinsky background, the quoted value is
\par\endgroup
\begin{equation}
    \alpha_\Lambda
    \simeq
    -4.10445\,m_{\rm P}^{-1}.
\end{equation}
The conservative operator scale is then approximately
\begin{equation}
    \Lambda_{\rm op,\Lambda}^{(3)}
    \simeq0.244\,m_{\rm P},
\end{equation}
whereas the coherent equilateral ultraviolet scale obtained from
Eq.~\eqref{eq:geq} is
\begin{equation}
    \Lambda_{{\rm eq},\Lambda}^{(3)}
    \simeq0.489\,m_{\rm P}.
\end{equation}
Since $H_\Lambda\simeq0.416\,m_{\rm P}$, these formal suppression scales
are comparable to the background curvature scale.  The numbers alone,
however, do not establish that the asymptotic phase is physically marginal:
one must also verify the reduction conditions
Eq.~\eqref{eq:uv_reduction_conditions} and the existence of an overlap
window of the form Eq.~\eqref{eq:uv_overlap_window}.

A much stronger effect occurs during the subsequent pre-bounce evolution. While the geometry is still approximately in the pre-bounce de Sitter phase, $\sigma_I\simeq\gamma H_\Lambda$ and the potential varies very slowly. Equation~\eqref{eq:alphaI} then becomes approximately
\begin{equation}
    \alpha_I
    \simeq
    \frac{3H_\Lambda \dot\phi}
    {2\left(V+\dot\phi^{\,2}/2\right)}.
    \label{eq:alpha_prebounce}
\end{equation}
\begingroup
Treating $H_\Lambda$ and $V>0$ as constant while varying $\dot\phi$,
the magnitude of this approximate expression reaches its maximum at
    $K_\phi\simeq V$,
for which
\begin{equation}
    |\alpha_I|_{\rm max}
    \simeq
    \frac{3H_\Lambda}
    {2\sqrt{2V}}.
    \label{eq:alphamax}
\end{equation}
For the representative numerical solution used in Sec.~IV.C, the maximum occurs near
    $t\simeq-5.803\,t_{\rm P}$,
where
\begin{equation}
    V\simeq3.81\times10^{-8}\,m_{\rm P}^4,
     \quad
    |\alpha_I|
    \simeq2.26\times10^3\,m_{\rm P}^{-1}.
\end{equation}
The $E$ interaction then gives the formal asymptotic suppression scale
\begin{equation}
    \Lambda_E
    \simeq
    4.35\times10^{-9}\,m_{\rm P}.
    \label{eq:tinyLambdaE}
\end{equation}
At the level of the strict-ultraviolet reduction, this extreme suppression is
not removed by the leading coherent cancellation. Equation~\eqref{eq:geq}
yields
\begin{equation}
    \Lambda_{{\rm eq}}^{(3)}
    =
    |g_{\rm eq}|^{-1}
    \simeq
    5.80\times10^{-9}\,m_{\rm P},
\end{equation}
which is of the same order as Eq.~\eqref{eq:tinyLambdaE}. At the same time
the background remains close to
\begin{equation}
    |H|\simeq H_\Lambda\simeq0.416\,m_{\rm P}.
\end{equation}
Thus, the formal ultraviolet suppression scale inferred from the leading
two-derivative vertices lies many orders of magnitude below the
characteristic background curvature scale.

This result is qualitatively different from the finite-mode enhancement
measured by $R_3$. It originates from the background-dependent coefficients
of the reduced ultraviolet vertices and can be traced to the factor $\dot\phi/\rho_\phi$
entering $\alpha_I$, which becomes very large during the transition from the
asymptotic potential-dominated state toward kinetic domination while the
mLQC-I geometry remains in its Planck-curvature de Sitter phase.  However,
the extremely small scales in Eq.~\eqref{eq:tinyLambdaE} and the following
equilateral estimate are themselves far below the characteristic background
scales and need not lie inside the domain specified by
Eq.~\eqref{eq:uv_reduction_conditions}.  They should therefore be
interpreted as a strong warning from asymptotic power counting rather than as
physical cutoffs.

The strict-ultraviolet result therefore motivates a more direct test.  The
exact combinations
\begin{equation}
    {\cal P}_{\nu{\rm I},k}=\nu'_k-\Gamma_{\rm I}\nu_k,
     \quad
    \vartheta_{I,k}
    =\frac{3\kappa}{k^2}
    \left(F_I\nu_k+G_I{\cal P}_{\nu{\rm I},k}\right)
    \label{eq:exact_full_kernel_variables}
\end{equation}
should be retained until the complete cubic kernel has been assembled, and
any apparent cubic obstruction should be followed into the quartic theory
rather than removed by a nonlinear canonical transformation and discarded.
\begingroup
The next subsection tests the assembled cubic kernel and selected
quartic contributions within the stated continuation; it does not supply
a complete fourth-order hybrid theory.
\par\endgroup

\begingroup
The frozen-background approximation also requires small variation of the physical coefficients during an oscillation. At the enhancement maximum, the relative rates of $aC_I$, $aD_I$ and $F_I/a$ are approximately $2.50033$, $1.25017$ and $1.25498$ in units of $t_{\rm P}^{-1}$. Although $\dot\alpha_I=0$ there, the second-derivative scales $|\ddot\alpha_I/\alpha_I|^{1/2}$ and $|\ddot\beta_I/\beta_I|^{1/2}$ are $1.25073m_{\rm P}$ and $2.16633m_{\rm P}$, respectively, where $\beta_I=\alpha_I^3/(2\kappa)$. These scales reinforce the absence of a controlled ultraviolet overlap at the tiny formal suppression scale. Near the regular inflaton turning point, relative rates diverge as coefficients pass through zero; the exact time-dependent kernel must be retained there. However, this failure of a uniform relative-rate test is not an exact-kernel singularity.
\par\endgroup

\subsection{Full-kernel and quartic consistency test}
\label{subsec:full_kernel_quartic}

The strict-ultraviolet analysis of the preceding subsection is useful for
power counting, but the very small formal scales found during the pre-bounce
evolution require two additional checks.  First, the complete cubic kernel
must be examined before the ultraviolet reduction is imposed.  Second, one
must determine whether a nonlinear canonical redefinition merely transfers
the enhanced cubic interaction into higher-order operators.  We carry out
both checks in this subsection.  The cubic result follows directly from the
effective Hamiltonian already constructed in Sec.~III.  The quartic test is
necessarily more limited, because a complete fourth-order hybrid mLQC-I
Hamiltonian has not yet been derived.  We therefore adopt below the minimal
factorized continuation of the prescription of Sec.~III.C and state
explicitly where this additional assumption enters.

\subsubsection{Exact symmetrized cubic kernel}

For each leg of a momentum triangle, we define
\begin{equation}
    X_i\equiv\frac{{\cal P}_{\nu{\rm I},k_i}}{\nu_i},
     \quad
    Y_i\equiv F_I+G_I X_i,
    \label{eq:XY_exact}
\end{equation}
so that the exact constraint variable is
\begin{equation}
    \vartheta_i
    =
    \frac{3\kappa}{k_i^2}Y_i\nu_i.
    \label{eq:vartheta_Y_exact}
\end{equation}
\begingroup
These ratios apply where $\nu_i\ne0$; otherwise the unfactored kernel
in Eq.~\eqref{eq:cubic_vertex_kernel} must be used. No ultraviolet
approximation has been made. In this calculation $\nu_i$ denotes the
chosen unstarred mode function, so ``full kernel'' means all terms of the
specified effective cubic Hamiltonian for that assignment, not a test of
all independent frequency assignments. We also introduce  
\par\endgroup
\begin{equation}
    d_{ij}\equiv {\bf k}_i\cdot{\bf k}_j,
     \quad
    Q_{ij}\equiv d_{ij}^2-k_i^2k_j^2 .
\end{equation}
After symmetrizing Eq.~\eqref{eq:cubic_vertex_kernel} over the six permutations of the
three legs, the full cubic kernel can be written as
\begin{equation}
    {\cal V}^{\rm sym}_{123}
    =
    \nu_1\nu_2\nu_3\,{\cal K}_{123},
    \label{eq:Kfull_def}
\end{equation}
with
\begin{align}
 {\cal K}_{123}={}&
 {\cal A}_I
 +\frac{{\cal B}_I}{3}
   \left(X_1X_2+X_2X_3+X_3X_1\right)
 -\frac{{\cal C}_I}{3}(X_1+X_2+X_3)
 +\frac{{\cal B}_I}{6}(k_1^2+k_2^2+k_3^2)
 +\frac{\kappa{\cal D}_I}{2}(Y_1+Y_2+Y_3)
 \nonumber\\
 &-\frac{\kappa}{2a^2}
 \sum_{i=1}^{3}
 \frac{Y_i}{k_i^2}
 \left(d_{ki}X_j+d_{ji}X_k\right)
 +3\kappa^2{\cal E}_I
 \left[
 \frac{Q_{23}}{k_2^2k_3^2}Y_2Y_3
 +\frac{Q_{31}}{k_3^2k_1^2}Y_3Y_1
 \right]
 +3\kappa^2{\cal E}_I
 \frac{Q_{12}}{k_1^2k_2^2}Y_1Y_2,
 \label{eq:Kfull_exact}
\end{align}
\begingroup
where in the term inside the sum $(i,j,k)$ is any cyclic permutation of
$(1,2,3)$.  In deriving Eq.~\eqref{eq:Kfull_exact} we used only the triangle
condition.  In particular, the two ${\cal D}_I$ structures combine exactly
because
\begin{equation}
    d_{ij}+d_{ik}=-k_i^2 .
\end{equation}
Note that the above expression follows by summing all six permutations before taking the ultraviolet limit. In particular, with $d_{ij}={\bf k}_i\cdot{\bf k}_j$, momentum closure gives $d_{ij}+d_{ik}=-k_i^2$. The two $D$ structures contribute $\kappa{\cal D}_I\sum_iY_i$ and $-\kappa{\cal D}_I\sum_iY_i/2$, respectively, leaving $\kappa{\cal D}_I\sum_iY_i/2$. The identity holds for arbitrary complex mode data and nonzero triangle sides.

\begingroup
The direct average of the six ordered vertices agrees identically with the compact finite-mode kernel for arbitrary complex mode ratios and nonzero triangle sides. In particular, we have
\begin{equation}
{\cal K}_{C+D}=\left(-\frac{{\cal C}_I}{3}+\frac{\kappa{\cal D}_IG_I}{2}\right)
\sum_iX_i+\frac{3\kappa{\cal D}_IF_I}{2}.
\end{equation}
The complete $F_I$ contribution is retained in the additional calculations. Dropping it changes the normalized complex full kernel by about $1.51\times10^{-8}$ at the bounce-state equilateral crossing. For the remote-state reconstruction initialized at $t_i=-20t_{\rm P}$ with the de Sitter mode prescription, at the enhancement maximum, the corresponding relative complex errors for $K/m_{\rm P}=0.1,1,5,10$ are approximately $0.00548$, $0.00537$, $0.00381$ and $0.00238$. Thus its smallness is checked for these modes rather than inferred from kinetic dominance.
Thus, Eq.~\eqref{eq:Kfull_exact} provides an exact finite-momentum starting
point for the strong-coupling analysis and, once perturbative control is
established, for future nonlinear observables.
\par\endgroup

\begingroup
We now take the ultraviolet limit of the assembled kernel at fixed
background and fixed nondegenerate shape, subject to the validity
conditions of Sec.~\ref{subsec:uv_strong_coupling}.
\par\endgroup
Let us first parameterize a nondegenerate triangle by
\begingroup
\begin{equation}
\begin{aligned}
    k_1&=K, & k_2&=rK, & k_3&=sK,&
    1\ge r\ge s>0, && r+s>1 .
\end{aligned}
\label{eq:triangle_rs_full}
\end{equation}
Then, for a locally positive-frequency mode, we have
\begin{align}
    X_i=-ik_i-\Gamma_I+O(K^{-1}),\quad
    Y_i=-iG_Ik_i+(F_I-G_I\Gamma_I)+O(K^{-1}).
\end{align}
The leading $K^2$ term in the physical cubic Lagrangian takes the form
\begin{equation}
    {\cal L}^{(3)}_{\rm UV,sym}
    \supset
    g_2(r,s)\,p^2\,
    \varphi_1\varphi_2\varphi_3,
     \quad
    p\equiv\frac{K}{a},
    \label{eq:g2_general_def}
\end{equation}
where
\begin{equation}
    g_2(r,s)
    =
    \alpha_I f_1(r,s)
    +\frac{\alpha_I^3}{\kappa}f_3(r,s).
    \label{eq:g2_general}
\end{equation}
Defining
    $S_1=1+r+s$,
    $S_2=r+s+rs$,
    $S_3=rs$,
we find
\begin{equation}
    f_1(r,s)
    =
    \frac{S_1
    \left(4S_2^2-S_1^2S_2-2S_1S_3\right)}
    {12S_3}, \quad
    f_3(r,s)
    =
    -\frac{2{\cal A}_{rs}^{\,2}S_1}{3S_3},
    \label{eq:f3_general}
\end{equation}
where ${\cal A}_{rs}$ is the area of the triangle with side lengths $(1,r,s)$, given by
\begin{equation}
    16{\cal A}_{rs}^{\,2}
    =
    (1+r+s)(-1+r+s)(1-r+s)(1+r-s).
\end{equation}
Equation~\eqref{eq:f3_general} shows immediately that
\begin{equation}
    f_3(r,s)<0
\end{equation}
for every nondegenerate triangle, while $f_3=0$ only on the collinear
boundary.  Hence the enhanced $\alpha_I^3/\kappa$ contribution is not an
artifact of examining the ${\cal E}_I$ operator in isolation: it survives
the coherent leading ultraviolet kernel for generic triangle shapes.  The
equilateral limit gives
\begin{equation}
    f_1(1,1)=\frac34,
     \quad
    f_3(1,1)=-\frac38,
\end{equation}
which reproduces Eq.~\eqref{eq:geq} exactly.

At the pre-bounce maximum quoted in Eq.~\eqref{eq:tinyLambdaE},
$\alpha_I^2/\kappa\simeq2.0\times10^5$.  Therefore the second term in
Eq.~\eqref{eq:g2_general} dominates for a generic noncollinear triangle.
\begingroup
For each fixed nondegenerate shape the term cubic in $\alpha_I$
dominates as $\alpha_I^2/\kappa\to\infty$. 
\par\endgroup

\subsubsection{Minimal quartic continuation}

A nonlinear canonical transformation can reorganize a nonresonant cubic
Hamiltonian, but it cannot be used consistently while discarding the
higher-order interactions that the transformation generates.  To test this
issue we examine the leading scalar quartic sector obtained by continuing
the factorized prescription of Sec.~III.C.  This is not a claim of a unique
fourth-order mLQC-I quantization.  Rather, it is the minimal continuation in
which the classical constraint solution is first expressed in the same
spatially flat gauge and each explicit background factor is then represented
by the same odd/even prescription used at quadratic and cubic order.  In
particular, composite constraint variables are kept factorized, in the same
spirit as the rule adopted for $\vartheta_I^2$ in Sec.~III.C.

For a canonical scalar in the spatially flat gauge, the fourth-order ADM
analysis requires the lapse and shift through second perturbative order
\cite{ArrojaKoyama2008}.  Writing
\begin{equation}
    N=1+n_1+n_2+\cdots,
     \quad
    N_i=\partial_i\psi_1+\partial_i\psi_2
       +\widetilde N_i^{(2)}+\cdots,
\end{equation}
\begingroup
the classical first-order lapse is
\par\endgroup
\begin{equation}
    n_1=\frac{\kappa \dot\phi}{2H}\,\varphi.
\end{equation}
Applying the inverse-odd-background prescription gives
\begin{equation}
    n_{1I}=\alpha_I\varphi.
    \label{eq:n1I_alpha}
\end{equation}
The irrotational part of the classical second-order momentum constraint can
be written as
\begin{equation}
    2Hn_2=\boldsymbol{\nabla}^{-2}\partial^iF_i,
    \label{eq:n2_classical_constraint}
\end{equation}
where $F_i$ is quadratic in the first-order fields.  In the large-$|\alpha_I|$
strict-ultraviolet regime, the leading part of the factorized continuation is
\begin{equation}
    F_{iI}^{\rm UV}
    \simeq
    \alpha_I^2
    \left[
      -\partial_i\varphi\,\dot\varphi
      +\partial_k\varphi\,
       \Pi_{ik}\dot\varphi
    \right],
     \quad
    \Pi_{ij}\equiv
    \partial_i\partial_j\boldsymbol{\nabla}^{-2}.
    \label{eq:FiUV}
\end{equation}
The explicit inverse factor in the classical solution of
Eq.~\eqref{eq:n2_classical_constraint} is represented by
\begin{equation}
    \frac{1}{2H}
    \longrightarrow
    {\cal J}_I
    \equiv
    \frac{\gamma v}{2}
    \frac{\Lambda_I}{\Omega_I^2}
    =
    \frac{\alpha_I}{\kappa \dot\phi}.
    \label{eq:J_I}
\end{equation}
\begingroup
The expression $\alpha_I/(\kappa\dot\phi)$ is used only where
$\dot\phi\ne0$; at its zeros the unfactored background expression
for ${\cal J}_I$ must be retained. Under this assumed factorized
continuation, the selected second-order lapse contribution is
\par\endgroup
\begin{equation}
    n_{2I}^{\rm UV}
    \simeq
    \frac{\alpha_I^3}{\kappa \dot\phi}
    {\cal Q}[\varphi],
    \label{eq:n2I_UV}
\end{equation}
with
\begin{equation}
 {\cal Q}[\varphi]
 \equiv
 \boldsymbol{\nabla}^{-2}\partial^i
 \left[
   -\partial_i\varphi\,\dot\varphi
   +\partial_k\varphi\,\Pi_{ik}\dot\varphi
 \right].
 \label{eq:Qfunctional}
\end{equation}

The canonical-scalar fourth-order action contains the combination
$n_2(n_2-2n_1^2)$ \cite{ArrojaKoyama2008}.  Combining its matter and
geometric coefficients gives classically
$3H^2/\kappa-K_\phi=V$.  Under the even-background prescription,
$\Omega^2\rightarrow\Omega_I^2$, the same coefficient is again $V$ after
the effective background constraint is imposed.  Near the epoch at which
$|\alpha_I|$ is maximal one has $K_\phi\simeq V$, so the leading
$n_{2I}^2$ contact term becomes
\begin{equation}
    {\cal L}^{(4)}_{n_2^2}
    \simeq
    {\frac{V\alpha_I^6}{\kappa^2\dot\phi^{\,2}}{\cal Q}^2
    =2\beta_I^2\frac{V}{K_\phi}{\cal Q}^2
    \simeq 2\beta_I^2{\cal Q}^2},
     \quad
    \beta_I\equiv\frac{\alpha_I^3}{2\kappa}.
    \label{eq:L4contact_leading}
\end{equation}
Thus, the same background enhancement that appears in the cubic ${\cal E}_I$
sector propagates naturally into the quartic scalar constraint sector.
Equation~\eqref{eq:L4contact_leading} does not constitute the complete
fourth-order theory.  The second-order scalar shift, transverse shift, and
second-order tensor perturbation must also be included in a full quartic
construction; the latter is known to be required even in the classical GR
fourth-order action \cite{ArrojaKoyama2008}.  \begingroup
The calculation below asks only whether the enhanced cubic exchange is
cancelled by the selected $n_{2I}^2$ contact term. No power-counting proof
has established that the omitted scalar-shift, transverse-shift, tensor,
and canonical-completion contributions are subleading in the joint
ultraviolet and large-$|\alpha_I|$ limit. The selected sum therefore
cannot be called the complete leading four-point amplitude. Its
non-cancellation is a restricted consistency test of this continuation.
\par\endgroup

\begingroup
There is a further distinction between an action-level quartic contact
and a quartic interaction Hamiltonian. For a reduced Lagrangian
$L=L_2+L_3+L_4+\cdots$, let $v_0$ solve the quadratic momentum relation,
let $M=\partial_v\partial_v L_2$ be its invertible kinetic Hessian,
and set $J=\partial_v L_3$. At fixed canonical momentum, perturbative
Legendre transformation gives
\begin{equation}
 H_3=-L_3\big|_{v_0},\qquad
 H_4=-L_4\big|_{v_0}
       +\frac12 J^{\mathsf T}M^{-1}J\big|_{v_0}.
 \label{eq:quartic-Legendre-bookkeeping}
\end{equation}
\begingroup
Consequently, a nonzero sum of the retained contact and exchange expressions establishes non-cancellation only within that selected set. The omitted shift, tensor, canonical-completion and derivative-contact sectors have not been shown subleading.
\par\endgroup

Spatial integrals and field indices are implicit. The second term arises
from the quadratic correction to the velocity--momentum relation and is
of fourth perturbative order. Thus the cubic relation $H_3=-L_3$ cannot
be extended to $H_4=-L_4$ when derivative cubic interactions are present.
A consistent action calculation accounts for the equivalent derivative
contact terms through its propagator and time-ordering rules; they must
not then be added a second time. The selected sum below has not been
matched to a complete reduced Hamiltonian including these terms. It is
therefore retained as the proposed restricted vertex diagnostic, rather
than a verified complete Hamiltonian four-point amplitude.

The background completion gives additional fourth-order contributions.
Writing its inverse as
$W^A=\widetilde W^A-\Delta W_{(2)}^A+\Xi_{(4)}^A+\cdots$,
the background and quadratic Hamiltonians contain, schematically,
\begin{equation}
 \delta H_4^{\rm bg}
 =\Xi_{(4)}^A\partial_A H_0
 +\frac12\Delta W_{(2)}^A\Delta W_{(2)}^B
                      \partial_A\partial_B H_0
 -\Delta W_{(2)}^A\partial_A H_2 .
 \label{eq:quartic-background-bookkeeping}
\end{equation}
Here the inverse-map coefficient $\Xi_{(4)}^A$ includes the effect of
inverting the quadratic shifts; it is not fixed merely by setting the
forward fourth-order shift to zero. The cubic corrections to the
inhomogeneous canonical map can also contribute through $H_2$.
Appendix~\ref{app:canonical-completion} establishes the absence of a new
cubic term for its minimal map, but supplies neither these complete
fourth-order corrections nor their effective quantization. These
requirements join the second-order scalar shift, transverse shift, and
tensor sectors already listed above. None has been shown negligible in
the limit used for the selected contact term.
\par\endgroup

\begingroup
\subsubsection{Scalar quartic Legendre contribution}\label{subsec:scalar-quartic-Legendre}
In conformal time set $q=\nu$, $w=\nu'$, $P=w-\Gamma_Iq$, and $\theta={\mathsf S}(F_Iq+G_IP)$, where ${\mathsf S}=3\kappa(-\partial^2)^{-1}$ excludes the homogeneous mode. With ${\mathsf P}_\perp$ the projection onto the inhomogeneous perturbations, differentiation of the complete scalar cubic functional gives
\bqn 
J&=&\frac{\delta L_3}{\delta w}
=-a{\mathsf P}_\perp\left[2{\cal B}_IqP-{\cal C}_Iq^2+a^{-2}\partial_iq\partial_i\theta+G_I{\mathsf S}{\cal Z}\right],\nb\\
{\cal Z}&=&-\frac{{\cal D}_I}{2}\partial^2(q^2)-a^{-2}\partial_i(P\partial_iq)+2{\cal E}_I\left[\partial_i\partial_j(q\partial_i\partial_j\theta)-\partial^2(q\partial^2\theta)\right].
\label{eq:full-scalar-Legendre-functional}
\eqn 
The scalar quartic Hamiltonian therefore contains $H_{4,\rm Leg}=\frac12\int d^3x\,J^2$, evaluated at the quadratic momentum--velocity relation. This functional expression includes all cross terms fixed by the displayed cubic theory; it does not determine the independent quartic action.

In the local physical ultraviolet variables, write $\upsilon=\dot\varphi$, $R_i=\partial_i\nabla^{-2}$ and $\Pi_{ij}=\partial_i\partial_j\nabla^{-2}$, and define
\begin{equation}
U={\mathsf P}_\perp[-\varphi\upsilon+\partial_i\varphi R_i\upsilon-R_i(\upsilon\partial_i\varphi)].
\end{equation}
On the inhomogeneous sector, the quadratic functional in the lapse contact obeys
\begin{equation}
{\cal Q}=-{\mathsf P}_\perp[\varphi\upsilon-\Pi_{ij}(\varphi\Pi_{ij}\upsilon)].
\end{equation}
Thus $J_{\rm UV}=\alpha_IU-2\beta_I{\cal Q}$ and
\begin{equation}
H_{4,\rm Leg}^{\rm UV}=\int d^3y\left[\frac{\alpha_I^2}{2}U^2-2\alpha_I\beta_IU{\cal Q}+2\beta_I^2{\cal Q}^2\right].
\end{equation}
Combining the last integrand with $-{\cal L}_{4,n_2^2}=-2r_V\beta_I^2{\cal Q}^2$, where $r_V=V/K_\phi$, leaves $2(1-r_V)\beta_I^2{\cal Q}^2$. These two contributions cancel at $r_V=1$; at the reconstructed peak $1-r_V\simeq0.0018153$. The small fractional residual does not imply small coupling, since $\beta_I$ is large. This is a Hamiltonian contact-sector cancellation, not a cancellation of the complete four-point amplitude. A consistent amplitude calculation must use matching exchange and derivative-contact rules, include the omitted quartic sectors, and avoid adding this term again to an action calculation that already accounts for its counterpart. 

The identity for ${\cal Q}$ follows directly in Fourier space. For an ordered product $\varphi_{\bf a}\upsilon_{\bf b}$, write $A=|{\bf a}|^2$, $B=|{\bf b}|^2$, $d={\bf a}\cdot{\bf b}$, and $P=|{\bf a}+{\bf b}|^2=A+B+2d$. For $B,P\ne0$, the coefficient in the bracket defining $-\mathcal Q$ is
\begin{equation}
1-\frac{(B+d)^2}{PB}=\frac{AB-d^2}{PB},
\end{equation}
whereas Eq.~\eqref{eq:Qfunctional} gives its negative. 

\par\endgroup

\subsubsection{Contact--exchange test}

\begingroup
In a frozen local background, consider massless ultraviolet legs in the
cosmological rest frame. This is a local vertex diagnostic, not a global
cosmological S-matrix. Frequencies are signed in the all-incoming
convention, and the spatial inverse operators below exclude zero internal
spatial momentum. Defining
\par\endgroup
\begin{equation}
    C_{ab}
    \equiv
    1-
    \left(
    \widehat{\bf k}_a\cdot\widehat{\bf k}_b
    \right)^2,
\end{equation}
we fidn that the symmetric quadratic Fourier kernel of
Eq.~\eqref{eq:Qfunctional} is
\begin{equation}
    {\cal Q}_{ab}
    =
    \frac{iC_{ab}}
    {2|{\bf k}_a+{\bf k}_b|^2}
    \left(
       \omega_b k_a^2+
       \omega_a k_b^2
    \right).
    \label{eq:Qab_kernel}
\end{equation}
The corresponding leading contact vertex is obtained from the three pairings
$(12)(34)$, $(13)(24)$, and $(14)(23)$.  The cubic ${\cal E}_I$ vertex can
likewise be written as
\begin{equation}
 {\cal G}_E(a,b,c)
 =
 -\frac{\beta_I}{3}
 \left(
 C_{ab}\omega_a\omega_b
 +C_{bc}\omega_b\omega_c
 +C_{ca}\omega_c\omega_a
 \right).
 \label{eq:GE_vertex}
\end{equation}
A useful analytic family is obtained by taking equal physical external
energies $E$ and choosing
\begin{align}
 {\bf p}_1&=E(\sin\theta,0,\cos\theta),\nonumber\\
 {\bf p}_2&=E(-\sin\theta,0,\cos\theta),
 \nonumber\\
 {\bf p}_3&=E(\sin\theta\cos\phi,
               \sin\theta\sin\phi,\cos\theta),
 \nonumber\\
 {\bf p}_4&=E(-\sin\theta\cos\phi,
               -\sin\theta\sin\phi,\cos\theta),
 \label{eq:fourpoint_family}
\end{align}
where ${\bf p}_{1,2}$ are incoming and ${\bf p}_{3,4}$ are outgoing.
In the all-incoming convention used for the vertices, the outgoing legs are
represented with reversed four-momenta, or equivalently with negative signed
frequencies.
Introducing
\begin{equation}
    x\equiv\sin^2\theta,
     \quad
    z\equiv\sin^2\frac{\phi}{2},
\end{equation}
\begingroup
and writing the selected scalar contact-plus-exchange quantity as
\par\endgroup
\begin{equation}
    {\cal M}^{(6)}
    =
    \beta_I^2E^2\,{\cal C}(x,z),
    \label{eq:M6_def}
\end{equation}
\begingroup
the separate contributions below use the approximation $r_V\equiv V/K_\phi=1$, with the chosen vertex and propagator conventions. The algebraic sum can be checked
independently of the unresolved completeness of the quartic construction:
\par\endgroup
\begin{align}
 {\cal C}_{\rm cont}&=16x^2,\;\;
 {\cal C}_{s}=-16x^3, \;\; {\cal C}_{t}=16xz(1-xz)^2,
\quad{\cal C}_{u}=16x(1-z)
                 \left[1-x(1-z)\right]^2.
 \label{eq:C_channels}
\end{align}
Their coherent sum simplifies to
\begin{equation}
 {
 {\cal C}(x,z)
 =
 16x
 \left[
 1-x+x(4-3x)z(1-z)
 \right].}
 \label{eq:C_total}
\end{equation}

\begingroup
Retaining the actual ratio $r_V$ multiplies only the selected contact contribution by $r_V$. Thus
\begin{equation}
{\cal C}_{\rm cont}=16r_Vx^2,\qquad
{\cal C}_{r_V}(x,z)=16x\left[1+(r_V-2)x+x(4-3x)z(1-z)\right].
\end{equation}
At the representative enhancement maximum $r_V\simeq0.9981847$, the contact coefficient differs from the $r_V=1$ approximation by about $-0.182\%$.
\par\endgroup

For every interior nondegenerate configuration,
$0<x<1$ and $0<z<1$, all factors in the bracket are nonnegative and at
least one is strictly positive.  Hence
\begin{equation}
    {\cal C}(x,z)>0
    \label{eq:C_positive}
\end{equation}
throughout this two-parameter family.  The coefficient approaches zero only
on degenerate boundaries where the angular ${\cal E}_I$ vertex itself is
suppressed or an internal homogeneous mode requires separate treatment.
The maximum occurs at
\begin{equation}
    x=\frac23,
     \quad
    z=\frac12,
     \quad
    {\cal C}_{\rm max}=\frac{64}{9}\simeq7.11.
    \label{eq:Cmax}
\end{equation}
For the representative point $x=z=1/2$ one obtains
${\cal C}=13/2$.

As a broader check within the same $r_V=1$ approximation, the reported exploratory sample contains $10^5$ random massless $2\to2$
configurations.  The configurations were generated in a center-of-momentum
frame and then transformed by random boosts with
$0.02<|{\bf v}|<0.97$ relative to the cosmological frame.  Configurations
with an exactly homogeneous internal spatial momentum or an internal
propagator arbitrarily close to its pole were excluded.  Defining
$\bar E=(E_1+E_2)/2$ and
${\cal M}^{(6)}=\beta_I^2\bar E^2{\cal C}$, the scan gave
\begin{center}
\begin{tabular}{c|cccccc}
 & min & 5\% & median & 95\% & 99\% & max\\
\hline
${\cal C}$
& $1.25\times10^{-3}$
& $0.472$
& $2.691$
& $4.996$
& $6.066$
& $7.084$
\end{tabular}
\end{center}
No negative coefficient was found in the $10^5$ accepted configurations.
The very small minimum occurs near a degenerate boosted geometry and should
not be interpreted as a positive lower bound on the full kinematic domain.
Near $t\simeq-5.803\,t_{\rm P}$,
Eq.~\eqref{eq:tinyLambdaE} gives
\begin{equation}
    \Lambda_E
    =
    |\beta_I|^{-1}
    \simeq4.35\times10^{-9}\,m_{\rm P}.
\end{equation}
Thus, a generic coefficient ${\cal C}=O(1)$ corresponds parametrically to
\begin{equation}
    |{\cal M}^{(6)}|
    \sim
    {\cal C}
    \left(\frac{E}{\Lambda_E}\right)^2.
    \label{eq:M6_scaling}
\end{equation}
\begingroup
For illustration, evaluating
Eq.~\eqref{eq:M6_scaling} at the edge of the reduction regime,
$p_{\rm red}\simeq0.416\,m_{\rm P}$, already gives values of order
$10^{16}$ for ${\cal C}=O(1)$; deeper in the ultraviolet the reduced
quantity grows as $E^2$.  The numerical value at $p=p_{\rm red}$ should not
itself be interpreted as a precision determination of a cutoff, since the
asymptotic reduction is only beginning to become valid there.
\begingroup
The combined results establish survival of the enhanced cubic coefficient
and failure of cancellation in a specified subset of four-point terms.
\par\endgroup  The enhanced
$\alpha_I^3/\kappa$ term survives the complete leading cubic kernel for
generic momentum triangles, and within the minimal factorized scalar
quartic continuation its four-point effect is not removed by the leading
$n_{2I}^2$ contact term.  It must ne noted that a final, robust value of
$\Lambda_{\rm sc}$ nevertheless requires the complete fourth-order hybrid
construction, including the second-order scalar shift, transverse shift,
and tensor sectors, together with any additional factor-ordering choices
that first appear at quartic order.  \begingroup
Until those terms and their joint power counting are established, the
quartic test is evidence of non-cancellation only between the retained
contact and exchange contributions. 
\par\endgroup

\subsection{{Dependence of formal ultraviolet suppression scales on the potential}}
\label{subsec:potential-dependence}

The mechanism described above is controlled mainly by the universal mLQC-I
pre-bounce background, but its quantitative strength depends on the
normalization of the inflaton potential.  This distinction is important.
The pre-bounce background evolution for a broad class of kinetic-energy-dominated
bounce data was shown in Ref.~\cite{Saeed:2024xhk} to depend only
weakly on the detailed form of the potential, including chaotic,
Starobinsky, generalized Starobinsky, polynomial, generalized T-model, and
natural-inflation examples.  The strong-coupling scale, however, inherits a
sensitive dependence on the value of the potential during the short
potential--kinetic crossover.
\begin{table*}[t]
\centering
\caption{Representative dependence of the pre-bounce enhancement on the
inflationary potential.  The chaotic and polynomial-chaotic ranges span the
Planck, ACT, and combined $(A_s,n_s)$ normalizations of
Ref.~\cite{BrownEtAl2026EPJC}; the PLP normalization is fixed as in
Ref.~\cite{YogeshEtAl2026APSOS}.  The quoted $\Lambda_E$ is the formal
$E$-sector suppression scale defined in Eq.~\eqref{eq:Lambda_individual},
not a determined physical cutoff $\Lambda_{\rm sc}$. The corrected chaotic and polynomial rows use the source-unit normalization. Each column gives its own minimum--maximum range; extrema in different columns need not belong to the same calibration.}
\label{tab:potential_dependence}
\begin{ruledtabular}
\begin{tabular}{lccc}
Potential & $V_s/m_{\rm P}^4$ & $|\alpha_I|_{\rm max}m_{\rm P}$ &
$\Lambda_{E,{\rm min}}/m_{\rm P}$ \\
\hline
Starobinsky
& $3.81\times10^{-8}$
& $2.26\times10^{3}$
& $4.35\times10^{-9}$ \\
PLP, $m=m_{\rm P}$
& $1.55\times10^{-12}$
& $3.55\times10^{5}$
& $1.12\times10^{-15}$ \\
{Chaotic}
& {$(1.6173$--$3.1820)\times10^{-13}$}
& {$(0.78286$--$1.0981)\times10^{6}$}
& {$(3.7965$--$10.477)\times10^{-17}$} \\
{Polynomial chaotic, $\phi_*>0$}
& {$(2.1485$--$4.2280)\times10^{-13}$}
& {$(0.67914$--$0.95272)\times10^{6}$}
& {$(5.8127$--$16.047)\times10^{-17}$} \\
{Polynomial chaotic, $\phi_*<0$}
& {$(3.2389$--$8.3502)\times10^{-14}$}
& {$(1.5282$--$2.4538)\times10^{6}$}
& {$(3.4023$--$14.084)\times10^{-18}$} \\
\end{tabular}
\end{ruledtabular}
\end{table*}

To make this dependence explicit, let $t_s$ denote the epoch at which
$|\alpha_I|$ reaches its pre-bounce maximum and define
\begin{equation}
 V_s\equiv V(\phi_s),
  \quad
 K_{\phi,s}\equiv K_\phi(t_s)
 =\frac12\dot\phi_s^{\,2}.
 \label{eq:Vs_definition}
\end{equation}
While the geometry is still close to the pre-bounce de Sitter phase,
$\sigma_I\simeq\gamma H_\Lambda$, and the potential changes slowly over the
crossover.  Equation~\eqref{eq:alphaI} may then be written as
\begin{equation}
 |\alpha_I|
 \simeq
 \frac{3H_\Lambda\sqrt{2K_\phi}}
 {2(K_\phi+V)}.
 \label{eq:alpha_KV}
\end{equation}
For fixed slowly varying $V$, the right-hand side is maximal at
\begin{equation}
 K_{\phi,s}\simeq V_s,
  \quad
 \rho_{\phi,s}\simeq2V_s,
 \label{eq:KsVs}
\end{equation}
which gives
\begin{equation}
 {
 |\alpha_I|_{\rm max}
 \simeq
 \frac{3H_\Lambda}{2\sqrt{2V_s}}
 }
 \label{eq:alphamax_Vs}
\end{equation}
and hence
\begin{equation}
 {
 \Lambda_{E,{\rm min}}
 \simeq
 \frac{32\sqrt{2}\,\kappa}{27}
 \frac{V_s^{3/2}}{H_\Lambda^3}
 }.
 \label{eq:LambdaE_Vs}
\end{equation}
Thus the occurrence of the enhancement and its numerical severity are two
separate questions.  The former is robust whenever the background reaches
the crossover while remaining close to the mLQC-I de Sitter phase, whereas
the latter is highly sensitive to the normalization through
$\Lambda_E\propto V_s^{3/2}$.

A useful reference scale follows from $|\alpha_I|_{\rm max}^2/\kappa\sim1$,
namely
\begin{equation}
 V_{\rm crit}
 \equiv
 \frac{9H_\Lambda^2}{8\kappa}
 \simeq{7.8\times10^{-3}}\,m_{\rm P}^4.
 \label{eq:Vcrit}
\end{equation}
\begingroup
The separate formal equality $\Lambda_{E,\min}=H_\Lambda$ would instead give
\begin{equation}
 V_{E=H}=\left(\frac{27}{32\sqrt2\kappa}\right)^{2/3}H_\Lambda^{8/3}.
\end{equation}

Therefore observationally relevant models with
$V_s\ll V_{\rm crit}$ naturally satisfy
$|\alpha_I|^2/\kappa\gg1$.  Conversely, an arbitrarily chosen sufficiently
large normalization could raise $V_s$ and weaken, or in principle remove,
the enhancement.  For this reason the potential parameters used below are
{specified for each example. The chaotic and polynomial ranges use the stated calibrations, whereas the Starobinsky row is a representative Paper-I benchmark and must not be relabeled as the reheating calibration.}

For the chaotic and polynomial-chaotic potentials we use the mass parameters
obtained from the measured pair $(A_s,n_s)$ in
Ref.~\cite{BrownEtAl2026EPJC}.  For the polynomial model,
\begin{equation}
 V(\phi)=\frac12m^2\phi^2
 \left(1-\alpha_1\phi+\alpha_2\phi^2\right)^2,
 \label{eq:poly_potential_compare}
\end{equation}
\begingroup
we use $\alpha_1m_{\rm P}=0.14$ and
$\alpha_2m_{\rm P}^2=6.644\times10^{-3}$ and include the two calibration
branches $\phi_*>0$ and $\phi_*<0$. The calibration equations of
Ref.~\cite{BrownEtAl2026EPJC} use the unreduced convention
$m_{\rm P}=G^{-1/2}$. Its numerical masses and polynomial coefficients
must therefore not be subjected to an additional reduced-to-unreduced
conversion. For example, for the quadratic potential the slow-roll
calibration yields
\begin{equation}
 \frac{m}{m_{\rm P}}=
 \frac{\sqrt{3\pi A_s}}{4}(1-n_s)
 \simeq1.2342\times10^{-6}
\end{equation}
for $A_s=2.0989\times10^{-9}$ and $n_s=0.9649$. A further division
by $\sqrt{8\pi}$ would change the intended normalization.
\par\endgroup  For these two
families we use the same representative kinetic-energy-dominated bounce condition
$\phi_B=m_{\rm P}$ and $\dot\phi_B>0$.  The power-law-plateau (PLP) example
uses
\begin{equation}
 V(\phi)=V_0\,m_{\rm P}^4\frac{\phi^2}{\phi^2+m^2},
  \quad
 m=m_{\rm P},
 \label{eq:PLP_compare}
\end{equation}
with $V_0\simeq1.63\times10^{-12}$ fixed by the scalar-amplitude
normalization in Ref.~\cite{YogeshEtAl2026APSOS}.

\begingroup
Table~\ref{tab:potential_dependence} retains the Starobinsky and PLP
benchmarks and adopts the corrected chaotic and polynomial ranges from
the numerical cross-check after removing the extra unit conversion.
The Starobinsky entry is the
representative background used throughout the preceding subsections.  The
PLP example uses the symmetry-related initial data
$\phi_B=-4\,m_{\rm P}$, $\dot\phi_B>0$; reversing both signs gives the same
magnitudes.  For the chaotic and polynomial cases, the ranges span the
Planck, ACT, and combined observational normalizations.  In every case
$|\alpha_I|_{\rm max}^2/\kappa\gg1$, while the value of
$\Lambda_{E,{\rm min}}$ varies by many orders of magnitude.


The two polynomial branches provide an instructive internal check.  The
$\phi_*<0$ branch is associated with a smaller observationally determined
mass parameter, and therefore with a smaller $V_s$.  Equation~\eqref{eq:LambdaE_Vs}
\begingroup
then predicts a lower $\Lambda_E$, consistent with the corrected
reported ranges.
The sign of
$\phi_*$ is not itself responsible for the nonlinear enhancement; rather, it
selects a different observational branch and hence a different potential
normalization.

The conclusion is therefore more precise than either strict potential
independence or model-specific strong coupling.  Within the kinetic-energy-dominated
\begingroup
mLQC-I backgrounds sampled here, the pre-bounce coefficient enhancement
persists in the corrected examples, while its quantitative severity is
strongly potential dependent through $V_s$. This supports robustness of
the formal enhancement, not a proof of a physical strong-coupling cutoff
for every potential. 

\FloatBarrier

\subsection{Comparison with the adiabaticity violation in the linear theory}
\label{subsec:adiabaticity-comparison}

It is useful to compare the nonlinear enhancement found above with the
nonadiabatic evolution of the linear perturbations identified in Paper~I.
For the quadratic MS equation, Paper~I used the WKB diagnostic
\begin{equation}
 {\cal J}_{\rm wkb}(t)
 =
 \left|
 \frac{3{\omega_k'}^2}{4\omega_k^4}
 -\frac{\omega_k''}{2\omega_k^3}
 \right|,
  \quad
 \omega_k^2=k^2+m_{\rm eff}^2,
 \label{eq:Jwkb_compare}
\end{equation}
\begingroup
Here primes denote conformal-time derivatives. With
$s=m_{\rm eff}^2$ and $q=k^2+s$, the algebraic expression is
\begin{equation}
 {\cal J}_{\rm alg}=
 \left|\frac{5(s')^2}{16q^3}-\frac{s''}{4q^2}\right|,
 \qquad s'=a\dot s,\quad s''=a^2(\ddot s+H\dot s).
\end{equation}
It is an oscillatory positive-frequency WKB diagnostic only where $q>0$.
Its continuation to $q<0$ can be plotted algebraically but describes an
exponential-mode interval, not an oscillatory adiabatic vacuum. At $q=0$
the expression is undefined. For finite
and intermediate wavenumbers, ${\cal J}_{\rm wkb}$ becomes large during a
pre-bounce interval, with stronger violations for smaller $k$.  Its sharp
behavior at a turning point, $\omega_k^2=0$, signals the breakdown of the WKB
expansion and is not itself a divergent particle number.  At the same time,
the strict ultraviolet tail remains adiabatic under the smooth exact linear
evolution, as established in Paper~I.

Figure~\ref{fig:adiabaticity-strong-coupling} compares this linear diagnostic
with the two background quantities most directly relevant to the ultraviolet
interaction analysis: $|\alpha_I|$, defined in
Eq.~\eqref{eq:alphaI}, and the reduction scale $p_{\rm red}$, defined in
Eq.~\eqref{eq:p_reduction}.  All curves are evaluated on the same
representative background used in the numerical analysis above.  To place
the quantities on a common dimensionless logarithmic scale, the figure shows
$|\alpha_I|m_{\rm P}$ and $p_{\rm red}/m_{\rm P}$ together with
${\cal J}_{\rm wkb}$ for $k/m_{\rm P}=0.1$, $1$, and $4$, with dashed
segments marking the algebraic continuation into $q<0$. The dotted and
dash-dotted curves are used for $|\alpha_I|m_{\rm P}$ and
$p_{\rm red}/m_{\rm P}$, respectively, so that the comparison remains clear
in black-and-white reproduction.

\begingroup
The independent background reconstruction has an inflaton turning point near
$t_0/t_{\rm P}=-11.410934899$, with $\dot\phi(t_0)=0$ and
$V_{,\phi}(t_0)\ne0$. Equation~\eqref{eq:p_reduction} therefore gives
an unbounded $p_{\rm red}$ on approach to $t_0$. This marks failure of
the $G_I{\cal P}_{\nu I}$-dominated reduction, not a demonstrated
singularity of the exact interaction kernel. The regenerated plot resolves this feature and distinguishes $q<0$
intervals from oscillatory WKB failure.
\par\endgroup

\begin{figure}[htp] 
 \centering
 \includegraphics[width=0.86\textwidth]{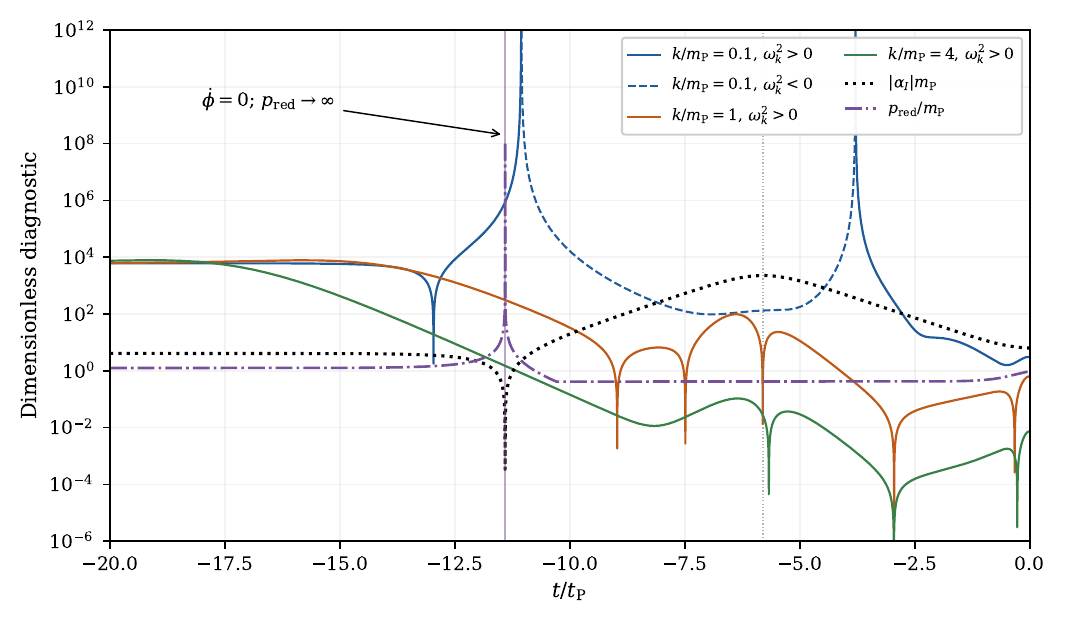}
 \caption{{Independent background reconstruction of the
 linear diagnostic and nonlinear background coefficients for the Starobinsky benchmark $m_*=3.08\times10^{-6}m_{\rm P}$, $\phi_B=-1.09m_{\rm P}$, $\gamma=0.2375$, and $a_B=1$. The horizontal axis is $t/t_{\rm P}$; the mode curves have $k/m_{\rm P}=0.1,1,4$. Solid colored
 curves show $\mathcal J_{\rm wkb}$ where $\omega_k^2>0$; dashed
 colored segments show its algebraic continuation where $\omega_k^2<0$.
 The black dotted curve is $|\alpha_I|m_{\rm P}$ and the purple
 dash-dotted curve is $p_{\rm red}/m_{\rm P}$. The reduction scale is
 unbounded at $\dot\phi=0$, near $t/t_{\rm P}=-11.410934899$; its
 finite plotted height reflects sampling, not a finite limiting value.
 WKB turning-point peaks do not represent divergent particle production.
 The inflaton turning point invalidates the $G_I$-dominated reduction,
 not necessarily the exact finite-mode kernel.}}
 \label{fig:adiabaticity-strong-coupling}
\end{figure}
\endgroup

\begingroup
Evaluating the benchmark at the maximum of $|\alpha_I|$, where
\par\endgroup
\begin{equation}
 t=t_s\simeq{-5.8028547104}\,t_{\rm P},
\end{equation}
{gives}
\begin{equation}
 |\alpha_I|\simeq2.26\times10^3\,m_{\rm P}^{-1},
  \quad
 p_{\rm red}\simeq0.416\,m_{\rm P},
 \label{eq:alpha_pred_adiabatic_compare}
\end{equation}
while the corresponding linear diagnostics are approximately
\bqn 
 {{\cal J}_{\rm alg}(0.1)}&\simeq&1.32\times10^2,
  \quad
 {\cal J}_{\rm wkb}(1)\simeq{0.88},
 \quad
 {\cal J}_{\rm wkb}(4)\simeq2.7\times10^{-2}.
 \label{eq:Jwkb_at_alphamax}
\eqn 
The strict
ultraviolet modes are more adiabatic still, even though the nonlinear
interaction coefficients become increasingly problematic there.

The two phenomena should therefore be regarded as correlated but not
identified.  The violation of adiabaticity is a property of the quadratic
frequency $\omega_k(t)$ and is driven by the rapid variation of the effective
mass during the pre-bounce transition.  The strong-coupling enhancement, by
contrast, originates from the nonlinear coefficient
\begin{equation}
 \alpha_I=\frac{3\sigma_I\dot\phi}{2\gamma\rho_\phi},
\end{equation}
which becomes very large because the quantum-geometrical de Sitter curvature
remains of Planck order, supported by the effective $\rho_\Lambda$, while
the inflaton energy density $\rho_\phi$ becomes small and
$\dot\phi/\rho_\phi$ grows rapidly.  \begingroup
The same background transition can therefore affect linear mode
evolution and enhance nonlinear coefficients. 
\par\endgroup

This distinction is conceptually important for the relation between Paper~I
and the present analysis.  Ultraviolet adiabatic regularity of the linear
state does not by itself guarantee weak coupling of the nonlinear theory.
\begingroup
Within the adopted nonlinear continuation, the strict-ultraviolet
quadratic modes can remain adiabatic while cubic coefficients become
large. This demonstrates the independence of the two diagnostics.
\par\endgroup

\FloatBarrier

\section{Initial conditions at the bounce as an alternative effective prescription}
\label{sec:bounce-initial-data}

The strong-coupling results of Sec.~\ref{sec:strong-coupling} were obtained
under the assumption that the effective perturbation theory is defined on the
contracting phase and that the perturbative state is evolved through the
pre-bounce regime.  There is, however, a logically distinct prescription.  One
may regard the quantum bounce as the earliest hypersurface on which an
effective field-theoretic description of the perturbations is imposed, specify
the perturbative state at $t=t_B$, and evolve only toward the expanding phase.
This prescription does not use the effective pre-bounce geometry to prepare
the perturbative state; instead, the state is specified directly on the bounce
hypersurface and evolved only toward the expanding phase.  The severe
pre-bounce enhancement
found above is then bypassed by construction.  The relevant question becomes
whether the perturbative theory is internally consistent on the bounce
hypersurface and immediately afterward.

\begingroup
A bounce-defined prescription is also examined in Paper~II from the
complementary viewpoint of correlation functions. Quantitative comparison
requires matching the finite-momentum mode data, background conventions,
and ultraviolet completion in the two calculations
\cite{PanSaeedWang2026PaperII}.  There the emphasis is on the primordial
\begingroup
bispectrum and the retained nonlinear-conversion correction to the
curvature power spectrum. Here we examine the cubic kernel on the bounce
hypersurface, a possible ultraviolet completion, and the requirements for
backreaction and subsequent evolution. Local bounce results alone do not
establish post-bounce interacting control.
\par\endgroup

\subsection{A bounce-defined quadratic state}

\begingroup
As a first diagnostic, consider the mode-by-mode instantaneous ground
state of the quadratic Hamiltonian in the canonical pair $(\nu,\pi_\nu)$
and conformal-time lapse $N=a$. This prescription depends on the chosen
canonical Hamiltonian and is not a uniquely selected cosmological vacuum.
\par\endgroup  From Eq.~\eqref{eq:H2_momentum}, define
\begin{equation}
 \omega_{kB}^2=k^2+m_{{\rm eff},B}^2,
\end{equation}
and impose
\begin{equation}
 {
 \nu_k(t_B)=\frac{1}{\sqrt{2\omega_{kB}}},
  \quad
 \nu'_k(t_B)=-i\omega_{kB}\nu_k(t_B).
 }
 \label{eq:bounce-instantaneous-vacuum}
\end{equation}
\begingroup
The prime denotes a conformal-time derivative, even though the initial
surface is labeled by proper time $t_B$. These data satisfy
$\nu_k\nu_k^{*\prime}-\nu_k^*\nu_k'=i$. For the shifted momentum,
the corresponding mode data are
\begin{equation}
 {\cal P}_{\nu I,k}(t_B)=
 (-i\omega_{kB}-\Gamma_B)\nu_k(t_B).
\end{equation}
Thus one must not impose ${\cal P}_{\nu I,k}=-i\omega_{kB}\nu_k$
in place of $\pi_{\nu,k}=-i\omega_{kB}\nu_k$ for this state.
\par\endgroup

For the representative Starobinsky background used throughout the numerical
\begingroup
analysis, we use $a_B=1$, $\phi_B=-1.09m_{\rm P}$,
$\dot\phi_B>0$, $\gamma=0.2375$, and
$\lambda^2=4\sqrt3\pi\gamma\ell_{\rm P}^2$. The local numerical
values correspond to
\begin{equation}
 V(\phi)=\frac{3m_\star^2}{4\kappa}
 \left(1-e^{-\sqrt{2\kappa/3}\,\phi}\right)^2,
 \qquad m_\star=3.08\times10^{-6}m_{\rm P}.
\end{equation}
Using the effective bounce constraint to set
$\dot\phi_B=\sqrt{2(\rho_B-V_B)}$, the quadratic mass and momentum
shift are
\par\endgroup
\bqn 
 m_{{\rm eff},B}^2&\simeq&0.81161\,m_{\rm P}^2,
  \quad
 m_{{\rm eff},B}\simeq0.90089\,m_{\rm P},
  \quad
 \Gamma_B\simeq0.92463\,m_{\rm P}.
 \label{eq:bounce-meff-numbers}
\eqn 
\begingroup
Thus $\omega_{kB}^2>0$ for every nonzero perturbative mode in this
representative background. The normalized oscillator data have energy
$\tfrac12(|\nu_k'|^2+\omega_{kB}^2|\nu_k|^2)=\omega_{kB}/2$,
as required for this quadratic ground state. Positivity is a property of
this background, not a result proved here for arbitrary bounce data or
inflationary potentials.
\par\endgroup

Equation~\eqref{eq:bounce-instantaneous-vacuum} is used here only as a clean
state-dependent diagnostic.  It should not yet be interpreted as the final
physical initial state.  In particular, a state suitable for a renormalized
stress tensor must possess the required ultraviolet regularity, and its
renormalized perturbative energy density must remain small compared with the
\begingroup
background. Applying the instantaneous prescription at every momentum
does not by itself establish the ultraviolet regularity required for a
renormalized stress tensor. It is used here for the finite modes tested;
the ultraviolet completion and backreaction are separate questions below.
A check on the bounce hypersurface also does not replace evolution of this
same state after the bounce.
\par\endgroup

\subsection{Exact cubic kernel on the bounce hypersurface}

We now evaluate the complete symmetrized kernel of
Sec.~\ref{subsec:momentum_strong_coupling} directly at $t=t_B$, using
Eq.~\eqref{eq:bounce-instantaneous-vacuum}.  No strict-ultraviolet reduction is
made.  In particular,
\begin{equation}
 X_i\equiv\frac{{\cal P}_{\nu{\rm I},k_i}}{\nu_i}
 =-i\omega_{k_iB}-\Gamma_B,
  \quad
 Y_i\equiv F_{I,B}+G_{I,B}X_i,
\end{equation}
are inserted into the exact symmetrized cubic kernel, while the quadratic
normalization is the symmetric quantity ${\cal Q}_{2,\triangle}$ of
Eq.~\eqref{eq:Q2-triangle}.

For equilateral configurations the resulting full-kernel ratio is
\begin{center}
\begin{tabular}{c|cccccc}
$K/m_{\rm P}$ & $0.1$ & $0.2$ & $0.5$ & $0.6483$ & $1$ & $5$ \\ \hline
${\cal R}_3^{\rm eq}(t_B)$
& $0.0874$ & $0.239$ & $0.773$ & $1.000$ & $1.372$ & $2.030$
\end{tabular}
\end{center}
A refined root search gives
\begin{equation}
 {
 K_{{\rm pert},B}^{(3)}\simeq0.64830\,m_{\rm P},
 }
 \label{eq:bounce-state-kpert}
\end{equation}
for the first equilateral crossing ${\cal R}_3=1$.  Since $a_B=1$, this is
also the corresponding physical momentum in the normalization used here.
\begingroup
The quoted crossing is independently reproduced as
$K/m_{\rm P}=0.6482958319$ using the exact finite-mode kernel and
background inputs specified in the preceding subsection. At this scale,
shape grids with $625$, $2500$, and $10000$ points all attain their
sampled maximum at the equilateral configuration. These grids use
$r\in[1/2,1]$ and
$s\in[\max(1-r,10^{-5}),r]$, including the collinear boundary when
 all legs are nonzero. 
At $K=m_{\rm P}$,
for example, we find
\begin{equation}
 {\cal R}_3^{\rm eq}\simeq1.372,
  \quad
 {\cal R}_3^{\rm flat}\simeq0.691,
  \quad
 {\cal R}_3^{\rm sq}\simeq0.0759,
 \label{eq:bounce-state-shapes}
\end{equation}
where the flattened and squeezed representatives are
$(K,K/2,K/2)$ and $(K,K,0.1K)$, respectively.

\begin{figure}[t]
 \centering
 \includegraphics[width=0.76\textwidth]{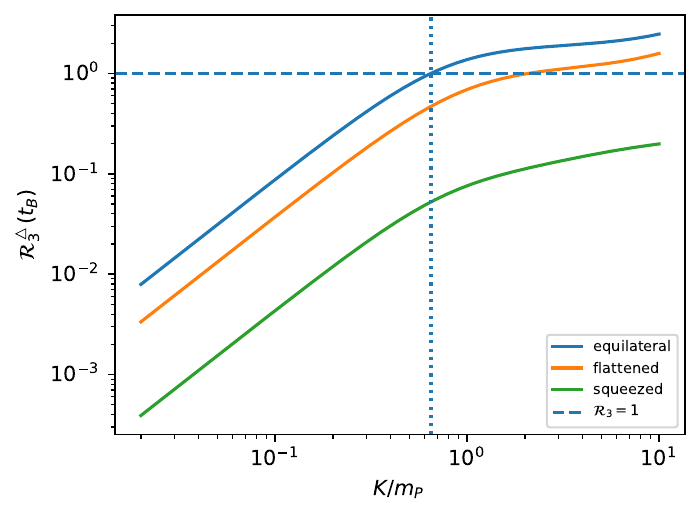}
 \caption{Exact cubic-to-quadratic ratio on the bounce hypersurface for the
 instantaneous quadratic ground state in Eq.~\eqref{eq:bounce-instantaneous-vacuum}. {The shapes are $(K,K,K)$, $(K,K/2,K/2)$, and $(K,K,0.1K)$, with $a_B=1$. The unity line marks the chosen diagnostic boundary; parametrically weak cubic corrections require ${\cal R}_3\ll1$.}
 The equilateral configuration is the most restrictive of the representative
 shapes and crosses ${\cal R}_3=1$ at
 $K\simeq0.6483\,m_{\rm P}$.  The plotted quantity is state dependent and is
 therefore a perturbativity diagnostic rather than a state-independent EFT
 cutoff.}
 \label{fig:bounce-initial-kernel}
\end{figure}

This result should be compared with the threshold
$k_{{\rm pert},B}^{(3)}\simeq0.21831\,m_{\rm P}$ obtained in
Sec.~IV.C when the modes are evolved to the bounce from the asymptotic
pre-bounce de Sitter state of Paper~I.  The shift of the threshold by a factor of
approximately three is a direct demonstration that ${\cal R}_3$ is
state dependent.  
\begingroup
For the specified states, the local bounce diagnostic therefore has a
larger below-unity momentum range. Although this comparison does not establish
that the bounce-defined state remains controlled during subsequent
evolution, it does not,
however, modify the background-dependent interaction coefficients themselves.
In particular, we have
\begin{equation}
 \alpha_B\simeq6.30175\,m_{\rm P}^{-1},
  \quad
 \frac{\alpha_B^2}{\kappa}\simeq1.58,
\end{equation}
so there is still no parametrically small nonlinear coefficient at the
bounce.  The exact finite-momentum kernel, rather than the strict-ultraviolet
operator estimate, is therefore essential for assessing bounce initial data.


\begingroup
An independent finite-time check evolves this same bounce-defined state
with the exact linear hybrid mode equation from $t_B=0$ to $10t_{\rm P}$
and evaluates the exact equilateral cubic diagnostic along that evolution.
Eight momenta were tested:
$k/m_{\rm P}=0.02,\,0.05,\,0.1,\,0.10951994,\,0.2,\,0.3,\,0.5$,
and $0.6482958319$. Representative results are
\begin{center}
\begin{tabular}{ccc}
$k/m_{\rm P}$ & ${\cal R}_3^{\rm eq}(t_B)$ &
$\max_{\text{sampled }t\in[0,10t_{\rm P}]}{\cal R}_3^{\rm eq}$ \\
\hline
$0.1$ & $0.087446$ & $0.087722$ \\
$0.2$ & $0.239245$ & $0.239991$ \\
$0.5$ & $0.773427$ & $0.775678$ \\
$0.648296$ & $1.000000$ & $1.002831$
\end{tabular}
\end{center}
The diagnostic initially rises slightly, with these sampled maxima occurring
near $t/t_{\rm P}=0.028$--$0.032$, before decreasing. Thus the local
bounce crossing is not itself a threshold ensuring that the diagnostic
stays below unity after the bounce. At the tested momentum whose bounce
value is approximately $0.1$, the subsequent sampled maximum is
$0.100316$; a tighter control level likewise requires a margin inside its
bounce crossing.
\par\endgroup

\begingroup
A further finite-domain scan tests $261$ shape entries at each of
$K/m_{\rm P}=0.1,0.2,0.5,0.64,0.6482958319$ over
$0\le t/t_{\rm P}\le10$. It combines a $16\times16$ grid with
$r\in[1/2,1]$, $s\in[\max(1-r,0.001),r]$, and explicit equilateral,
flattened, and squeezed entries $(r,s)=(1,1),(1/2,1/2),(1,0.1),
(1,0.01),(1,0.001)$. Entries can coincide at shared grid boundaries.
The time sampling uses spacing $0.0005t_{\rm P}$ through $0.1t_{\rm P}$
and $0.01t_{\rm P}$ thereafter. At every tested $K$, the sampled maximum
is equilateral. The maxima are approximately $0.0877224$, $0.239991$,
$0.775678$, $0.991257$, and $1.002831$, respectively, occurring near
$t/t_{\rm P}=0.028$--$0.032$. Thus none of the sampled nonequilateral shapes strengthens the early-time
constraint at these five momenta.

For the equilateral family, a subsequent root search refines the crossing of
$\max_{0\le t/t_{\rm P}\le10}{\cal R}^{\rm eq}_3(K,t)=1$ to
\begin{equation}
 K_{\mathrm{pert},[0,10]}^{(3),\mathrm{eq}}
 \simeq 0.64626\,m_{\rm P}.
 \label{eq:equilateral-evolved-threshold}
\end{equation}
The peak occurs at
$t\simeq0.02802t_{\rm P}$, while the bounce value at this momentum is
${\cal R}^{\rm eq}_3(K,0)\simeq0.99718$. This is about
$0.31\%$ below the bounce-only crossing.
\par\endgroup

\begingroup
\subsubsection{Independent frequency assignments and separate-term thresholds}
For each leg we evaluate both the mode and its complex conjugate before assembling the cubic vertex. Simultaneous conjugation of all legs conjugates the full kernel, so the four assignments $+++$, $-++$, $+-+$ and $++-$ cover the eight moduli. At the bounce, the first equilateral unity crossings are $K=0.6482958319m_{\rm P}$ for $+++$ and $K=0.9248868678m_{\rm P}$ for the mixed assignments. Evolving the same bounce state over $0\le t/t_{\rm P}\le10$ gives the respective equilateral crossings $0.6462578064m_{\rm P}$ and $0.9231409356m_{\rm P}$. Refined bounce shape samples and a finite evolution scan including all four assignments find no stronger channel near the first crossing. The additional bounce momentum sample spans $10^{-5}\le K/m_{\rm P}\le0.6482958319$; the evolution shape tests cover $K/m_{\rm P}=0.1,0.2,0.5,0.64,0.6462578064,0.6482958319$ over the same finite time interval. These additional channel tests support the quoted local window on the tested domain.

For the equilateral bounce state, the first separate-monomial crossing occurs at $K\simeq0.4048092m_{\rm P}$, whereas the full coherent kernel crosses at $0.6482958m_{\rm P}$. Combining the $C+D$ terms before testing separate sectors moves the first grouped-sector crossing to $1.0638995m_{\rm P}$. The absolute sum of all eight monomial contributions reaches unity at $0.1689122m_{\rm P}$. At the coherent crossing, the separate $C$, $D$-Laplacian and $D$-gradient ratios are approximately $1.53450$, $1.61643$ and $0.80822$, respectively. These differences explicitly demonstrate the effects of grouping and interference. 
\par\endgroup

\subsection{Ultraviolet regularity of bounce-defined initial data}
\label{subsec:bounce-uv}

The instantaneous state in Eq.~\eqref{eq:bounce-instantaneous-vacuum} was
introduced as a transparent finite-mode diagnostic.  Before using it as part
of a physical initial-state prescription, one must verify compatibility with
the ultraviolet requirements of adiabatic renormalization.  This issue can be
addressed rather directly by using the fourth-order bounce construction of
Paper~I. Writing
\begin{equation}
 s(\eta)\equiv m_{\rm eff}^2(\eta)
 =U_{\rm eff}(\eta)-\frac{a''}{a},
\end{equation}
the fourth-order adiabatic frequency has the ultraviolet expansion
\begin{equation}
 W_k^{(4)}
 =k+\frac{s}{2k}
 -\frac{s^2+s''}{8k^3}
 +{\cal O}(k^{-5}),
 \label{eq:bounce-W4}
\end{equation}
with
\begin{equation}
 W_k^{(4)\prime}
 =\frac{s'}{2k}
 -\frac{2ss'+s'''}{8k^3}
 +{\cal O}(k^{-5}).
 \label{eq:bounce-W4prime}
\end{equation}
\begingroup
For sufficiently large $k$, choose a smooth positive real representative
of this asymptotic frequency and its real conformal-time derivative.
The truncated large-$k$ series is not assumed to remain positive at small
$k$. The corresponding ultraviolet adiabatic initial data are
\par\endgroup
\begin{equation}
 \nu_k(\eta_B)=\frac{1}{\sqrt{2W_{k,B}^{(4)}}},
  \quad
 \nu_k'(\eta_B)=
 \left[-\frac{W_{k,B}^{(4)\prime}}{2W_{k,B}^{(4)}}
 -iW_{k,B}^{(4)}\right]\nu_k(\eta_B).
 \label{eq:bounce-adiabatic-state}
\end{equation}

\begingroup
For the specified quadratic energy observable, define $U_{\rm eff}=s+a''/a$ and
\begin{equation}
 {I_k^{\rm bare}=|\nu_k'-(a'/a)\nu_k|^2+(k^2+U_{\rm eff})|\nu_k|^2.}
\end{equation}
 Its bare density convention is $(4\pi^2a^4)^{-1}\int dk\,k^2 I_k^{\rm bare}$.
For this quadratic energy integrand and its matching
fourth-order subtraction $C^{(Q)}$ used in Paper~I, the stated
ultraviolet matching condition is
\par\endgroup
\begin{equation}
 I_k^{\rm bare}-C^{(Q)}(k,\eta_B)={\cal O}(k^{-5}).
 \label{eq:bounce-adiabatic-finite}
\end{equation}
\begingroup
If the energy integral has radial measure $k^2dk$, this remainder gives
an integrable ultraviolet tail. 
\par\endgroup

There is a useful additional simplification at the bounce itself.  Since
$a_B=1$ and $a'_B=0$, the instantaneous state of
Eq.~\eqref{eq:bounce-instantaneous-vacuum} gives
\begin{equation}
 I_{k,B}^{\rm inst}
 =\frac12\left[
 \omega_{kB}+\frac{k^2+U_{{\rm eff},B}}{\omega_{kB}}
 \right],
  \quad
 \omega_{kB}^2=k^2+s_B.
\end{equation}
Using $s_B=U_{{\rm eff},B}-a_B''/a_B$ and expanding for large $k$,
\begin{equation}
 I_{k,B}^{\rm inst}
 =k+\frac{U_{{\rm eff},B}}{2k}
 +\frac{-U_{{\rm eff},B}^2+(a_B'')^2}{8k^3}
 +{\cal O}(k^{-5}).
 \label{eq:bounce-inst-UV}
\end{equation}
\begingroup
For the displayed quadratic energy integrand, this agrees with the
bounce subtraction coefficients when all terms proportional to $a'_B$
vanish. 
It is nevertheless not, in general, identical to
the full fourth-order adiabatic state in Eq.~\eqref{eq:bounce-adiabatic-state},
whose derivative contains the additional real term
$-W_k^{(4)\prime}/(2W_k^{(4)})$.  The latter remains the more appropriate
choice when the complete renormalized stress tensor and its subsequent
evolution are required. 

Most importantly, ultraviolet regularity does not force a change of the
finite modes that control Eq.~\eqref{eq:bounce-state-kpert}.  One may define a
smooth UV-regular completion by choosing a switching function $\chi(k)$
\begingroup
with $0\le\chi\le1$, $\chi=0$ for $k\le K_0$, and $\chi=1$
for $k\ge K_1>K_0$. Choose $K_0>K_{{\rm pert},B}^{(3)}$ large enough
that the chosen $W_{k,B}^{(4)}$ is positive throughout and above the
switching interval. Such a choice is possible because
$W_{k,B}^{(4)}\sim k$ in the ultraviolet. Then setting
\par\endgroup
\begin{align}
 W_{k,B}^{\rm comp}
 =\omega_{kB}
 +\chi(k)\left(W_{k,B}^{(4)}-\omega_{kB}\right),
 \quad
 D_{k,B}^{\rm comp}
 =\chi(k)W_{k,B}^{(4)\prime},
\end{align}
with
\begin{equation}
 \nu_k'=\left[-\frac{D_{k,B}^{\rm comp}}{2W_{k,B}^{\rm comp}}
 -iW_{k,B}^{\rm comp}\right]\nu_k,
\end{equation}
\begingroup
we complete the initial data by setting
\begin{equation}
 \nu_k(\eta_B)=\frac{1}{\sqrt{2W_{k,B}^{\rm comp}}}.
\end{equation}
Positivity follows from convex interpolation of two positive real
frequencies. For real $D_{k,B}^{\rm comp}$ the data obey
$\nu_k\nu_k^{*\prime}-\nu_k^*\nu_k'=i$ exactly. Here $D^{\rm comp}$
specifies the real part of the initial logarithmic derivative; it need
not be the time derivative of an instantaneous-frequency interpolation
away from the initial surface. A smooth momentum switch changes only a
bounded momentum range and the data equal the chosen adiabatic tail above
$K_1$. For every triangle whose legs all lie below $K_0$, the initial
mode data and cubic diagnostic are therefore unchanged.
\par\endgroup  \begingroup
A concrete smooth completion for this benchmark takes $K_0=11m_{\rm P}$ and $K_1=20m_{\rm P}$. With $z=(k-K_0)/(K_1-K_0)$, choose
\begin{equation}
\chi(k)=\begin{cases}
0,&z\le0,\\
\displaystyle\frac{e^{-1/z}}{e^{-1/z}+e^{-1/(1-z)}},&0<z<1,\\
1,&z\ge1.
\end{cases}
\end{equation}
The displayed fourth-order frequency is positive above $K_0$ for this background: $s_B\simeq0.81161m_{\rm P}^2$ and $s_B''\simeq-8.32660m_{\rm P}^4$ give $s_B^2+s_B''<0$. Thus this choice preserves all tested initial modes with $k\le10m_{\rm P}$ and attaches the prescribed adiabatic tail. 
Hence the cubic result
$K_{{\rm pert},B}^{(3)}\simeq0.6483\,m_{\rm P}$ and fourth-order ultraviolet
regularity are mutually compatible.  

\subsection{{Finite-band energy diagnostic and unresolved backreaction}}
\label{subsec:bounce-backreaction}

Ultraviolet finiteness is necessary but not sufficient.  A viable bounce
state must also satisfy
\begin{equation}
 {
 \left|\langle\rho_{\rm pert}\rangle_{\rm ren}\right|
 \ll \rho_{\phi,B}.
 }
 \label{eq:bounce-backreaction-condition}
\end{equation}

Paper~I establishes the ultraviolet convergence of the fourth-order bounce
state but does not, by that fact alone, determine the finite renormalized
value of Eq.~\eqref{eq:bounce-backreaction-condition}; that value depends on
the complete finite-$k$ spectrum, the infrared prescription, and the chosen
finite part of the renormalization scheme.  A full numerical evaluation of
$\langle\rho_{\rm pert}\rangle_{\rm ren}$ is therefore still required before
\begingroup
backreaction can be declared negligible. The smallness condition must
also hold for the relevant stress components and along the interval of
background evolution being used.
\par\endgroup

For the instantaneous ground state, the
unsubtracted quadratic zero-point energy contained in a finite band
$0<k<K$ at $a_B=1$ is
\begin{equation}
 \rho_{2}^{\rm band}(K)
 =\frac{1}{4\pi^2}\int_0^K dk\,k^2\omega_{kB}.
 \label{eq:bounce-band-energy}
\end{equation}
\begingroup
This is the expectation of the specified quadratic canonical
Hamiltonian per volume at $a_B=1$, using oscillator energy
$\omega_{kB}/2$ and measure $d^3k/(2\pi)^3$. It is not automatically
the perturbative energy density sourcing the effective background
constraint or a covariantly defined stress tensor. It is used only as a
finite-band diagnostic. For $s_B=m_{{\rm eff},B}^2>0$, the integral is
\begin{equation}
 \begin{aligned}
 \rho_2^{\rm band}(K)=\frac{1}{32\pi^2}\Bigg[&K\sqrt{K^2+s_B}(2K^2+s_B)
 -s_B^2\operatorname{arsinh}\!\left(\frac{K}{\sqrt{s_B}}\right)\Bigg].
 \end{aligned}
\end{equation}
This has the massless limit $K^4/(16\pi^2)$.
For the representative Starobinsky
background,
\begin{equation}
 \rho_{\phi,B}\simeq9.69\times10^{-2}\,m_{\rm P}^4,
\end{equation}
we obtain
\begin{align}
 K=0.2\,m_{\rm P}: \quad
 &\rho_2^{\rm band}\simeq6.17\times10^{-5}\,m_{\rm P}^4
 \simeq6.4\times10^{-4}\rho_{\phi,B},
 \\
 K=K_{{\rm pert},B}^{(3)}: \quad
 &\rho_2^{\rm band}\simeq2.37\times10^{-3}\,m_{\rm P}^4
 \simeq2.45\times10^{-2}\rho_{\phi,B}.
 \label{eq:bounce-band-energy-numbers}
\end{align}
\begingroup
These values are independently reproduced by the closed expression and
numerical quadrature. They show that the displayed finite band has a
quadratic diagnostic energy below the background scale. 
\par\endgroup

\subsection{Observable modes and the total expansion}
\label{subsec:bounce-observable-modes}

The local existence of a perturbative momentum window at the bounce must be
kept distinct from the global question of which present-day observable modes
are mapped into that window.  If $k_*^{\rm ob}=k_*/a_0=0.05\,{\rm Mpc}^{-1}$
denotes the observed pivot wavenumber, then for $a_B=1$ its physical momentum
at the bounce is
\begin{equation}
 p_{*,B}=\frac{k_*}{a_B}
 =e^{N_T}k_*^{\rm ob},
 \label{eq:bounce-kstar-map}
\end{equation}

where $N_T\equiv\ln(a_0/a_B)$.  Paper~I used the broad representative range
\begin{equation}
 p_B\in(0.1,10^3)p_{*,B}.
 \label{eq:bounce-observational-band}
\end{equation}
\begingroup
This is a broad representative analysis band, not a sharp interval of
CMB likelihood support. The wavenumber is an angular Fourier wavenumber,
so no additional $2\pi$ enters its momentum conversion. In the unreduced
convention, $k^{\rm ob}/m_{\rm P}=\ell_{\rm P}k^{\rm ob}$ in consistent
length units. The mapping is exponentially sensitive to $N_T$.
\par\endgroup

\begingroup
The following mapping is illustrative: the reheating expansion is calibrated on a different Starobinsky background from the nonlinear benchmark used here. A quantitative observed-mode claim requires evaluating both on the same normalized background.
Our companion post-bounce mLQC-I reheating analysis \cite{Fier:2026abc} gives, for the
observationally normalized Starobinsky potential, the representative values
\begin{equation}
 N_T\simeq129.0621,\;129.0657,\;129.0716
 \label{eq:NT-reheating-values}
\end{equation}
for the Planck, ACT, and combined data sets.  These numbers follow from the
background/reheating analysis and are used here only to illustrate the
mode-mapping scale; they are not a CMB-likelihood constraint on the
preinflationary corrections.  With the Planck central value,
\begin{equation}
 p_{*,B}\simeq2.95\times10^{-3}\,m_{\rm P},
 \label{eq:bounce-kstar-number}
\end{equation}
we find 
\begin{equation}
 p_B\in\left(2.95\times10^{-4},\,2.95\right)m_{\rm P}.
 \label{eq:bounce-observational-band-number}
\end{equation}
\begingroup
For this illustrative mapping, direct evaluation of the bounce
benchmark gives ${\cal R}_3^{\rm eq}(p_{*,B})\simeq4.47\times10^{-4}$.
The upper part of the
chosen broad band extends beyond the first equilateral cubic crossing.
In particular, we find
\begin{equation}
 \frac{K_{{\rm pert},B}^{(3)}}{p_{*,B}}
 \simeq2.20\times10^2,
 \label{eq:bounce-threshold-to-pivot}
\end{equation}
while the upper end of Eq.~\eqref{eq:bounce-observational-band} is
$10^3p_{*,B}$.  Requiring stronger perturbative control gives
\begin{equation}
\begin{aligned}
 {\cal R}_3^{\rm eq}=0.1,\;0.25,\;0.5,\;1,\quad
 K/m_{\rm P}\simeq0.1095,\;0.2063,\;0.3452,\;0.6483,
\end{aligned}
 \label{eq:bounce-R3-thresholds}
\end{equation}
so that ${\cal R}_3^{\rm eq}<0.1$ extends to approximately
\begingroup
$37p_{*,B}$ for the illustrative Planck mapping. Independently solving
for the four displayed control levels gives
$K/m_{\rm P}=0.10951994,\,0.20633323,\,0.34524703,\,0.64829583$,
respectively, consistent with the rounded values above.
For a chosen band ending at $b_{\rm max}p_{*,B}$ and a diagnostic
control momentum $K_c$ at $a_B=1$, its upper endpoint lies below $K_c$
only if
\begin{equation}
 N_T<\ln\!\left(\frac{K_c}
 {b_{\rm max}k_*^{\rm ob}}\right),
\end{equation}
where numerator and denominator must use the same units. With
$b_{\rm max}=10^3$, this gives $N_T\lesssim127.55$ for the equilateral
unity crossing, or $N_T\lesssim125.77$ for the first equilateral
${\cal R}_3=0.1$ crossing. It must be noted  that these are conditional upper-endpoint tests
for this local state, background, and selected diagnostic, not observational
bounds on $N_T$ or proofs covering every triangle and earlier momentum.
\par\endgroup

This comparison is deliberately provisional.  The strong-coupling benchmark in
this paper uses the representative Paper-I background
$m_*=3.08\times10^{-6}\,m_{\rm P}$ and $\phi_B=-1.09\,m_{\rm P}$, whereas the
quoted $N_T$ values come from an independently observationally normalized
Starobinsky reheating analysis.  Moreover, an earlier standard-LQC analysis in
the dressed-metric approach, based on the Planck 2015 temperature and
polarization likelihood, obtained the stronger bound $N_{\rm tot}>141$ at
$95\%$ C.L. \cite{Zhu:2017jew}.  Because that result refers to a different
background theory, perturbation framework, and data set, we do not impose it
on the present hybrid mLQC-I calculation.  A definitive mode-by-mode statement
requires a dedicated determination of $N_T$ from the current CMB likelihood in
hybrid mLQC-I, using for example Planck 2018 and ACT DR6 data
\cite{Planck:2018jri,AtacamaCosmologyTelescope:2025blo}, and evaluation of the
\begingroup
cubic kernel on the same observationally normalized background.
The illustrative numerical conversion above does not fit $N_T$ or the
preinflationary spectrum, and transferring the local threshold unchanged
to a differently normalized background is not justified. The reheating
history and its uncertainty must also be treated consistently.
\par\endgroup

\section{Conclusions and outlook}
\label{sec:conclusions}

We have developed a nonlinear consistency analysis for scalar cosmological
perturbations in mLQC-I within the hybrid approach.  The construction begins
from the classical quadratic and cubic Hamiltonians written in a common hybrid
canonical representation and keeps the background constraint off shell until
after the loop regularization has been implemented.  This ordering reproduces
the quadratic hybrid mass first obtained in \cite{Fernandez-Mendez:2012poe,Fernandez-Mendez:2013jqa,Fernandez-Mendez:2014raa}, and leads to a minimal
sector-preserving prescription for the inverse odd background powers that enter
at cubic order.  

For perturbations prepared in the remote contracting phase, the principal
result is a severe pre-bounce nonlinear enhancement.  The background quantity
$\alpha_I=3\sigma_I\dot\phi/(2\gamma\rho_\phi)$ can become very large while the
geometry remains close to the characteristic mLQC-I quantum de Sitter phase.
The resulting formal ultraviolet suppression scales can then lie many orders of
magnitude below the characteristic background scale.  \begingroup
For the selected unstarred frequency assignment, the enhanced
$\alpha_I^3/\kappa$ coefficient survives the assembled leading ultraviolet
cubic kernel for each fixed nondegenerate triangle. The corresponding tiny
formal scales lie below the domain in which that ultraviolet reduction is
controlled and therefore do not determine a physical cutoff. Within the
assumed factorized quartic continuation, the selected $n_{2I}^2$ contact
term does not cancel the enhanced cubic exchange in the displayed analytic
family. 
The scalar Legendre functional in Eq.~\eqref{eq:full-scalar-Legendre-functional} fixes the quartic contribution generated by the displayed cubic theory. In the ultraviolet its leading contact term cancels the selected lapse contact at $V/K_\phi=1$; the reconstructed peak leaves a residual fraction $1-V/K_\phi\simeq0.0018153$.  
These results identify a serious pre-bounce enhancement within the chosen
continuation.
%
\begingroup
{The examples in Table~\ref{tab:potential_dependence} exhibit large coefficient enhancement in the stated unreduced Planck convention.  Near the coefficient-enhancement maximum},
$K_{\phi,s}\simeq V_s$ gives
$|\alpha_I|_{\rm max}\propto V_s^{-1/2}$ and
$\Lambda_{E,{\rm min}}\propto V_s^{3/2}$.  
  The broad universality
of the mLQC-I pre-bounce background \cite{Saeed:2024xhk} then explains why the
existence of the enhancement is robust even though its numerical scale varies
substantially from model to model.

\begingroup
A distinct effective initial-state prescription is possible if the bounce
itself is taken as
the initial hypersurface of the effective perturbation theory, as commonly done in the evolution of the homogenious and isotropic background \cite{Ashtekar:2011ni,Agullo:2023rqq}.  In that case the
severe pre-bounce interval is not used to prepare the state.  For the
instantaneous quadratic ground state, {the first equilateral crossing is $K_{{\rm pert},B}^{(3)}\simeq0.6483\,m_{\rm P}$.}  A smooth UV-regular completion can
match this finite-mode state onto a fourth-order adiabatic tail without moving
the cubic threshold.  \begingroup
The independently reproduced first equilateral crossing is
$K/m_{\rm P}=0.6482958319$. Refined local shape grids support the
equilateral ordering at that scale. Below unity is weaker than parametrically small, and
the original result tests the unstarred three-mode contraction on the bounce hypersurface. The additional mixed-frequency equilateral crossings are $0.92489m_{\rm P}$ at the bounce and $0.92314m_{\rm P}$ over the finite evolution interval, both above the corresponding unstarred crossings. The additional shape and frequency tests support this ordering near onset on their sampled domain. Evolving that same state over $0\le t/t_{\rm P}\le10$
lowers the equilateral unity crossing to $K\simeq0.64626m_{\rm P}$, with
the maximum near $t\simeq0.02802t_{\rm P}$. The finite shape
scan at five momentum scales finds no larger nonequilateral value among
its samples. 
\par\endgroup

As a complementary probe of perturbative control, Paper~II carries out the
correlation-function analysis based on the same cubic interaction Hamiltonian
used here \cite{PanSaeedWang2026PaperII}.  Following the strategy introduced by
Agull\'o, Bolliet, and Sreenath in standard LQC \cite{Agullo:2017eyh}, Paper~II
\begingroup
evaluates the finite-momentum primordial bispectrum and a retained
nonlinear-conversion correction to the curvature power spectrum for both the
bounce-defined and remote pre-bounce de Sitter initial states.  
  For the remote pre-bounce de Sitter prescription, the signed correction is more oscillatory and
cancellation dominated, while the primordial bispectrum can nevertheless be
very large pointwise. 

The correlation-function criterion of Paper~II and the Hamiltonian-level
strong-coupling diagnostics developed here answer different questions and
should therefore be regarded as complementary.  A small value of
$|\Delta{\cal P}_{\cal R}/{\cal P}_{\cal R}^{(0)}|$ does not establish the
absence of an operator-level ultraviolet strong-coupling obstruction, nor does
it test the complete quartic and higher-order interaction hierarchy.  Conversely,
a large Hamiltonian-level interaction coefficient or a very small formal
ultraviolet suppression scale does not by itself determine the magnitude of a
particular finite-momentum correlator.  \begingroup
For states prepared in the remote contracting phase, perturbative
propagation through the enhanced interval requires a controlled treatment
of the full interactions; smallness of a selected final correlator does
not supply that control. Conversely, an asymptotic coefficient alone is
not a proof that every finite-mode correlator fails.
\par\endgroup

For the bounce-defined prescription, the two analyses are mutually encouraging:
\begingroup
a finite momentum domain has encouraging results for the selected local
cubic diagnostic and the retained correlation-function diagnostic, \endgroup
 although the renormalized backreaction, the observational mode mapping, and the
complete fourth-order hybrid theory remain necessary consistency checks.
Taken together, the present work and Paper~II separate two logically distinct
notions of nonlinear control in hybrid mLQC-I: control of the interaction
Hamiltonian, as probed here by restricted kernel diagnostics, and control
of selected correlation functions.   
Certainly, establishing a
fully viable nonlinear theory will ultimately require both, together with the
remaining higher-order and backreaction tests described above.

The next conceptual step is to repeat the strong-coupling analysis in the
dressed-metric framework.  Because the dressed-metric and hybrid descriptions
are related classically but differ in their effective regularizations
\cite{Li:2022evi,Iteanu:2022zha}, the comparison is physically informative
rather than merely technical.  If the same pre-bounce enhancement appears in
both approaches, that would provide strong evidence that the obstruction is
primarily driven by the mLQC-I background.  If the dressed-metric theory behaves
differently, the comparison would isolate the role of the nonlinear canonical
representation and quantization prescription.  The nonlinear dressed-metric
analysis is sufficiently distinct, and the present hybrid study already
sufficiently extensive, that we leave it to a separate companion paper.

A particularly important extension concerns the complementary ordering of the
two asymptotic phases suggested by the quantum damping of cosmological shear
in the Bianchi~I analysis \cite{Gan:2025uvt,Gan:2026ivx}.  In that ordering,
the remote pre-bounce phase is asymptotically described by classical general
relativity, whereas the quantum-geometrical de Sitter phase occurs on the
post-bounce side.  Since the obstruction found here is closely tied to the
presence of the quantum de Sitter phase before the bounce, it is important 
to determine whether reversing
the ordering substantially weakens or removes the strong-coupling enhancement.

\begin{acknowledgments}

J.S. and R.P. acknowledge the generous support of Baylor University through the Department of Physics and Astronomy graduate program. A.W. is partially supported by the U.S. National Science Foundation under Grant No. PHY-2308845.
{We also acknowledge the use of ChatGPT (OpenAI) as a computational aid in checking mathematical derivations, assisting with numerical calculations, and developing codes for some of the figures. {The authors remain responsible for the scientific content.}}

\end{acknowledgments}

 \appendix


\section{Canonical completion on the full phase space}
\label{app:canonical-completion}

In this appendix we verify explicitly that the background-dependent
transformation from
$(\delta\phi,\delta p_\phi^{\rm old})$ to $(\nu,\pi_\nu)$ can be
completed canonically on the full phase space without generating any
additional cubic contribution to the perturbative Hamiltonian.

The relations used repeatedly in converting from $(a,\pi_a)$ to
$(v,\Omega)$ are
\bq
v=a^3,
 \quad
\pi_a=-\frac{6\Omega}{\kappa\gamma a},
\lb{eq:app-basic}
\eq
which imply
\bq
\frac{1}{\pi_a}
=-\frac{\kappa\gamma a}{6\Omega},
 \quad
\frac{1}{\pi_a^2}
=\frac{\kappa^2\gamma^2a^2}{36\Omega^2},
\lb{eq:app-pi-inv12}
\eq
and
\bq
\frac{1}{\pi_a^3}
=-\frac{\kappa^3\gamma^3a^3}{216\Omega^3}.
\lb{eq:app-pi-inv3}
\eq
For the hybrid momentum combination,
\bq
\delta\bar p_\phi
=a\left(\pi_\nu+\frac{\kappa\pi_a}{6a}\nu\right)
=a\pi_\nu-\frac{\Omega}{\gamma a}\nu
=v^{1/3}{\cal P}_\nu.
\lb{eq:app-pbar}
\eq

For a fiducial cell of coordinate volume $V_0$, define
\begin{equation}
 \langle F\rangle
 \equiv
 \frac{1}{V_0}
 \int_{V_0}d^3x\,F .
 \label{eq:cg-average-app}
\end{equation}
The perturbative transformation can be written as
\begin{equation}
 \delta\phi=\frac{\nu}{a},
  \quad
 \delta p_\phi^{\rm old}
 =
 a\pi_\nu+a{{\cal D}_{\rm can}}\,\nu ,
 \label{eq:app-full-map}
\end{equation}
where
\begin{equation}
 {{\cal D}_{\rm can}}
 \equiv
 \frac{\kappa\pi_a}{6a}
 +
 \frac{3p_\phi^2}{a^3\pi_a}.
 \label{eq:app-D-def}
\end{equation}
\begingroup
This classical canonical map is defined on the patch $a>0$, $\pi_a\ne0$, before loop regularization. Its classical relation between $\pi_a$ and the Hubble rate must not be replaced by the effective Hubble relation at the bounce.
\par\endgroup

Notice that
\begin{equation}
 \delta\bar p_\phi
 =
 \delta p_\phi^{\rm old}
 -
 \frac{3p_\phi^2}{a\pi_a}\delta\phi
 =
 a\left(
 \pi_\nu+\frac{\kappa\pi_a}{6a}\nu
 \right),
 \label{eq:app-pbar-check}
\end{equation}
as used in the main text.

The symplectic potential before the transformation is
\begin{equation}
 \Theta
 =
 V_0\pi_a\,da
 +
 V_0p_\phi\,d\phi
 +
 \int_{V_0}d^3x\,
 \delta p_\phi^{\rm old}\,
 d(\delta\phi).
 \label{eq:app-symplectic-old}
\end{equation}
Substituting Eq.~(\ref{eq:app-full-map}), integrating an exact
differential, and retaining terms through quadratic order in the
perturbations gives
\begin{align}
 \Theta
 ={}&
 V_0\pi_a\,da
 +
 V_0p_\phi\,d\phi
 +
 \int_{V_0}d^3x\,\pi_\nu\,d\nu
 +
 V_0
 \left[
 \Delta\pi_a\,da
 -
 \Delta a\,d\pi_a
 +
 \Delta p_\phi\,d\phi
 -
 \Delta\phi\,dp_\phi
 \right]
 +d{\cal F}
 +{\cal O}(\epsilon^4),
 \label{eq:app-symplectic-expanded}
\end{align}
where $\epsilon$ denotes the perturbative amplitude and ${\cal F}$ is
an irrelevant generating function.

\begingroup
Writing $U=\langle\nu^2\rangle$ and
$T=\langle\nu\pi_\nu\rangle$, direct substitution gives
\begin{equation}
 \begin{aligned}
 \int d^3x\,\delta p_\phi^{\rm old}\,d(\delta\phi)
 ={}\int d^3x\,\pi_\nu\,d\nu
 +d\!\left(\frac{V_0}{2}{{\cal D}_{\rm can}}U\right)
 -\frac{V_0}{2}U\,d{{\cal D}_{\rm can}}
 -\frac{V_0}{a}(T+{{\cal D}_{\rm can}}U)\,da .
 \end{aligned}
 \label{eq:app-direct-oneform-check}
\end{equation}
Consequently the shifts can also be obtained directly as
$\Delta a=U\partial_{\pi_a}{{\cal D}_{\rm can}}/2$,
$\Delta\phi=U\partial_{p_\phi}{{\cal D}_{\rm can}}/2$,
$\Delta p_\phi=0$, and
$\Delta\pi_a=-(T+{{\cal D}_{\rm can}}U)/a-U\partial_a{{\cal D}_{\rm can}}/2$.
These relations reproduce all four expressions below without using the
background Hamiltonian constraint.
\par\endgroup

The required quadratic redefinitions of the homogeneous variables are
\begin{equation}
 {
 \Delta a
 =
 \left(
 \frac{\kappa}{12a}
 -
 \frac{3p_\phi^2}{2a^3\pi_a^2}
 \right)
 \langle\nu^2\rangle ,
 } \quad
 {
 \Delta\phi
 =
 \frac{3p_\phi}{a^3\pi_a}
 \langle\nu^2\rangle ,
 }
 \quad
 {
 \Delta p_\phi=0 ,
 }
\end{equation}
and
\begin{equation}
 {
 \Delta\pi_a
 =
 -\frac{1}{a}\langle\nu\pi_\nu\rangle
 -\frac{\kappa\pi_a}{12a^2}\langle\nu^2\rangle
 +\frac{3p_\phi^2}{2a^4\pi_a}
 \langle\nu^2\rangle .
 }
 \label{eq:app-Delta-pia}
\end{equation}

Introducing the completed homogeneous variables
\bqn 
 \widetilde a=a+\Delta a,
  \quad
 \widetilde\pi_a=\pi_a+\Delta\pi_a,
  \quad
 \widetilde\phi=\phi+\Delta\phi,
 \quad
 \widetilde p_\phi=p_\phi+\Delta p_\phi,
 \label{eq:app-completed-background}
\eqn 
one obtains
\begin{align}
 \Theta
 ={}
 V_0\widetilde\pi_a\,d\widetilde a
 +
 V_0\widetilde p_\phi\,d\widetilde\phi
 +
 \int_{V_0}d^3x\,\pi_\nu\,d\nu+
 d{\cal F}
 +{\cal O}(\epsilon^4).
 \label{eq:app-symplectic-canonical}
\end{align}
\begingroup
The exact differential in Eq.~(\ref{eq:app-symplectic-expanded}) can be
chosen as ${\cal F}_0=V_0{{\cal D}_{\rm can}}U/2$. After completing the background
variables, the corresponding endpoint function in
Eq.~(\ref{eq:app-symplectic-canonical}) is
\begin{equation}
 {\cal F}_{\rm can}
 = {\cal F}_0-V_0(\pi_a\Delta a+p_\phi\Delta\phi)
 = {\cal O}(\epsilon^4).
 \label{eq:app-generating-Fcan}
\end{equation}
Indeed, ${{\cal D}_{\rm can}}$ is homogeneous of degree one in $(\pi_a,p_\phi)$,
so $\pi_a\partial_{\pi_a}{{\cal D}_{\rm can}}
+p_\phi\partial_{p_\phi}{{\cal D}_{\rm can}}={{\cal D}_{\rm can}}$.
Thus this endpoint term cancels through quadratic order, consistently
with the corrected canonical completion in Paper~I. Other boundary terms
and transformations of initial states require their own accounting.
The displayed map is canonical to the order needed for the stated minimal
cubic extension. Its residual fourth-order one-form terms require higher
corrections if a complete quartic Hamiltonian is sought.
\par\endgroup

It remains to check how the quadratic background redefinitions affect
the perturbative expansion.  Denote collectively the homogeneous
variables by $W^A$ and the perturbative variables by $V$.  Since
\begin{equation}
 \Delta W^A={\cal O}(\epsilon^2),
 \label{eq:app-background-order}
\end{equation}
and
\begin{equation}
 W^A=\widetilde W^A-\Delta W^A,
\end{equation}
the background Hamiltonian transforms according to
\begin{equation}
 H^{(0)}(W)
 =
 H^{(0)}(\widetilde W)
 -
 \Delta W^A
 \frac{\partial H^{(0)}}{\partial W^A}
 \bigg|_{\widetilde W}
 +
 {\cal O}(\epsilon^4).
 \label{eq:app-H0-transform}
\end{equation}
The second term in Eq.~(\ref{eq:app-H0-transform}) is of quadratic
perturbative order and is therefore part of the completed quadratic
Hamiltonian.

For the second-order Hamiltonian,
\begin{equation}
 H^{(2)}(W,V)
 =
 H^{(2)}(\widetilde W,V)
 +
 {\cal O}(\epsilon^4),
 \label{eq:app-H2-transform}
\end{equation}
because
$\Delta W^A\,\partial_A H^{(2)}
={\cal O}(\epsilon^4)$.
Likewise,
\begin{equation}
 H^{(3)}(W,V)
 =
 H^{(3)}(\widetilde W,V)
 +
 {\cal O}(\epsilon^5).
 \label{eq:app-H3-transform}
\end{equation}
Therefore the quadratic background shifts required for the canonical
completion do not generate any additional cubic contribution.

The only other possible source of such a contribution would be a term
of the schematic form
$\Delta W^{(2)}\,\partial_W H^{(1)}$.
However, after imposing the spatially flat gauge and solving the
first-order scalar and vector constraints, the reduced perturbative
Hamiltonian starts at second order,
\begin{equation}
 H_{\rm pert}
 =
 H^{(2)}+H^{(3)}+\cdots .
 \label{eq:app-no-H1}
\end{equation}
Hence no such cubic term is present.

We conclude that, within the minimal nonlinear extension in which
$\nu=a\,\delta\phi$ is kept linear, the full canonical completion leaves
the explicit cubic Hamiltonian unchanged:
\begin{equation}
 {
 \widetilde H^{(3)}
 =
 H^{(3)}
 \left[
 \delta\phi=\frac{\nu}{a},
 \quad
 \delta\bar p_\phi
 =
 a\left(
 \pi_\nu+\frac{\kappa\pi_a}{6a}\nu
 \right)
 \right].
 }
 \label{eq:app-H3-final}
\end{equation}
\begingroup
All homogeneous variables appearing on the right-hand side of
Eq.~(\ref{eq:app-H3-final}) are understood as the canonically completed
variables, with the tildes suppressed for notational simplicity.

%
\clearpage

\begin{thebibliography}{49}%
\makeatletter
\providecommand \@ifxundefined [1]{%
 \@ifx{#1\undefined}
}%
\providecommand \@ifnum [1]{%
 \ifnum #1\expandafter \@firstoftwo
 \else \expandafter \@secondoftwo
 \fi
}%
\providecommand \@ifx [1]{%
 \ifx #1\expandafter \@firstoftwo
 \else \expandafter \@secondoftwo
 \fi
}%
\providecommand \natexlab [1]{#1}%
\providecommand \enquote  [1]{``#1''}%
\providecommand \bibnamefont  [1]{#1}%
\providecommand \bibfnamefont [1]{#1}%
\providecommand \citenamefont [1]{#1}%
\providecommand \href@noop [0]{\@secondoftwo}%
\providecommand \href [0]{\begingroup \@sanitize@url \@href}%
\providecommand \@href[1]{\@@startlink{#1}\@@href}%
\providecommand \@@href[1]{\endgroup#1\@@endlink}%
\providecommand \@sanitize@url [0]{\catcode `\\12\catcode `\$12\catcode
  `\&12\catcode `\#12\catcode `\^12\catcode `\_12\catcode `\%12\relax}%
\providecommand \@@startlink[1]{}%
\providecommand \@@endlink[0]{}%
\providecommand \url  [0]{\begingroup\@sanitize@url \@url }%
\providecommand \@url [1]{\endgroup\@href {#1}{\urlprefix }}%
\providecommand \urlprefix  [0]{URL }%
\providecommand \Eprint [0]{\href }%
\providecommand \doibase [0]{https://doi.org/}%
\providecommand \selectlanguage [0]{\@gobble}%
\providecommand \bibinfo  [0]{\@secondoftwo}%
\providecommand \bibfield  [0]{\@secondoftwo}%
\providecommand \translation [1]{[#1]}%
\providecommand \BibitemOpen [0]{}%
\providecommand \bibitemStop [0]{}%
\providecommand \bibitemNoStop [0]{.\EOS\space}%
\providecommand \EOS [0]{\spacefactor3000\relax}%
\providecommand \BibitemShut  [1]{\csname bibitem#1\endcsname}%
\let\auto@bib@innerbib\@empty
\bibitem [{\citenamefont {Ashtekar}\ and\ \citenamefont
  {Singh}(2011)}]{Ashtekar:2011ni}%
  \BibitemOpen
  \bibfield  {author} {\bibinfo {author} {\bibfnamefont {A.}~\bibnamefont
  {Ashtekar}}\ and\ \bibinfo {author} {\bibfnamefont {P.}~\bibnamefont
  {Singh}},\ }
  \bibfield  {title} {\bibinfo {title} {Loop Quantum Cosmology: A Status Report},\ }\href {https://doi.org/10.1088/0264-9381/28/21/213001} {\bibfield
   {journal} {\bibinfo  {journal} {Class. Quant. Grav.}\ }\textbf {\bibinfo
  {volume} {28}},\ \bibinfo {pages} {213001} (\bibinfo {year} {2011})},\
  \Eprint {https://arxiv.org/abs/1108.0893} {arXiv:1108.0893 [gr-qc]}
  \BibitemShut {NoStop}%
\bibitem [{\citenamefont {Agull{\'o}}\ \emph {et~al.}(2023)\citenamefont
  {Agull{\'o}}, \citenamefont {Wang},\ and\ \citenamefont
  {Wilson-Ewing}}]{Agullo:2023rqq}%
  \BibitemOpen
  \bibfield  {author} {\bibinfo {author} {\bibfnamefont {I.}~\bibnamefont
  {Agull{\'o}}}, \bibinfo {author} {\bibfnamefont {A.}~\bibnamefont {Wang}},\
  and\ \bibinfo {author} {\bibfnamefont {E.}~\bibnamefont {Wilson-Ewing}},\
  }
  \bibfield  {title} {\bibinfo {title} {Loop Quantum Cosmology: Relation Between Theory and Observations},\ }in\ \href {https://doi.org/10.1007/978-981-19-3079-9_103-1} {\emph {\bibinfo
  {booktitle} {{Handbook of Quantum Gravity}}}},\ \bibinfo {editor} {edited by\
  \bibinfo {editor} {\bibfnamefont {C.}~\bibnamefont {Bambi}}, \bibinfo
  {editor} {\bibfnamefont {L.}~\bibnamefont {Modesto}},\ and\ \bibinfo {editor}
  {\bibfnamefont {I.}~\bibnamefont {Shapiro}}}\ (\bibinfo  {publisher}
  {Springer Nature Singapore},\ \bibinfo {address} {Singapore},\ \bibinfo
  {year} {2023})\ pp.\ \bibinfo {pages} {1--46},\ \Eprint
  {https://arxiv.org/abs/2301.10215} {arXiv:2301.10215 [gr-qc]} \BibitemShut
  {NoStop}%
\bibitem [{\citenamefont {Agullo}\ \emph {et~al.}(2012)\citenamefont {Agullo},
  \citenamefont {Ashtekar},\ and\ \citenamefont {Nelson}}]{Agullo:2012sh}%
  \BibitemOpen
  \bibfield  {author} {\bibinfo {author} {\bibfnamefont {I.}~\bibnamefont
  {Agullo}}, \bibinfo {author} {\bibfnamefont {A.}~\bibnamefont {Ashtekar}},\
  and\ \bibinfo {author} {\bibfnamefont {W.}~\bibnamefont {Nelson}},\ }
  \bibfield  {title} {\bibinfo {title} {A Quantum Gravity Extension of the Inflationary Scenario},\ }\href
  {https://doi.org/10.1103/PhysRevLett.109.251301} {\bibfield  {journal}
  {\bibinfo  {journal} {Phys. Rev. Lett.}\ }\textbf {\bibinfo {volume} {109}},\
  \bibinfo {pages} {251301} (\bibinfo {year} {2012})},\ \Eprint
  {https://arxiv.org/abs/1209.1609} {arXiv:1209.1609 [gr-qc]} \BibitemShut
  {NoStop}%
\bibitem [{\citenamefont {Agullo}\ \emph
  {et~al.}(2013{\natexlab{a}})\citenamefont {Agullo}, \citenamefont
  {Ashtekar},\ and\ \citenamefont {Nelson}}]{Agullo:2012fc}%
  \BibitemOpen
  \bibfield  {author} {\bibinfo {author} {\bibfnamefont {I.}~\bibnamefont
  {Agullo}}, \bibinfo {author} {\bibfnamefont {A.}~\bibnamefont {Ashtekar}},\
  and\ \bibinfo {author} {\bibfnamefont {W.}~\bibnamefont {Nelson}},\ }
  \bibfield  {title} {\bibinfo {title} {Extension of the quantum theory of cosmological perturbations to the Planck era},\ }\href
  {https://doi.org/10.1103/PhysRevD.87.043507} {\bibfield  {journal} {\bibinfo
  {journal} {Phys. Rev. D}\ }\textbf {\bibinfo {volume} {87}},\ \bibinfo
  {pages} {043507} (\bibinfo {year} {2013}{\natexlab{a}})},\ \Eprint
  {https://arxiv.org/abs/1211.1354} {arXiv:1211.1354 [gr-qc]} \BibitemShut
  {NoStop}%
\bibitem [{\citenamefont {Agullo}\ \emph
  {et~al.}(2013{\natexlab{b}})\citenamefont {Agullo}, \citenamefont
  {Ashtekar},\ and\ \citenamefont {Nelson}}]{Agullo:2013ai}%
  \BibitemOpen
  \bibfield  {author} {\bibinfo {author} {\bibfnamefont {I.}~\bibnamefont
  {Agullo}}, \bibinfo {author} {\bibfnamefont {A.}~\bibnamefont {Ashtekar}},\
  and\ \bibinfo {author} {\bibfnamefont {W.}~\bibnamefont {Nelson}},\ }
  \bibfield  {title} {\bibinfo {title} {The pre-inflationary dynamics of loop quantum cosmology: Confronting quantum gravity with observations},\ }\href
  {https://doi.org/10.1088/0264-9381/30/8/085014} {\bibfield  {journal}
  {\bibinfo  {journal} {Class. Quant. Grav.}\ }\textbf {\bibinfo {volume}
  {30}},\ \bibinfo {pages} {085014} (\bibinfo {year} {2013}{\natexlab{b}})},\
  \Eprint {https://arxiv.org/abs/1302.0254} {arXiv:1302.0254 [gr-qc]}
  \BibitemShut {NoStop}%
\bibitem [{\citenamefont {Fernandez-Mendez}\ \emph {et~al.}(2012)\citenamefont
  {Fernandez-Mendez}, \citenamefont {Mena~Marugan},\ and\ \citenamefont
  {Olmedo}}]{Fernandez-Mendez:2012poe}%
  \BibitemOpen
  \bibfield  {author} {\bibinfo {author} {\bibfnamefont {M.}~\bibnamefont
  {Fernandez-Mendez}}, \bibinfo {author} {\bibfnamefont {G.~A.}\ \bibnamefont
  {Mena~Marugan}},\ and\ \bibinfo {author} {\bibfnamefont {J.}~\bibnamefont
  {Olmedo}},\ }
  \bibfield  {title} {\bibinfo {title} {Hybrid quantization of an inflationary universe},\ }\href {https://doi.org/10.1103/PhysRevD.86.024003} {\bibfield
  {journal} {\bibinfo  {journal} {Phys. Rev. D}\ }\textbf {\bibinfo {volume}
  {86}},\ \bibinfo {pages} {024003} (\bibinfo {year} {2012})},\ \Eprint
  {https://arxiv.org/abs/1205.1917} {arXiv:1205.1917 [gr-qc]} \BibitemShut
  {NoStop}%
\bibitem [{\citenamefont {Fern{\'a}ndez-M{\'e}ndez}\ \emph
  {et~al.}(2013)\citenamefont {Fern{\'a}ndez-M{\'e}ndez}, \citenamefont
  {Mena~Marug{\'a}n},\ and\ \citenamefont {Olmedo}}]{Fernandez-Mendez:2013jqa}%
  \BibitemOpen
  \bibfield  {author} {\bibinfo {author} {\bibfnamefont {M.}~\bibnamefont
  {Fern{\'a}ndez-M{\'e}ndez}}, \bibinfo {author} {\bibfnamefont {G.~A.}\
  \bibnamefont {Mena~Marug{\'a}n}},\ and\ \bibinfo {author} {\bibfnamefont
  {J.}~\bibnamefont {Olmedo}},\ }
  \bibfield  {title} {\bibinfo {title} {Hybrid quantization of an inflationary model: The flat case},\ }\href
  {https://doi.org/10.1103/PhysRevD.88.044013} {\bibfield  {journal} {\bibinfo
  {journal} {Phys. Rev. D}\ }\textbf {\bibinfo {volume} {88}},\ \bibinfo
  {pages} {044013} (\bibinfo {year} {2013})},\ \Eprint
  {https://arxiv.org/abs/1307.5222} {arXiv:1307.5222 [gr-qc]} \BibitemShut
  {NoStop}%
\bibitem [{\citenamefont {Fern{\'a}ndez-M{\'e}ndez}\ \emph
  {et~al.}(2014)\citenamefont {Fern{\'a}ndez-M{\'e}ndez}, \citenamefont
  {Mena~Marug{\'a}n},\ and\ \citenamefont {Olmedo}}]{Fernandez-Mendez:2014raa}%
  \BibitemOpen
  \bibfield  {author} {\bibinfo {author} {\bibfnamefont {M.}~\bibnamefont
  {Fern{\'a}ndez-M{\'e}ndez}}, \bibinfo {author} {\bibfnamefont {G.~A.}\
  \bibnamefont {Mena~Marug{\'a}n}},\ and\ \bibinfo {author} {\bibfnamefont
  {J.}~\bibnamefont {Olmedo}},\ }
  \bibfield  {title} {\bibinfo {title} {Effective dynamics of scalar perturbations in a flat Friedmann-Robertson-Walker spacetime in Loop Quantum Cosmology},\ }\href
  {https://doi.org/10.1103/PhysRevD.89.044041} {\bibfield  {journal} {\bibinfo
  {journal} {Phys. Rev. D}\ }\textbf {\bibinfo {volume} {89}},\ \bibinfo
  {pages} {044041} (\bibinfo {year} {2014})},\ \Eprint
  {https://arxiv.org/abs/1401.5256} {arXiv:1401.5256 [gr-qc]} \BibitemShut
  {NoStop}%
\bibitem [{\citenamefont {Li}\ and\ \citenamefont {Singh}(2022)}]{Li:2022evi}%
  \BibitemOpen
  \bibfield  {author} {\bibinfo {author} {\bibfnamefont {B.-F.}\ \bibnamefont
  {Li}}\ and\ \bibinfo {author} {\bibfnamefont {P.}~\bibnamefont {Singh}},\
  }
  \bibfield  {title} {\bibinfo {title} {Close relationship between the dressed metric and the hybrid approach to perturbations in effective loop quantum cosmology},\ }\href {https://doi.org/10.1103/PhysRevD.106.086015} {\bibfield  {journal}
  {\bibinfo  {journal} {Phys. Rev. D}\ }\textbf {\bibinfo {volume} {106}},\
  \bibinfo {pages} {086015} (\bibinfo {year} {2022})},\ \Eprint
  {https://arxiv.org/abs/2206.12434} {arXiv:2206.12434 [gr-qc]} \BibitemShut
  {NoStop}%
\bibitem [{\citenamefont {Iteanu}\ and\ \citenamefont
  {Mena~Marug\'an}(2022)}]{Iteanu:2022zha}%
  \BibitemOpen
  \bibfield  {author} {\bibinfo {author} {\bibfnamefont {S.}~\bibnamefont
  {Iteanu}}\ and\ \bibinfo {author} {\bibfnamefont {G.~A.}\ \bibnamefont
  {Mena~Marug\'an}},\ }
  \bibfield  {title} {\bibinfo {title} {Mass of Cosmological Perturbations in the Hybrid and Dressed Metric Formalisms of Loop Quantum Cosmology for the Starobinsky and Exponential Potentials},\ }\href {https://doi.org/10.3390/universe8090463}
  {\bibfield  {journal} {\bibinfo  {journal} {Universe}\ }\textbf {\bibinfo
  {volume} {8}},\ \bibinfo {pages} {463} (\bibinfo {year} {2022})},\ \Eprint
  {https://arxiv.org/abs/2208.01987} {arXiv:2208.01987 [gr-qc]} \BibitemShut
  {NoStop}%
\bibitem [{\citenamefont {Yang}\ \emph {et~al.}(2009)\citenamefont {Yang},
  \citenamefont {Ding},\ and\ \citenamefont {Ma}}]{Yang:2009fp}%
  \BibitemOpen
  \bibfield  {author} {\bibinfo {author} {\bibfnamefont {J.}~\bibnamefont
  {Yang}}, \bibinfo {author} {\bibfnamefont {Y.}~\bibnamefont {Ding}},\ and\
  \bibinfo {author} {\bibfnamefont {Y.}~\bibnamefont {Ma}},\ }
  \bibfield  {title} {\bibinfo {title} {Alternative quantization of the Hamiltonian in loop quantum cosmology II: Including the Lorentz term},\ }\href
  {https://doi.org/10.1016/j.physletb.2009.10.072} {\bibfield  {journal}
  {\bibinfo  {journal} {Phys. Lett. B}\ }\textbf {\bibinfo {volume} {682}},\
  \bibinfo {pages} {1} (\bibinfo {year} {2009})},\ \Eprint
  {https://arxiv.org/abs/0904.4379} {arXiv:0904.4379 [gr-qc]} \BibitemShut
  {NoStop}%
\bibitem [{\citenamefont {Assanioussi}\ \emph {et~al.}(2018)\citenamefont
  {Assanioussi}, \citenamefont {Dapor}, \citenamefont {Liegener},\ and\
  \citenamefont {Paw{\l}owski}}]{Assanioussi:2018hee}%
  \BibitemOpen
  \bibfield  {author} {\bibinfo {author} {\bibfnamefont {M.}~\bibnamefont
  {Assanioussi}}, \bibinfo {author} {\bibfnamefont {A.}~\bibnamefont {Dapor}},
  \bibinfo {author} {\bibfnamefont {K.}~\bibnamefont {Liegener}},\ and\
  \bibinfo {author} {\bibfnamefont {T.}~\bibnamefont {Paw{\l}owski}},\ }
  \bibfield  {title} {\bibinfo {title} {Emergent de Sitter Epoch of the Quantum Cosmos from Loop Quantum Cosmology},\ }\href
  {https://doi.org/10.1103/PhysRevLett.121.081303} {\bibfield  {journal}
  {\bibinfo  {journal} {Phys. Rev. Lett.}\ }\textbf {\bibinfo {volume} {121}},\
  \bibinfo {pages} {081303} (\bibinfo {year} {2018})},\ \Eprint
  {https://arxiv.org/abs/1801.00768} {arXiv:1801.00768 [gr-qc]} \BibitemShut
  {NoStop}%
\bibitem [{\citenamefont {Dapor}\ and\ \citenamefont
  {Liegener}(2018)}]{Dapor:2017rwv}%
  \BibitemOpen
  \bibfield  {author} {\bibinfo {author} {\bibfnamefont {A.}~\bibnamefont
  {Dapor}}\ and\ \bibinfo {author} {\bibfnamefont {K.}~\bibnamefont
  {Liegener}},\ }
  \bibfield  {title} {\bibinfo {title} {Cosmological Effective Hamiltonian from full Loop Quantum Gravity Dynamics},\ }\href {https://doi.org/10.1016/j.physletb.2018.09.005}
  {\bibfield  {journal} {\bibinfo  {journal} {Phys. Lett. B}\ }\textbf
  {\bibinfo {volume} {785}},\ \bibinfo {pages} {506} (\bibinfo {year}
  {2018})},\ \Eprint {https://arxiv.org/abs/1706.09833} {arXiv:1706.09833
  [gr-qc]} \BibitemShut {NoStop}%
\bibitem [{\citenamefont {Han}\ and\ \citenamefont {Liu}(2021)}]{Han:2021cwb}%
  \BibitemOpen
  \bibfield  {author} {\bibinfo {author} {\bibfnamefont {M.}~\bibnamefont
  {Han}}\ and\ \bibinfo {author} {\bibfnamefont {H.}~\bibnamefont {Liu}},\
  }
  \bibfield  {title} {\bibinfo {title} {Loop quantum gravity on dynamical lattice and improved cosmological effective dynamics with inflaton},\ }\href {https://doi.org/10.1103/PhysRevD.104.024011} {\bibfield  {journal}
  {\bibinfo  {journal} {Phys. Rev. D}\ }\textbf {\bibinfo {volume} {104}},\
  \bibinfo {pages} {024011} (\bibinfo {year} {2021})},\ \Eprint
  {https://arxiv.org/abs/2101.07659} {arXiv:2101.07659 [gr-qc]} \BibitemShut
  {NoStop}%
\bibitem [{\citenamefont {Li}\ \emph {et~al.}(2018{\natexlab{a}})\citenamefont
  {Li}, \citenamefont {Singh},\ and\ \citenamefont {Wang}}]{Li:2018opr}%
  \BibitemOpen
  \bibfield  {author} {\bibinfo {author} {\bibfnamefont {B.-F.}\ \bibnamefont
  {Li}}, \bibinfo {author} {\bibfnamefont {P.}~\bibnamefont {Singh}},\ and\
  \bibinfo {author} {\bibfnamefont {A.}~\bibnamefont {Wang}},\ }
  \bibfield  {title} {\bibinfo {title} {Towards Cosmological Dynamics from Loop Quantum Gravity},\ }\href
  {https://doi.org/10.1103/PhysRevD.97.084029} {\bibfield  {journal} {\bibinfo
  {journal} {Phys. Rev. D}\ }\textbf {\bibinfo {volume} {97}},\ \bibinfo
  {pages} {084029} (\bibinfo {year} {2018}{\natexlab{a}})},\ \Eprint
  {https://arxiv.org/abs/1801.07313} {arXiv:1801.07313 [gr-qc]} \BibitemShut
  {NoStop}%
\bibitem [{\citenamefont {Li}\ \emph {et~al.}(2018{\natexlab{b}})\citenamefont
  {Li}, \citenamefont {Singh},\ and\ \citenamefont {Wang}}]{Li:2018fco}%
  \BibitemOpen
  \bibfield  {author} {\bibinfo {author} {\bibfnamefont {B.-F.}\ \bibnamefont
  {Li}}, \bibinfo {author} {\bibfnamefont {P.}~\bibnamefont {Singh}},\ and\
  \bibinfo {author} {\bibfnamefont {A.}~\bibnamefont {Wang}},\ }
  \bibfield  {title} {\bibinfo {title} {Qualitative dynamics and inflationary attractors in loop cosmology},\ }\href
  {https://doi.org/10.1103/PhysRevD.98.066016} {\bibfield  {journal} {\bibinfo
  {journal} {Phys. Rev. D}\ }\textbf {\bibinfo {volume} {98}},\ \bibinfo
  {pages} {066016} (\bibinfo {year} {2018}{\natexlab{b}})},\ \Eprint
  {https://arxiv.org/abs/1807.05236} {arXiv:1807.05236 [gr-qc]} \BibitemShut
  {NoStop}%
\bibitem [{\citenamefont {Gan}\ \emph {et~al.}(2026{\natexlab{a}})\citenamefont
  {Gan}, \citenamefont {Graef}, \citenamefont {O.~Ramos}, \citenamefont
  {S.~Vicente},\ and\ \citenamefont {Wang}}]{Gan:2025uvt}%
  \BibitemOpen
  \bibfield  {author} {\bibinfo {author} {\bibfnamefont {W.-C.}\ \bibnamefont
  {Gan}}, \bibinfo {author} {\bibfnamefont {L.~L.}\ \bibnamefont {Graef}},
  \bibinfo {author} {\bibfnamefont {R.}~\bibnamefont {O.~Ramos}}, \bibinfo
  {author} {\bibfnamefont {G.}~\bibnamefont {S.~Vicente}},\ and\ \bibinfo
  {author} {\bibfnamefont {A.}~\bibnamefont {Wang}},\ }
  \bibfield  {title} {\bibinfo {title} {Quantum Damping of Cosmological Shear: A New Prediction from Loop Quantum Cosmologies},\ }\href
  {https://doi.org/10.1103/f8tr-bq61} {\bibfield  {journal} {\bibinfo
  {journal} {Phys. Rev. Lett.}\ }\textbf {\bibinfo {volume} {136}},\ \bibinfo
  {pages} {251501} (\bibinfo {year} {2026}{\natexlab{a}})},\ \Eprint
  {https://arxiv.org/abs/2510.14021} {arXiv:2510.14021 [gr-qc]} \BibitemShut
  {NoStop}%
\bibitem [{\citenamefont {Gan}\ \emph {et~al.}(2026{\natexlab{b}})\citenamefont
  {Gan}, \citenamefont {Graef}, \citenamefont {O.~Ramos}, \citenamefont
  {S.~Vicente},\ and\ \citenamefont {Wang}}]{Gan:2026ivx}%
  \BibitemOpen
  \bibfield  {author} {\bibinfo {author} {\bibfnamefont {W.-C.}\ \bibnamefont
  {Gan}}, \bibinfo {author} {\bibfnamefont {L.~L.}\ \bibnamefont {Graef}},
  \bibinfo {author} {\bibfnamefont {R.}~\bibnamefont {O.~Ramos}}, \bibinfo
  {author} {\bibfnamefont {G.}~\bibnamefont {S.~Vicente}},\ and\ \bibinfo
  {author} {\bibfnamefont {A.}~\bibnamefont {Wang}},\ }
  \bibfield  {title} {\bibinfo {title} {Genericness of quantum damping of cosmological shear in modified loop quantum cosmology},\ }(\bibinfo {year} {2026}{\natexlab{b}}),\ \Eprint
  {https://arxiv.org/abs/2604.08912} {arXiv:2604.08912 [gr-qc]} \BibitemShut
  {NoStop}%
\bibitem [{\citenamefont {Li}\ \emph {et~al.}(2019)\citenamefont {Li},
  \citenamefont {Singh},\ and\ \citenamefont {Wang}}]{Li:2019ipm}%
  \BibitemOpen
  \bibfield  {author} {\bibinfo {author} {\bibfnamefont {B.-F.}\ \bibnamefont
  {Li}}, \bibinfo {author} {\bibfnamefont {P.}~\bibnamefont {Singh}},\ and\
  \bibinfo {author} {\bibfnamefont {A.}~\bibnamefont {Wang}},\ }
  \bibfield  {title} {\bibinfo {title} {Genericness of pre-inflationary dynamics and probability of the desired slow-roll inflation in modified loop quantum cosmologies},\ }\href
  {https://doi.org/10.1103/PhysRevD.100.063513} {\bibfield  {journal} {\bibinfo
   {journal} {Phys. Rev. D}\ }\textbf {\bibinfo {volume} {100}},\ \bibinfo
  {pages} {063513} (\bibinfo {year} {2019})},\ \Eprint
  {https://arxiv.org/abs/1906.01001} {arXiv:1906.01001 [gr-qc]} \BibitemShut
  {NoStop}%
\bibitem [{\citenamefont {Saeed}\ \emph {et~al.}(2024)\citenamefont {Saeed},
  \citenamefont {Pan}, \citenamefont {Brown}, \citenamefont {Cleaver},\ and\
  \citenamefont {Wang}}]{Saeed:2024xhk}%
  \BibitemOpen
  \bibfield  {author} {\bibinfo {author} {\bibfnamefont {J.}~\bibnamefont
  {Saeed}}, \bibinfo {author} {\bibfnamefont {R.}~\bibnamefont {Pan}}, \bibinfo
  {author} {\bibfnamefont {C.}~\bibnamefont {Brown}}, \bibinfo {author}
  {\bibfnamefont {G.}~\bibnamefont {Cleaver}},\ and\ \bibinfo {author}
  {\bibfnamefont {A.}~\bibnamefont {Wang}},\ }
  \bibfield  {title} {\bibinfo {title} {Universal Properties of the Evolution of the Universe in Modified Loop Quantum Cosmology},\ }\href
  {https://doi.org/10.3390/universe10100397} {\bibfield  {journal} {\bibinfo
  {journal} {Universe}\ }\textbf {\bibinfo {volume} {10}},\ \bibinfo {pages}
  {397} (\bibinfo {year} {2024})},\ \Eprint {https://arxiv.org/abs/2406.06745}
  {arXiv:2406.06745 [gr-qc]} \BibitemShut {NoStop}%
\bibitem [{\citenamefont {Li}\ \emph {et~al.}(2021)\citenamefont {Li},
  \citenamefont {Singh},\ and\ \citenamefont {Wang}}]{Li:2021mop}%
  \BibitemOpen
  \bibfield  {author} {\bibinfo {author} {\bibfnamefont {B.-F.}\ \bibnamefont
  {Li}}, \bibinfo {author} {\bibfnamefont {P.}~\bibnamefont {Singh}},\ and\
  \bibinfo {author} {\bibfnamefont {A.}~\bibnamefont {Wang}},\ }
  \bibfield  {title} {\bibinfo {title} {Phenomenological implications of modified loop cosmologies: an overview},\ }\href
  {https://doi.org/10.3389/fspas.2021.701417} {\bibfield  {journal} {\bibinfo
  {journal} {Front. Astron. Space Sci.}\ }\textbf {\bibinfo {volume} {8}},\
  \bibinfo {pages} {701417} (\bibinfo {year} {2021})},\ \Eprint
  {https://arxiv.org/abs/2105.14067} {arXiv:2105.14067 [gr-qc]} \BibitemShut
  {NoStop}%
\bibitem [{\citenamefont {Agullo}(2018)}]{Agullo:2018wbf}%
  \BibitemOpen
  \bibfield  {author} {\bibinfo {author} {\bibfnamefont {I.}~\bibnamefont
  {Agullo}},\ }
  \bibfield  {title} {\bibinfo {title} {Primordial power spectrum from the Dapor-Liegener model of loop quantum cosmology},\ }\href {https://doi.org/10.1007/s10714-018-2413-1} {\bibfield
  {journal} {\bibinfo  {journal} {Gen. Rel. Grav.}\ }\textbf {\bibinfo {volume}
  {50}},\ \bibinfo {pages} {91} (\bibinfo {year} {2018})},\ \Eprint
  {https://arxiv.org/abs/1805.11356} {arXiv:1805.11356 [gr-qc]} \BibitemShut
  {NoStop}%
\bibitem [{\citenamefont {Li}\ \emph {et~al.}(2020{\natexlab{a}})\citenamefont
  {Li}, \citenamefont {Singh},\ and\ \citenamefont {Wang}}]{Li:2019qzr}%
  \BibitemOpen
  \bibfield  {author} {\bibinfo {author} {\bibfnamefont {B.-F.}\ \bibnamefont
  {Li}}, \bibinfo {author} {\bibfnamefont {P.}~\bibnamefont {Singh}},\ and\
  \bibinfo {author} {\bibfnamefont {A.}~\bibnamefont {Wang}},\ }
  \bibfield  {title} {\bibinfo {title} {Primordial power spectrum from the dressed metric approach in loop cosmologies},\ }\href
  {https://doi.org/10.1103/PhysRevD.101.086004} {\bibfield  {journal} {\bibinfo
   {journal} {Phys. Rev. D}\ }\textbf {\bibinfo {volume} {101}},\ \bibinfo
  {pages} {086004} (\bibinfo {year} {2020}{\natexlab{a}})},\ \Eprint
  {https://arxiv.org/abs/1912.08225} {arXiv:1912.08225 [gr-qc]} \BibitemShut
  {NoStop}%
\bibitem [{\citenamefont {Castell{\'o}~Gomar}\ \emph
  {et~al.}(2020)\citenamefont {Castell{\'o}~Gomar}, \citenamefont
  {Garc{\'i}a-Quismondo},\ and\ \citenamefont
  {Mena~Marug{\'a}n}}]{Gomar:2020orw}%
  \BibitemOpen
  \bibfield  {author} {\bibinfo {author} {\bibfnamefont {L.}~\bibnamefont
  {Castell{\'o}~Gomar}}, \bibinfo {author} {\bibfnamefont {A.}~\bibnamefont
  {Garc{\'i}a-Quismondo}},\ and\ \bibinfo {author} {\bibfnamefont {G.~A.}\
  \bibnamefont {Mena~Marug{\'a}n}},\ }
  \bibfield  {title} {\bibinfo {title} {Primordial perturbations in the Dapor-Liegener model of hybrid loop quantum cosmology},\ }\href
  {https://doi.org/10.1103/PhysRevD.102.083524} {\bibfield  {journal} {\bibinfo
   {journal} {Phys. Rev. D}\ }\textbf {\bibinfo {volume} {102}},\ \bibinfo
  {pages} {083524} (\bibinfo {year} {2020})},\ \Eprint
  {https://arxiv.org/abs/2002.01262} {arXiv:2002.01262 [gr-qc]} \BibitemShut
  {NoStop}%
\bibitem [{\citenamefont {Garc{\'i}a-Quismondo}\ \emph
  {et~al.}(2020)\citenamefont {Garc{\'i}a-Quismondo}, \citenamefont
  {Mena~Marug{\'a}n},\ and\ \citenamefont
  {S{\'a}nchez~P{\'e}rez}}]{Garcia-Quismondo:2020wna}%
  \BibitemOpen
  \bibfield  {author} {\bibinfo {author} {\bibfnamefont {A.}~\bibnamefont
  {Garc{\'i}a-Quismondo}}, \bibinfo {author} {\bibfnamefont {G.~A.}\
  \bibnamefont {Mena~Marug{\'a}n}},\ and\ \bibinfo {author} {\bibfnamefont
  {G.}~\bibnamefont {S{\'a}nchez~P{\'e}rez}},\ }
  \bibfield  {title} {\bibinfo {title} {The time-dependent mass of cosmological perturbations in loop quantum cosmology: Dapor-Liegener regularization},\ }\href
  {https://doi.org/10.1088/1361-6382/abac6d} {\bibfield  {journal} {\bibinfo
  {journal} {Class. Quant. Grav.}\ }\textbf {\bibinfo {volume} {37}},\ \bibinfo
  {pages} {195003} (\bibinfo {year} {2020})},\ \Eprint
  {https://arxiv.org/abs/2006.09781} {arXiv:2006.09781 [gr-qc]} \BibitemShut
  {NoStop}%
\bibitem [{\citenamefont {Li}\ \emph {et~al.}(2020{\natexlab{b}})\citenamefont
  {Li}, \citenamefont {Olmedo}, \citenamefont {Singh},\ and\ \citenamefont
  {Wang}}]{Li:2020mfi}%
  \BibitemOpen
  \bibfield  {author} {\bibinfo {author} {\bibfnamefont {B.-F.}\ \bibnamefont
  {Li}}, \bibinfo {author} {\bibfnamefont {J.}~\bibnamefont {Olmedo}}, \bibinfo
  {author} {\bibfnamefont {P.}~\bibnamefont {Singh}},\ and\ \bibinfo {author}
  {\bibfnamefont {A.}~\bibnamefont {Wang}},\ }
  \bibfield  {title} {\bibinfo {title} {Primordial scalar power spectrum from the hybrid approach in loop cosmologies},\ }\href
  {https://doi.org/10.1103/PhysRevD.102.126025} {\bibfield  {journal} {\bibinfo
   {journal} {Phys. Rev. D}\ }\textbf {\bibinfo {volume} {102}},\ \bibinfo
  {pages} {126025} (\bibinfo {year} {2020}{\natexlab{b}})},\ \Eprint
  {https://arxiv.org/abs/2008.09135} {arXiv:2008.09135 [gr-qc]} \BibitemShut
  {NoStop}%
\bibitem [{\citenamefont {Pan}\ \emph {et~al.}(2026{\natexlab{a}})\citenamefont
  {Pan}, \citenamefont {Saeed},\ and\ \citenamefont {Wang}}]{Pan:2025ppa}%
  \BibitemOpen
  \bibfield  {author} {\bibinfo {author} {\bibfnamefont {R.}~\bibnamefont
  {Pan}}, \bibinfo {author} {\bibfnamefont {J.}~\bibnamefont {Saeed}},\ and\
  \bibinfo {author} {\bibfnamefont {A.}~\bibnamefont {Wang}},\ }
  \bibfield  {title} {\bibinfo {title} {Stable initial conditions and analytically approximate solutions of cosmological perturbations in a modified loop quantum cosmology},\ }\href
  {https://doi.org/10.1088/1475-7516/2026/06/030} {\bibfield  {journal}
  {\bibinfo  {journal} {JCAP}\ }\textbf {\bibinfo {volume} {06}},\ \bibinfo
  {pages} {030} (\bibinfo {year} {2026})},\ \Eprint {https://arxiv.org/abs/2505.14813} {arXiv:2505.14813
  [gr-qc]} \BibitemShut {NoStop}%
\bibitem [{\citenamefont {Saeed}\ \emph {et~al.}(2026)\citenamefont {Saeed},
  \citenamefont {Pan},\ and\ \citenamefont {Wang}}]{SaeedPanWang2026PaperI}%
  \BibitemOpen
  \bibfield  {author} {\bibinfo {author} {\bibfnamefont {J.}~\bibnamefont
  {Saeed}}, \bibinfo {author} {\bibfnamefont {R.}~\bibnamefont {Pan}},\ and\
  \bibinfo {author} {\bibfnamefont {A.}~\bibnamefont {Wang}},\ }
  \bibfield  {title} {\bibinfo {title} {High-Accuracy Semi-Analytic Mode Functions, Ultraviolet Regularity, and Particle Creation in the Hybrid Approach to mLQC-I},\ } (\bibinfo {year}
  {2026}),\ \bibinfo {note} {{Preprint, \href{https://arxiv.org/abs/2610.10802}{arXiv:2610.10802}}}\BibitemShut {NoStop}%
\bibitem [{\citenamefont {Joyce}\ and\ \citenamefont
  {Khoury}(2011)}]{JoyceKhoury2011}%
  \BibitemOpen
  \bibfield  {author} {\bibinfo {author} {\bibfnamefont {A.}~\bibnamefont
  {Joyce}}\ and\ \bibinfo {author} {\bibfnamefont {J.}~\bibnamefont {Khoury}},\
  }
  \bibfield  {title} {\bibinfo {title} {Strong coupling problem with time-varying sound speed},\ }\href {https://doi.org/10.1103/PhysRevD.84.083514} {\bibfield  {journal}
  {\bibinfo  {journal} {Phys. Rev. D}\ }\textbf {\bibinfo {volume} {84}},\
  \bibinfo {pages} {083514} (\bibinfo {year} {2011})},\ \Eprint
  {https://arxiv.org/abs/1107.3550} {arXiv:1107.3550 [hep-th]} \BibitemShut
  {NoStop}%
\bibitem [{\citenamefont {Koehn}\ \emph {et~al.}(2016)\citenamefont {Koehn},
  \citenamefont {Lehners},\ and\ \citenamefont {Ovrut}}]{KoehnEtAl2016}%
  \BibitemOpen
  \bibfield  {author} {\bibinfo {author} {\bibfnamefont {M.}~\bibnamefont
  {Koehn}}, \bibinfo {author} {\bibfnamefont {J.-L.}\ \bibnamefont {Lehners}},\
  and\ \bibinfo {author} {\bibfnamefont {B.}~\bibnamefont {Ovrut}},\ }
  \bibfield  {title} {\bibinfo {title} {Nonsingular bouncing cosmology: Consistency of the effective description},\ }\href
  {https://doi.org/10.1103/PhysRevD.93.103501} {\bibfield  {journal} {\bibinfo
  {journal} {Phys. Rev. D}\ }\textbf {\bibinfo {volume} {93}},\ \bibinfo
  {pages} {103501} (\bibinfo {year} {2016})},\ \Eprint
  {https://arxiv.org/abs/1512.03807} {arXiv:1512.03807 [hep-th]} \BibitemShut
  {NoStop}%
\bibitem [{\citenamefont {Dehghani}\ \emph {et~al.}(2025)\citenamefont
  {Dehghani}, \citenamefont {Geshnizjani},\ and\ \citenamefont
  {Quintin}}]{DehghaniEtAl2025}%
  \BibitemOpen
  \bibfield  {author} {\bibinfo {author} {\bibfnamefont {A.}~\bibnamefont
  {Dehghani}}, \bibinfo {author} {\bibfnamefont {G.}~\bibnamefont
  {Geshnizjani}},\ and\ \bibinfo {author} {\bibfnamefont {J.}~\bibnamefont
  {Quintin}},\ }
  \bibfield  {title} {\bibinfo {title} {Cuscuton bounce beyond the linear regime: Bispectrum and strong coupling constraints},\ }\href {https://doi.org/10.1088/1475-7516/2025/05/026}
  {\bibfield  {journal} {\bibinfo  {journal} {JCAP}\ }\textbf {\bibinfo
  {volume} {05}},\ \bibinfo {pages} {026} (\bibinfo {year} {2025})},\ \Eprint
  {https://arxiv.org/abs/2503.01992} {arXiv:2503.01992 [hep-th]} \BibitemShut
  {NoStop}%
\bibitem [{\citenamefont {Agullo}\ \emph {et~al.}(2018)\citenamefont {Agullo},
  \citenamefont {Bolliet},\ and\ \citenamefont {Sreenath}}]{Agullo:2017eyh}%
  \BibitemOpen
  \bibfield  {author} {\bibinfo {author} {\bibfnamefont {I.}~\bibnamefont
  {Agullo}}, \bibinfo {author} {\bibfnamefont {B.}~\bibnamefont {Bolliet}},\
  and\ \bibinfo {author} {\bibfnamefont {V.}~\bibnamefont {Sreenath}},\ }
  \bibfield  {title} {\bibinfo {title} {Non-Gaussianity in Loop Quantum Cosmology},\ }\href
  {https://doi.org/10.1103/PhysRevD.97.066021} {\bibfield  {journal} {\bibinfo
  {journal} {Phys. Rev. D}\ }\textbf {\bibinfo {volume} {97}},\ \bibinfo
  {pages} {066021} (\bibinfo {year} {2018})},\ \Eprint
  {https://arxiv.org/abs/1712.08148} {arXiv:1712.08148 [gr-qc]} \BibitemShut
  {NoStop}%
\bibitem [{\citenamefont {Agullo}(2015)}]{Agullo:2015aba}%
  \BibitemOpen
  \bibfield  {author} {\bibinfo {author} {\bibfnamefont {I.}~\bibnamefont
  {Agullo}},\ }
  \bibfield  {title} {\bibinfo {title} {Loop quantum cosmology, non-Gaussianity, and CMB power asymmetry},\ }\href {https://doi.org/10.1103/PhysRevD.92.064038} {\bibfield
  {journal} {\bibinfo  {journal} {Phys. Rev. D}\ }\textbf {\bibinfo {volume}
  {92}},\ \bibinfo {pages} {064038} (\bibinfo {year} {2015})},\ \Eprint
  {https://arxiv.org/abs/1507.04703} {arXiv:1507.04703 [gr-qc]} \BibitemShut
  {NoStop}%
\bibitem [{\citenamefont {Zhu}\ \emph {et~al.}(2018)\citenamefont {Zhu},
  \citenamefont {Wang}, \citenamefont {Kirsten}, \citenamefont {Cleaver},\ and\
  \citenamefont {Sheng}}]{Zhu:2017onp}%
  \BibitemOpen
  \bibfield  {author} {\bibinfo {author} {\bibfnamefont {T.}~\bibnamefont
  {Zhu}}, \bibinfo {author} {\bibfnamefont {A.}~\bibnamefont {Wang}}, \bibinfo
  {author} {\bibfnamefont {K.}~\bibnamefont {Kirsten}}, \bibinfo {author}
  {\bibfnamefont {G.}~\bibnamefont {Cleaver}},\ and\ \bibinfo {author}
  {\bibfnamefont {Q.}~\bibnamefont {Sheng}},\ }
  \bibfield  {title} {\bibinfo {title} {Primordial non-Gaussianity and power asymmetry with quantum gravitational effects in loop quantum cosmology},\ }\href
  {https://doi.org/10.1103/PhysRevD.97.043501} {\bibfield  {journal} {\bibinfo
  {journal} {Phys. Rev. D}\ }\textbf {\bibinfo {volume} {97}},\ \bibinfo
  {pages} {043501} (\bibinfo {year} {2018})},\ \Eprint
  {https://arxiv.org/abs/1709.07479} {arXiv:1709.07479 [astro-ph.CO]}
  \BibitemShut {NoStop}%
\bibitem [{\citenamefont {Wu}\ \emph {et~al.}(2018)\citenamefont {Wu},
  \citenamefont {Zhu},\ and\ \citenamefont {Wang}}]{Wu:2018sbr}%
  \BibitemOpen
  \bibfield  {author} {\bibinfo {author} {\bibfnamefont {Q.}~\bibnamefont
  {Wu}}, \bibinfo {author} {\bibfnamefont {T.}~\bibnamefont {Zhu}},\ and\
  \bibinfo {author} {\bibfnamefont {A.}~\bibnamefont {Wang}},\ }
  \bibfield  {title} {\bibinfo {title} {Nonadiabatic evolution of primordial perturbations and non-Gaussianity in hybrid approach of loop quantum cosmology},\ }\href
  {https://doi.org/10.1103/PhysRevD.98.103528} {\bibfield  {journal} {\bibinfo
  {journal} {Phys. Rev. D}\ }\textbf {\bibinfo {volume} {98}},\ \bibinfo
  {pages} {103528} (\bibinfo {year} {2018})},\ \Eprint
  {https://arxiv.org/abs/1809.03172} {arXiv:1809.03172 [gr-qc]} \BibitemShut
  {NoStop}%
\bibitem [{\citenamefont {Agull{\'o}}\ \emph
  {et~al.}(2021{\natexlab{a}})\citenamefont {Agull{\'o}}, \citenamefont
  {Kranas},\ and\ \citenamefont {Sreenath}}]{Agullo:2020cvg}%
  \BibitemOpen
  \bibfield  {author} {\bibinfo {author} {\bibfnamefont {I.}~\bibnamefont
  {Agull{\'o}}}, \bibinfo {author} {\bibfnamefont {D.}~\bibnamefont {Kranas}},\
  and\ \bibinfo {author} {\bibfnamefont {V.}~\bibnamefont {Sreenath}},\ }
  \bibfield  {title} {\bibinfo {title} {Large scale anomalies in the CMB and non-Gaussianity in bouncing cosmologies},\ }\href
  {https://doi.org/10.1088/1361-6382/abc521} {\bibfield  {journal} {\bibinfo
  {journal} {Class. Quant. Grav.}\ }\textbf {\bibinfo {volume} {38}},\ \bibinfo
  {pages} {065010} (\bibinfo {year} {2021}{\natexlab{a}})},\ \Eprint
  {https://arxiv.org/abs/2006.09605} {arXiv:2006.09605 [astro-ph.CO]}
  \BibitemShut {NoStop}%
\bibitem [{\citenamefont {Agull{\'o}}\ \emph
  {et~al.}(2021{\natexlab{b}})\citenamefont {Agull{\'o}}, \citenamefont
  {Kranas},\ and\ \citenamefont {Sreenath}}]{Agullo:2020fbw}%
  \BibitemOpen
  \bibfield  {author} {\bibinfo {author} {\bibfnamefont {I.}~\bibnamefont
  {Agull{\'o}}}, \bibinfo {author} {\bibfnamefont {D.}~\bibnamefont {Kranas}},\
  and\ \bibinfo {author} {\bibfnamefont {V.}~\bibnamefont {Sreenath}},\ }
  \bibfield  {title} {\bibinfo {title} {Anomalies in the CMB from a cosmic bounce},\ }\href
  {https://doi.org/10.1007/s10714-020-02778-9} {\bibfield  {journal} {\bibinfo
  {journal} {Gen. Rel. Grav.}\ }\textbf {\bibinfo {volume} {53}},\ \bibinfo
  {pages} {17} (\bibinfo {year} {2021}{\natexlab{b}})},\ \Eprint
  {https://arxiv.org/abs/2005.01796} {arXiv:2005.01796 [astro-ph.CO]}
  \BibitemShut {NoStop}%
\bibitem [{\citenamefont {Agullo}\ \emph {et~al.}(2021)\citenamefont {Agullo},
  \citenamefont {Kranas},\ and\ \citenamefont {Sreenath}}]{Agullo:2021oqk}%
  \BibitemOpen
  \bibfield  {author} {\bibinfo {author} {\bibfnamefont {I.}~\bibnamefont
  {Agullo}}, \bibinfo {author} {\bibfnamefont {D.}~\bibnamefont {Kranas}},\
  and\ \bibinfo {author} {\bibfnamefont {V.}~\bibnamefont {Sreenath}},\ }
  \bibfield  {title} {\bibinfo {title} {Anomalies in the Cosmic Microwave Background and Their Non-Gaussian Origin in Loop Quantum Cosmology},\ }\href
  {https://doi.org/10.3389/fspas.2021.703845} {\bibfield  {journal} {\bibinfo
  {journal} {Front. Astron. Space Sci.}\ }\textbf {\bibinfo {volume} {8}},\
  \bibinfo {pages} {703845} (\bibinfo {year} {2021})},\ \Eprint
  {https://arxiv.org/abs/2105.12993} {arXiv:2105.12993 [gr-qc]} \BibitemShut
  {NoStop}%
\bibitem [{\citenamefont {{Roshna K}}\ and\ \citenamefont
  {Sreenath}(2023)}]{Roshna:2023}%
  \BibitemOpen
  \bibfield  {author} {\bibinfo {author} {\bibnamefont {{Roshna K}}}\ and\
  \bibinfo {author} {\bibfnamefont {V.}~\bibnamefont {Sreenath}},\ }
  \bibfield  {title} {\bibinfo {title} {Estimation of imprints of the bounce in loop quantum cosmology on the bispectra of cosmic microwave background},\ }\href
  {https://doi.org/10.1088/1475-7516/2023/08/014} {\bibfield  {journal}
  {\bibinfo  {journal} {JCAP}\ }\textbf {\bibinfo {volume} {08}},\ \bibinfo
  {pages} {014} (\bibinfo {year} {2023})},\ \Eprint {https://arxiv.org/abs/2301.05406} {arXiv:2301.05406
  [astro-ph.CO]} \BibitemShut {NoStop}%
\bibitem [{\citenamefont {van Tent}\ \emph {et~al.}(2023)\citenamefont {van
  Tent}, \citenamefont {Delgado},\ and\ \citenamefont {Durrer}}]{vanTent:2023}%
  \BibitemOpen
  \bibfield  {author} {\bibinfo {author} {\bibfnamefont {B.}~\bibnamefont {van
  Tent}}, \bibinfo {author} {\bibfnamefont {P.~C.~M.}\ \bibnamefont
  {Delgado}},\ and\ \bibinfo {author} {\bibfnamefont {R.}~\bibnamefont
  {Durrer}},\ }
  \bibfield  {title} {\bibinfo {title} {Constraining the Bispectrum from Bouncing Cosmologies with Planck},\ }\href {https://doi.org/10.1103/PhysRevLett.130.191002}
  {\bibfield  {journal} {\bibinfo  {journal} {Phys. Rev. Lett.}\ }\textbf
  {\bibinfo {volume} {130}},\ \bibinfo {pages} {191002} (\bibinfo {year}
  {2023})},\ \Eprint {https://arxiv.org/abs/2212.05977} {arXiv:2212.05977
  [astro-ph.CO]} \BibitemShut {NoStop}%
\bibitem [{\citenamefont {Pan}\ \emph {et~al.}(2026{\natexlab{b}})\citenamefont
  {Pan}, \citenamefont {Saeed},\ and\ \citenamefont
  {Wang}}]{PanSaeedWang2026PaperII}%
  \BibitemOpen
  \bibfield  {author} {\bibinfo {author} {\bibfnamefont {R.}~\bibnamefont
  {Pan}}, \bibinfo {author} {\bibfnamefont {J.}~\bibnamefont {Saeed}},\ and\
  \bibinfo {author} {\bibfnamefont {A.}~\bibnamefont {Wang}},\ }
  \bibfield  {title} {\bibinfo {title} {Primordial Non-Gaussianity and Correlation-Function Perturbativity in mLQC-I: The Hybrid Approach},\ } (\bibinfo {year}
  {2026}{\natexlab{b}}),\ 
  \ \Eprint
  {https://arxiv.org/abs/2611.18899} {arXiv:2611.18899 [gr-qc]}\BibitemShut {NoStop}%
\bibitem [{\citenamefont {Garc{\'\i}a-Quismondo}\ and\ \citenamefont
  {Mena~Marug{\'a}n}(2019)}]{Garcia-Quismondo:2019dwa}%
  \BibitemOpen
  \bibfield  {author} {\bibinfo {author} {\bibfnamefont {A.}~\bibnamefont
  {Garc{\'\i}a-Quismondo}}\ and\ \bibinfo {author} {\bibfnamefont {G.~A.}\
  \bibnamefont {Mena~Marug{\'a}n}},\ }
  \bibfield  {title} {\bibinfo {title} {The Martin-Benito-Mena Marugan-Olmedo prescription for the Dapor-Liegener model of Loop Quantum Cosmology},\ }\href
  {https://doi.org/10.1103/PhysRevD.99.083505} {\bibfield  {journal} {\bibinfo
  {journal} {Phys. Rev. D}\ }\textbf {\bibinfo {volume} {99}},\ \bibinfo
  {pages} {083505} (\bibinfo {year} {2019})},\ \Eprint
  {https://arxiv.org/abs/1903.00265} {arXiv:1903.00265 [gr-qc]} \BibitemShut
  {NoStop}%
\bibitem [{\citenamefont {Arroja}\ and\ \citenamefont
  {Koyama}(2008)}]{ArrojaKoyama2008}%
  \BibitemOpen
  \bibfield  {author} {\bibinfo {author} {\bibfnamefont {F.}~\bibnamefont
  {Arroja}}\ and\ \bibinfo {author} {\bibfnamefont {K.}~\bibnamefont
  {Koyama}},\ }
  \bibfield  {title} {\bibinfo {title} {Non-gaussianity from the trispectrum in general single field inflation},\ }\href {https://doi.org/10.1103/PhysRevD.77.083517} {\bibfield
  {journal} {\bibinfo  {journal} {Phys. Rev. D}\ }\textbf {\bibinfo {volume}
  {77}},\ \bibinfo {pages} {083517} (\bibinfo {year} {2008})},\ \Eprint
  {https://arxiv.org/abs/0802.1167} {arXiv:0802.1167 [hep-th]} \BibitemShut
  {NoStop}%
\bibitem [{\citenamefont {Brown}\ \emph {et~al.}(2026)\citenamefont {Brown},
  \citenamefont {Fier}, \citenamefont {Phillips}, \citenamefont {Cleaver},\
  and\ \citenamefont {Wang}}]{BrownEtAl2026EPJC}%
  \BibitemOpen
  \bibfield  {author} {\bibinfo {author} {\bibfnamefont {C.}~\bibnamefont
  {Brown}}, \bibinfo {author} {\bibfnamefont {J.}~\bibnamefont {Fier}},
  \bibinfo {author} {\bibfnamefont {B.}~\bibnamefont {Phillips}}, \bibinfo
  {author} {\bibfnamefont {G.}~\bibnamefont {Cleaver}},\ and\ \bibinfo {author}
  {\bibfnamefont {A.}~\bibnamefont {Wang}},\ }
  \bibfield  {title} {\bibinfo {title} {Effects of the ekpyrotic mechanism on inflationary phase in loop quantum cosmologies},\ }\href
  {https://doi.org/10.1140/epjc/s10052-026-16275-x} {\bibfield  {journal}
  {\bibinfo  {journal} {Eur. Phys. J. C}\ }\textbf {\bibinfo {volume} {86}},\
  \bibinfo {pages} {1006} (\bibinfo {year} {2026})}\BibitemShut {NoStop}%
\bibitem [{\citenamefont {Yogesh}\ \emph {et~al.}(2026)\citenamefont {Yogesh},
  \citenamefont {Li}, \citenamefont {Gangopadhyay},\ and\ \citenamefont
  {Wang}}]{YogeshEtAl2026APSOS}%
  \BibitemOpen
  \bibfield  {author} {\bibinfo {author} {\bibnamefont {Yogesh}}, \bibinfo
  {author} {\bibfnamefont {B.-F.}\ \bibnamefont {Li}}, \bibinfo {author}
  {\bibfnamefont {M.~R.}\ \bibnamefont {Gangopadhyay}},\ and\ \bibinfo {author}
  {\bibfnamefont {A.}~\bibnamefont {Wang}},\ }
  \bibfield  {title} {\bibinfo {title} {Dynamics of reheating in loop quantum cosmology},\ }\href
  {https://doi.org/10.1103/fz2y-fnzp} {\bibfield  {journal} {\bibinfo
  {journal} {APS Open Science}\ }\textbf {\bibinfo {volume} {1}},\ \bibinfo
  {pages} {000120} (\bibinfo {year} {2026})}\BibitemShut {NoStop}%
\bibitem [{\citenamefont {Fier}\ \emph {et~al.}(2026)\citenamefont {Fier},
  \citenamefont {Pan}, \citenamefont {Saeed},\ and\ \citenamefont
  {Wang}}]{Fier:2026abc}%
  \BibitemOpen
  \bibfield  {author} {\bibinfo {author} {\bibfnamefont {J.}~\bibnamefont
  {Fier}}, \bibinfo {author} {\bibfnamefont {R.}~\bibnamefont {Pan}}, \bibinfo
  {author} {\bibfnamefont {J.}~\bibnamefont {Saeed}},\ and\ \bibinfo {author}
  {\bibfnamefont {A.}~\bibnamefont {Wang}},\ }
  \bibfield  {title} {\bibinfo {title} {{Dynamics of Reheating in Modified Loop Quantum Cosmologies: mLQC-I}},\ } (\bibinfo {year} {2026}),\ \bibinfo {note} {{Preprint, \href{https://arxiv.org/abs/2610.12639}{arXiv:2610.12639}}}\BibitemShut {NoStop}%
\bibitem [{\citenamefont {Zhu}\ \emph {et~al.}(2017)\citenamefont {Zhu},
  \citenamefont {Wang}, \citenamefont {Cleaver}, \citenamefont {Kirsten},\ and\
  \citenamefont {Sheng}}]{Zhu:2017jew}%
  \BibitemOpen
  \bibfield  {author} {\bibinfo {author} {\bibfnamefont {T.}~\bibnamefont
  {Zhu}}, \bibinfo {author} {\bibfnamefont {A.}~\bibnamefont {Wang}}, \bibinfo
  {author} {\bibfnamefont {G.}~\bibnamefont {Cleaver}}, \bibinfo {author}
  {\bibfnamefont {K.}~\bibnamefont {Kirsten}},\ and\ \bibinfo {author}
  {\bibfnamefont {Q.}~\bibnamefont {Sheng}},\ }
  \bibfield  {title} {\bibinfo {title} {Pre-inflationary universe in loop quantum cosmology},\ }\href
  {https://doi.org/10.1103/PhysRevD.96.083520} {\bibfield  {journal} {\bibinfo
  {journal} {Phys. Rev. D}\ }\textbf {\bibinfo {volume} {96}},\ \bibinfo
  {pages} {083520} (\bibinfo {year} {2017})},\ \Eprint
  {https://arxiv.org/abs/1705.07544} {arXiv:1705.07544 [gr-qc]} \BibitemShut
  {NoStop}%
\bibitem [{\citenamefont {Akrami}\ \emph {et~al.}(2020)\citenamefont {Akrami}
  \emph {et~al.}}]{Planck:2018jri}%
  \BibitemOpen
  \bibfield  {author} {\bibinfo {author} {\bibfnamefont {Y.}~\bibnamefont
  {Akrami}} \emph {et~al.} (\bibinfo {collaboration} {Planck}),\ }
  \bibfield  {title} {\bibinfo {title} {Planck 2018 results. X. Constraints on inflation},\ }\href
  {https://doi.org/10.1051/0004-6361/201833887} {\bibfield  {journal} {\bibinfo
   {journal} {Astron. Astrophys.}\ }\textbf {\bibinfo {volume} {641}},\
  \bibinfo {pages} {A10} (\bibinfo {year} {2020})},\ \Eprint
  {https://arxiv.org/abs/1807.06211} {arXiv:1807.06211 [astro-ph.CO]}
  \BibitemShut {NoStop}%
\bibitem [{\citenamefont {Louis}\ \emph {et~al.}(2025)\citenamefont {Louis}
  \emph {et~al.}}]{AtacamaCosmologyTelescope:2025blo}%
  \BibitemOpen
  \bibfield  {author} {\bibinfo {author} {\bibfnamefont {T.}~\bibnamefont
  {Louis}} \emph {et~al.} (\bibinfo {collaboration} {Atacama Cosmology
  Telescope}),\ }
  \bibfield  {title} {\bibinfo {title} {The Atacama Cosmology Telescope: DR6 power spectra, likelihoods and {\ensuremath{\Lambda}}CDM parameters},\ }\href {https://doi.org/10.1088/1475-7516/2025/11/062}
  {\bibfield  {journal} {\bibinfo  {journal} {JCAP}\ }\textbf {\bibinfo
  {volume} {11}},\ \bibinfo {pages} {062} (\bibinfo {year} {2025})},\ \Eprint
  {https://arxiv.org/abs/2503.14452} {arXiv:2503.14452 [astro-ph.CO]}
  \BibitemShut {NoStop}%
\end{thebibliography}
\end{document}